\pdfoutput=1
\documentclass[acmsmall,authorversion,nonacm]{acmart}

\usepackage{tikz}
\usetikzlibrary{positioning, fit, calc, intersections, shapes.geometric, arrows}
\usepackage{paralist}
\usepackage{colortbl}
\usepackage{tabu}
\usepackage{amsmath}
\usepackage{amssymb}
\usepackage{fancyvrb}
\usepackage{subfig}
\usepackage{multirow}
\usepackage{adjustbox}
\usepackage[b]{esvect}
\usepackage{wrapfig}
\usepackage{unison}

\newcommand{\columnWidth}{3cm}
\newcommand{\noMathVar}[2]{\operatorname{#1}(#2)}
\newcommand{\var}[2]{$\noMathVar{#1}{#2}$}

\newcommand{\ZZ}{\ensuremath{\mathbb Z}}
\newcommand{\RR}{\ensuremath{\mathbb R}}
\newcommand{\summationSup}[3]{\displaystyle\sum_{\mathclap{#1}}^{#2}#3}
\newcommand{\finiteDomainNumbers}{\ensuremath{\mathbb D}}

\newcommand{\textSummation}[2]{\textstyle\sum_{#1}#2}
\newcommand{\exitWeightSymbol}{\operatorname{weight}}
\newcommand{\exitWeight}[1]{\exitWeightSymbol(#1)}
\newcommand{\exitInstructionSymbol}{\operatorname{exit}}
\newcommand{\exitInstruction}[1]{\exitInstructionSymbol(#1)}
\newcommand{\entryInstructionSymbol}{\operatorname{entry}}
\newcommand{\entryInstruction}[1]{\entryInstructionSymbol(#1)}

\acmJournal{CSUR}

\newcommand{\sics}{%
\affiliation{%
  \institution{RISE SICS}
  \streetaddress{Box 1263}
  \city{Kista}
  \postcode{164 40}
  \country{Sweden}}}

\newcommand{\kth}{%
\affiliation{%
  \institution{KTH Royal Institute of Technology}
  \department{School of Electrical Engineering and Computer
    Science}
  \streetaddress{Electrum 229}
  \city{Kista}
  \postcode{164 40}
  \country{Sweden}}}

\begin{document}

% Page heads
\markboth{R.~Casta\~{n}eda Lozano and C.~Schulte}{Survey on
  Combinatorial Register Allocation and Instruction Scheduling}

\title{Survey on Combinatorial Register Allocation and Instruction
  Scheduling}

\author{Roberto Casta\~{n}eda Lozano}
\orcid{0000-0002-2806-7333}
\sics{}
\kth{}
\email{roberto.castaneda@ri.se}

\author{Christian Schulte}
\orcid{0000-0002-6283-7004}
\kth{}
\sics{}
\email{cschulte@kth.se}

\begin{abstract}
  Register allocation (mapping variables to processor registers or
  memory) and instruction scheduling (reordering instructions to
  increase instruction-level parallelism) are essential tasks for
  generating efficient assembly code in a compiler.
  In the last three decades, combinatorial optimization has emerged as
  an alternative to traditional, heuristic algorithms for these two
  tasks.
  Combinatorial optimization approaches can deliver optimal solutions
  according to a model, can precisely capture trade-offs between
  conflicting decisions, and are more flexible at the expense of
  increased compilation time.

  This paper provides an exhaustive literature review and a
  classification of combinatorial optimization approaches to register
  allocation and instruction scheduling, with a focus on the
  techniques that are most applied in this context: integer
  programming, constraint programming, partitioned Boolean quadratic
  programming, and enumeration.
  Researchers in compilers and combinatorial optimization can benefit
  from identifying developments, trends, and challenges in the area;
  compiler practitioners may discern opportunities and grasp the
  potential benefit of applying combinatorial optimization.
\end{abstract}

%
% The code below is generated by the tool at
% http://dl.acm.org/ccs.cfm
%
 \begin{CCSXML}
<ccs2012>
<concept>
<concept_id>10002944.10011122.10002945</concept_id>
<concept_desc>General and reference~Surveys and overviews</concept_desc>
<concept_significance>500</concept_significance>
</concept>
<concept>
<concept_id>10003752.10003790.10003795</concept_id>
<concept_desc>Theory of computation~Constraint and logic programming</concept_desc>
<concept_significance>500</concept_significance>
</concept>
<concept>
<concept_id>10003752.10003809.10003716</concept_id>
<concept_desc>Theory of computation~Mathematical optimization</concept_desc>
<concept_significance>500</concept_significance>
</concept>
<concept>
<concept_id>10003752.10003809.10011254</concept_id>
<concept_desc>Theory of computation~Algorithm design techniques</concept_desc>
<concept_significance>300</concept_significance>
</concept>
<concept>
<concept_id>10011007.10011006.10011008.10011009.10011015</concept_id>
<concept_desc>Software and its engineering~Constraint and logic languages</concept_desc>
<concept_significance>500</concept_significance>
</concept>
<concept>
<concept_id>10011007.10011006.10011041.10011043</concept_id>
<concept_desc>Software and its engineering~Retargetable compilers</concept_desc>
<concept_significance>500</concept_significance>
</concept>
<concept>
<concept_id>10011007.10011006.10011008.10011009.10011020</concept_id>
<concept_desc>Software and its engineering~Assembly languages</concept_desc>
<concept_significance>300</concept_significance>
</concept>
</ccs2012>
\end{CCSXML}

\ccsdesc[500]{General and reference~Surveys and overviews}
\ccsdesc[500]{Software and its engineering~Retargetable compilers}
\ccsdesc[300]{Software and its engineering~Assembly languages}
\ccsdesc[500]{Theory of computation~Constraint and logic programming}
\ccsdesc[500]{Theory of computation~Mathematical optimization}
\ccsdesc[300]{Theory of computation~Algorithm design techniques}

%
% End generated code
%

\keywords{Combinatorial optimization, register allocation, instruction scheduling}

\thanks{%
This article is a revised and extended version of a technical
report~\cite{Castaneda2014b}.}

\maketitle

\section{Introduction}\label{sec:introduction}

Compiler back-ends take an intermediate representation (IR) of a
program and generate assembly code for a particular processor.
The main tasks in a back-end are instruction selection, register
allocation, and instruction scheduling.
Instruction selection implements abstract operations with processor
instructions.
Register allocation maps temporaries (program and compiler-generated
variables in the IR) to processor registers or to memory.
Instruction scheduling reorders instructions to improve the total
latency or throughput.
This survey is concerned with combinatorial approaches (explained in
this section) for register allocation and instruction scheduling.
Combinatorial instruction selection approaches are reviewed
elsewhere~\cite{HjortBlindell2016}.

Register allocation and instruction scheduling are of paramount
importance to optimizing
compilers~\cite{Fisher1983,Nandivada2007,Hennessy2011}.
In general, problems for these tasks are computationally complex
(NP-hard) and interdependent: the solution to one of them affects the
other~\cite{Goodman1988}.
Solving instruction scheduling first tends to increase the
\emph{register pressure} (number of temporaries that need to be stored
simultaneously), which may degrade the result of register allocation.
Conversely, solving register allocation first tends to increase the
reuse of registers, which introduces additional dependencies between
instructions and may degrade the result of instruction
scheduling~\cite{Govindarajan2007}.

\begin{figure}
  \centering
  \scalebox{1}{\input{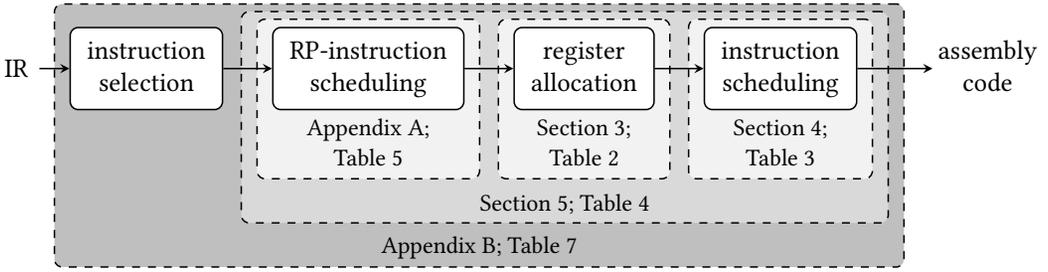}%
\pgfdeclarelayer{layer0}%
\pgfdeclarelayer{layer1}%
\pgfdeclarelayer{layer2}%
\pgfdeclarelayer{layer3}%
\pgfdeclarelayer{layer4}%
\pgfdeclarelayer{layer5}%
\pgfsetlayers{layer0,layer1,layer2,layer3,layer4,layer5}%
\begin{tikzpicture}%
% Unison
  \begin{pgfonlayer}{layer5}
    \node [input, inner xsep=0.5mm, outer xsep=1mm] (input) {IR};
    \node [problem, right=0.4cm of input] (isel) {%
      \begin{tabular}{c}
        instruction\\
        selection
      \end{tabular}
    };
    \node [problem, right=0.65cm of isel] (rpisched) {%
      \begin{tabular}{c}
        RP-instruction\\
        scheduling
      \end{tabular}
    };
    \node [problem, right=0.65cm of rpisched] (regalloc) {%
      \begin{tabular}{c}
        register\\
        allocation
      \end{tabular}
    };
    \node [problem, right=0.65cm of regalloc] (isched) {%
      \begin{tabular}{c}
        instruction\\
        scheduling
      \end{tabular}
    };
    \node [output, right=1cm of isched, inner xsep=0mm, outer xsep=-1mm] (output) {%
      \begin{tabular}{c}
        assembly\\
        code
      \end{tabular}
    };
  \end{pgfonlayer}

  \begin{pgfonlayer}{layer4}
    \node [problem scope, fill=tempcolor3, fit= (rpisched)] (rpisched-scope) {};
    \node [problem scope label, above=of rpisched-scope.south] (rpisched-label)
          {\begin{tabular}{c}Appendix~\ref{app:rp-instruction-scheduling};\\Table~\ref{tab:rp-instruction-scheduling}\end{tabular}};
    \node [problem scope, fill=tempcolor3, fit= (regalloc)] (regalloc-scope) {};
    \node [problem scope label, above=of regalloc-scope.south] (regalloc-label)
          {\begin{tabular}{c}Section~\ref{sec:register-allocation};\\Table~\ref{tab:register-allocation}\end{tabular}};
    \node [problem scope, fill=tempcolor3, fit= (isched)] (isched-scope) {};
    \node [problem scope label, above=of isched-scope.south] (isched-label)
          {\begin{tabular}{c}Section~\ref{sec:instruction-scheduling};\\Table~\ref{tab:instruction-scheduling}\end{tabular}};
  \end{pgfonlayer}

  \begin{pgfonlayer}{layer3}
    \node [problem scope, fill=tempcolor1, fit= (rpisched-scope) (isched-scope), inner ysep=3.3mm, yshift=1.6mm] (regalloc-isched-scope) {};
    \node [problem scope label, above=of regalloc-isched-scope.south, yshift=0.5mm] (regalloc-isched-label)
          {Section~\ref{sec:integrated}; Table~\ref{tab:integrated}};
  \end{pgfonlayer}

  \begin{pgfonlayer}{layer2}
    \node [problem scope, fill=tempcolor4, fit= (isel) (regalloc-isched-scope), inner ysep=3.3mm, yshift=1.6mm] (isel-regalloc-isched-scope) {};
    \node [problem scope label, above=of isel-regalloc-isched-scope.south, yshift=0.5mm] (isel-regalloc-isched-label)
          {Appendix~\ref{app:full-integration}; Table~\ref{tab:fully-integrated}};
  \end{pgfonlayer}

  \begin{pgfonlayer}{layer5}
    \begin{scope}[flow]
      \draw (input) -- (isel);
      \draw (isel) -- (rpisched);
      \draw (rpisched) -- (regalloc);
      \draw (regalloc) -- (isched);
      \draw (isched) -- (output);
    \end{scope}
  \end{pgfonlayer}

\end{tikzpicture}}
  \caption{Compiler back-end with section and table references.}
  \label{fig:backend}
\end{figure}

\paragraph{Heuristic approaches}

Traditional back-ends solve each problem in isolation with custom
heuristic algorithms, which take a sequence of greedy decisions based
on local optimization criteria.
This arrangement makes traditional back-ends fast but precludes
solving the problems optimally and complicates exploiting irregular
architectures.
Classic heuristic algorithms are graph coloring~\cite{Chaitin1981} for
register allocation and list scheduling~\cite{Rau1993} for instruction
scheduling.
A typical scheme to partially account for the interdependencies
between instruction scheduling and register allocation in this setup
is to solve a register pressure (\emph{RP})-aware version of
instruction scheduling before register allocation~\cite{Goodman1988}
as shown in Figure~\ref{fig:backend}.
Heuristic algorithms that further approximate this integration have
also been
proposed~\cite{Bradlee1991,Pinter1993,Motwani1995,Berson1996}.

\paragraph{Combinatorial approaches}

Numerous approaches that use combinatorial optimization techniques to
overcome the limitations in traditional back-ends have been presented
starting in the 1980s~\cite{Kremer1997}.
Combinatorial approaches can solve compiler back-end problems
optimally according to a model at the expense of increased compilation
time, and their declarative nature provides more flexibility.
The accuracy with which a combinatorial approach models its problem is
key as the computed solutions are only optimal with respect to the
model rather than the problem itself.
Recent progress in optimization technology and improved understanding
of the structure of back-end problems allow us today to solve
optimally register allocation and instruction scheduling problems of
practical size in the order of seconds, as this survey illustrates.
Furthermore, combinatorial approaches can precisely capture the
interdependencies between different back-end problems to generate even
better solutions, although doing it efficiently remains a major
computational challenge.
Combinatorial approaches might never fully replace traditional
approaches due to their high computation cost, however they can act as
a complement rather than a replacement. Given that combinatorial
approaches precisely capture interdependencies, they can be used to
experiment with new ideas as well as evaluate and possibly improve
existing heuristics used in traditional approaches. For example,
Ericsson uses UNISON (see Section~\ref{sec:unison} for a discussion),
for that purpose as can be seen from an entry on their research
blog~\cite{ericsson-blog}.

For consistency and ease of comparison, this survey focuses on
combinatorial techniques that use a general-purpose modeling language.
These include integer programming~\cite{Nemhauser1999}, constraint
programming~\cite{CPH:2006}, and partitioned Boolean quadratic
programming~\cite{Scholz2002}.
A uniform treatment of integer programming and constraint programming
is offered by Hooker~\cite{Hooker:2012}.
For completeness, the survey also includes the most prominent
special-purpose enumeration techniques, which are often founded on
methods such as dynamic programming~\cite{Cormen2009} and
branch-and-bound search~\cite{Nemhauser1999}.

\paragraph{Contributions}

This paper reviews and classifies combinatorial optimization
approaches to register allocation and instruction scheduling.
It is primarily addressed to researchers in compilers and
combinatorial optimization who can benefit from identifying
developments, trends, and challenges in the area; but may also help
compiler practitioners to discern opportunities and grasp the
potential benefit of applying combinatorial optimization.
To serve these goals, the survey contributes:
\begin{itemize}
\item an overview of combinatorial optimization techniques used for
  register allocation and instruction scheduling with a focus on the
  most relevant aspects for these problems
  (Section~\ref{sec:combinatorial-optimization});
\item an exhaustive literature review of combinatorial approaches for
  register allocation (Section~\ref{sec:register-allocation}),
  instruction scheduling (Section~\ref{sec:instruction-scheduling}),
  and the integration of both problems (Section~\ref{sec:integrated});
  and
\item a classification of the reviewed approaches
  (Tables~\ref{tab:register-allocation},
  \ref{tab:instruction-scheduling}, and~\ref{tab:integrated}) based on
  technique, scope, problem coverage, approximate scalability, and
  evaluation method.
\end{itemize}

In addition, Appendices~\ref{app:rp-instruction-scheduling}
and~\ref{app:full-integration} review and classify
(Tables~\ref{tab:rp-instruction-scheduling}
and~\ref{tab:fully-integrated}) register pressure-aware instruction
scheduling and integrated approaches to the three compiler back-end
tasks.
The paper complements available surveys of register
allocation~\cite{Johnsson1973,Rajlich1984,Nandivada2007b,PereiraSurvey2008,Protzenko2009},
instruction
scheduling~\cite{Rau1993,Allan1995,DeDinechin2004,Govindarajan2007,Rong2007},
and integrated code generation~\cite{Kessler2010}, whose focus tends
to be on heuristic approaches.

\section{Combinatorial Optimization}\label{sec:combinatorial-optimization}

Combinatorial optimization is a collection of \emph{complete}
techniques to solve combinatorial problems.
\emph{Combinatorial} refers to the problems' nature that the value
combinations in their solutions must satisfy properties that are
mutually interdependent. Not all combinatorial optimization problems
are NP-hard, even though general scheduling and register allocation
problems are. Relaxations of these problems, for example by dropping
the optimality requirement, might also be solvable in polynomial time.

Complete techniques automatically explore the full solution space and
guarantee to eventually find the optimal solution to a combinatorial
problem -- or prove that there is no solution at all.
For consistency and ease of comparison among different approaches,
this survey focuses on those combinatorial optimization techniques
that provide support for describing the problem at hand with a
general-purpose modeling language.
This category comprises a wide range of techniques often presenting
complementary strengths as illustrated in this survey.
Those that are most commonly applied to code generation are Integer
Programming (IP), Constraint Programming (CP), and, to a lesser
extent, Partitioned Boolean Quadratic Programming (PBQP).
This section reviews the modeling and solving aspects of these
techniques, as well as the common solving methods in special-purpose
enumeration techniques.

Section~\ref{sec:modeling} presents the modeling language provided by
IP, CP, and PBQP.
Section~\ref{sec:solving} describes the main solving methods of each
combinatorial technique with a focus on methods applied by the
reviewed approaches.

\subsection{Modeling}\label{sec:modeling}

Combinatorial models consist, regardless of the particular
optimization technique discussed in this survey, of \emph{variables},
\emph{constraints}, and an \emph{objective function}.
\emph{Variables} capture decisions that are combined to form a
solution to a problem.
Variables can take values from different domains (for example,
integers~$\ZZ$ or subsets of integers such as Booleans~$\set{0,1}$).
The variables in a model are denoted here as $x_1, x_2, \ldots, x_n$.
\emph{Constraints} are relations over the values for the variables
that must hold for a solution to a problem.
The set of constraints in a model defines all legal combinations of
values for its variables.
The types of constraints that can be used depend on each combinatorial
optimization technique.
The \emph{objective function} is an expression on the model variables
to be minimized by the solving method.
We assume without loss of generality that the objective function is to
be minimized.
The term \emph{model} in this survey refers to combinatorial models
unless otherwise stated.

\newcommand{\technique}[4]{#1 & #2 & #3 & #4\\}
\newcommand{\techniqueNotes}[1]{& \multicolumn{3}{c}{(#1)}\\}

\begin{table}[b]%
  \setlength{\tabcolsep}{0.2cm}
  \setlength{\tabulinesep}{0.1cm}
  \caption{Modeling elements for different techniques.\label{tab:techniques}}
    \begin{tabu}{%
        l%
        c%
        c%
        c%
      }\hline
      \rowfont{\bfseries}
      \technique{technique}{variables}{constraints}{objective function}
      \hline
      \technique
          {IP}
          {$x_i \in \ZZ$}
          {$\summationSup{i = 1}{n}{a_i x_i} \le b$}
          {$\summationSup{i = 1}{n}{c_i x_i}$}
          \techniqueNotes{$a_i, b, c_i\in\ZZ$ are constant coefficients}
      \technique
          {CP}
          {$x_i \in \finiteDomainNumbers{}$}
          {any $r(x_1, x_2, \ldots, x_n)$}
          {any $f(x_1, x_2, \ldots, x_n)$}
          \techniqueNotes{$\finiteDomainNumbers{}\subset\ZZ$ is a
            finite subset of the integers}
      \technique
          {PBQP}
          {$x_i \in \finiteDomainNumbers{}$}
          {\mbox{none}}
          {$\summationSup{i = 1}{n}{c(x_i)} + \summationSup{i,j = 1}{n}{C(x_i, x_j)}$}
          \techniqueNotes{$\finiteDomainNumbers{}\subset\ZZ$ is a
            finite int.~subset; $c(x_i)$ is the cost of $x_i$; $C(x_i, x_j)$ is the cost of $x_i \land x_j$}
  \end{tabu}%
\end{table}

\paragraph{Integer Programming (IP)}

IP is a special case of Linear Programming (LP)~\cite{Vanderbei2013}
where the variables range over integer values, the constraints are
linear inequalities (which can also express linear equalities), and
the objective function is linear as shown in
Table~\ref{tab:techniques}.
Most compiler applications use bounded variables (with known lower and
upper bounds that are parametric with respect to the specific problem
being solved) and variables which range over $\set{0,1}$ (called
\mbox{0-1} variables).
IP models are often called \emph{formulations} in the literature.
For an overview, see for example the classic introduction by Nemhauser
and Wolsey~\cite{Nemhauser1999}.

\paragraph{Constraint Programming (CP)}

CP models can be seen as a generalization of bounded IP models where
the variables take values from a finite subset
$\finiteDomainNumbers\subset\ZZ$ of the integers (including 0-1
variables), and the constraints and the objective function are
expressed by general relations.
CP typically supports a rich set of constraints
over~$\finiteDomainNumbers$ including arithmetic and logical
constraints but also constraints to model more specific subproblems
such as assignment, scheduling, graphs, and bin-packing.  Often, these
more specific constraints are referred to as \emph{global constraints}
that express recurrent substructures involving several
variables. Global constraints are convenient for modeling, but more
importantly, are key to solving as these constraints have
constraint-specific efficient and powerful implementations.
The solution to a CP model is an assignment of values to the variables
such that all constraints are satisfied.
More information on CP can be found in a handbook edited by
Rossi~\etal{}~\cite{CPH:2006}.

\paragraph{Partitioned Boolean Quadratic Programming (PBQP)}

PBQP is a special case of the Quadratic Assignment
Problem~\cite{Koopmans1957} that was specifically developed to solve
compiler problems with constraints involving up to two variables at a
time~\cite{Scholz2002,Eckstein2003,Eckstein2003b}.
As such, it is not as widely spread as other combinatorial
optimization techniques such as IP and CP, but this section presents
it at the same level for uniformity.
As with CP, variables range over a finite subset
$\finiteDomainNumbers\subset\ZZ$ of the integers.
However, PBQP models do not explicitly formulate constraints but
define problems by a quadratic cost function.
Each single variable assignment $x_i$ is given a cost $c(x_i)$ and
each pair of variable assignments $x_i \land x_j$ is given a cost
$C(x_i, x_j)$.
Single assignments and pairs of assignments can then be forbidden by
setting their cost to a conceptually infinite value.
The objective function is the combination of the cost of each single
assignment and the cost of each pair of assignments as shown in
Table~\ref{tab:techniques}.
PBQP is described by Scholz~\etal{}~\cite{Scholz2002,Hames2006}, more
background information can be found in Eckstein's doctoral
dissertation~\cite{Eckstein2003c}.

\subsection{Solving Methods}\label{sec:solving}

\paragraph{Integer Programming}

The most common approach for IP solvers is to exploit \emph{linear
  relaxations} and \emph{branch-and-bound search}.
State-of-the-art solvers, however, exploit numerous other
methods~\cite{Nemhauser1999}.
A first step computes the optimal solution to a relaxed LP problem,
where the variables can take any value from the set~$\RR$ of real
numbers.
LP relaxations can be derived directly from the IP models as these
only contain linear constraints, and are computed efficiently.
If all the variables in the solution to the LP problem are integers
(they are said to be integral), the optimal solution to the LP
relaxation is also optimal for the original IP model.
Otherwise, the basic approach is to use branch-and-bound search that
decomposes the problem into alternative subproblems in which a
non-integral variable is assigned different integer values and the
process is repeated. Modern solvers use a number of improvements such
as cutting-plane methods, in particular Gomory cuts, that add linear
inequalities to remove non-integer parts of the search
space~\cite{Nemhauser1999}.
LP relaxations provide lower bounds on the objective function which
are used to prove optimality.
Solutions found during solving provide upper bounds which are used to
discard subproblems that cannot produce better solutions.

\paragraph{Constraint Programming}

CP solvers typically proceed by interleaving \emph{constraint
  propagation} and \emph{branch-and-bound search}.
Constraint propagation reduces the search space by discarding values
for variables that cannot be part of any solution.
Constraint propagation discards values for each constraint in the
model iteratively until no more values can be
discarded~\cite{CPH:prop}.
Global constraints play a key role in solving as they are implemented
by particularly efficient and effective propagation
algorithms~\cite{CPH:global}.
A key application area for CP is scheduling, in particular variants of
cumulative scheduling problems where the tasks to be scheduled cannot
exceed the capacity of a resource used by the
tasks~\cite{Baptiste2001,CPH:scheduling}. These problems are
captured by global scheduling constraints and implemented
by efficient algorithms providing strong propagation.
When no further propagation is possible, search tries several
alternatives on which constraint propagation and search is repeated.
The alternatives in search typically follow a heuristic to reduce the
search space.
As with IP solving, valid solutions found during solving are exploited
by branch-and-bound search to reduce the search
space~\cite{CPH:search}.

\paragraph{Partitioned Boolean Quadratic Programming}

Optimal PBQP solvers interleave \emph{reduction} and
\emph{branch-and-bound search}~\cite{Hames2006}.
Reduction transforms the original problem by iteratively applying a
set of rules that eliminate one \emph{reducible} variable at a time.
Reducible variables are those related to at most two other variables
by non-zero costs.
If at the end of reduction the objective function becomes trivial
(that is, only the costs of single assignments $c(x_i)$ remain), a
solution is obtained.
Otherwise, branch-and-bound search derives a set of alternative PBQP
subproblems on which the process is recursively repeated.
The branch-and-bound method maintains lower and upper bounds on the
objective function to prove optimality and discard subproblems as the
search goes.

\paragraph{Properties and expressiveness}

The solving methods for IP, CP, and PBQP all rely on branch-and-bound
search. All techniques are in principle designed to be
\emph{complete}, that is, to find the best solution with respect to
the model and objective function and to prove its optimality. However,
all three approaches also support \emph{anytime behavior}: the search
finds solutions with increasing quality and can be interrupted at any
time. The more time is allocated for solving, the better the found
solution is.

The three techniques offer different trade-offs between the
expressiveness of their respective modeling languages and their
typical strength and weaknesses in solving.

IP profits from its regular and simple modeling language in its
solving methods that exploit its regularity. For example, Gomory cuts
generated during solving are linear inequalities themselves. IP is in
general good at proving optimality due to its simple language and rich
collection of global methods, in particular relaxation and
cutting-plane methods. However, the restricted expressiveness of the
modeling language can sometimes result in large models, both in the
number of variables as well as in the number of constraints. A typical
example are scheduling problems which need to capture the order among
tasks to be scheduled. Ordering requires disjunctions which are
difficult to express concisely and can reduce the strength of the
relaxation methods.

CP has somewhat complementary properties. CP is good in capturing
structure in problems, typically by global constraints, due to its
more expressive language.  The individual structures are efficiently
exploited for propagation algorithms specialized for a particular
global constraint. However, CP has limited search capabilities
compared to IP. For example, there is no natural equivalent to a
Gomory cut as the language is diverse and not regular and there is no
general concept of relaxation. Recent approaches try to alleviate this
restriction using methods from SAT (Boolean
satisfiability)~\cite{lazy-clause} techniques. CP is in general less
effective at optimization. It might find a first solution quickly, but
proving optimality can be challenging.

PBQP is the least explored technique and has been mostly applied to
problems in compilation. Its trade-offs are not obvious as it does not
offer any constraints but captures the problem by the objective
function. In a sense, it offers a hybrid approach as optimality for
the objective function can be relaxed and hence the approach turns
into a heuristic. Or it is used together with branch and bound search
which makes it complete while retaining anytime behavior.

\paragraph{Special-purpose enumeration}

Special-purpose enumeration techniques define and explore a search
tree where each node represents a partial solution to the problem.
The focus of these techniques is usually in exploiting
problem-specific properties to reduce the amount of nodes that need to
be explored, rather than relying in a general-purpose framework such
as IP, CP, or PBQP.
Typical methods include merging equivalent partial
solutions~\cite{Kessler2001} in a similar manner to \emph{dynamic
  programming}~\cite{Cormen2009}, detection of \emph{dominated}
decisions that are not essential in optimal
solutions~\cite{Ramamoorthy1972}, branch-and-bound
search~\cite{Shobaki2004}, computation of lower
bounds~\cite{Rim1994,Langevin1996}, and feasibility checks similar to
constraint propagation in CP~\cite{Shobaki2004}.
Developing special-purpose enumeration techniques incurs a significant
cost but provides high flexibility in implementing and combining
different solving methods.
For example, while CP typically explores search trees in a depth-first
search fashion, merging equivalent partial solutions requires
breadth-first search~\cite{Kessler2001}.

\section{Register Allocation}\label{sec:register-allocation}

Register allocation takes as input a function where instructions of a
particular processor have been selected.
Functions are usually represented by their control-flow graph (CFG).
A basic block in the CFG is a straight-line sequence of
\emph{instructions} without branches from or into the middle of the
sequence.
Instructions use and define \emph{temporaries}.
Temporaries are storage locations holding values corresponding to
program and compiler-generated variables in the IR.

A \emph{program point} is located between two consecutive instructions.
A temporary~$t$ is \emph{live} at a program point if $t$ holds a value
that might be used in the future.
The \emph{live range} of a temporary~$t$ is the set of program points
where $t$ is live.
Two temporaries holding different values \emph{interfere} if their
live ranges overlap.

\begin{figure}
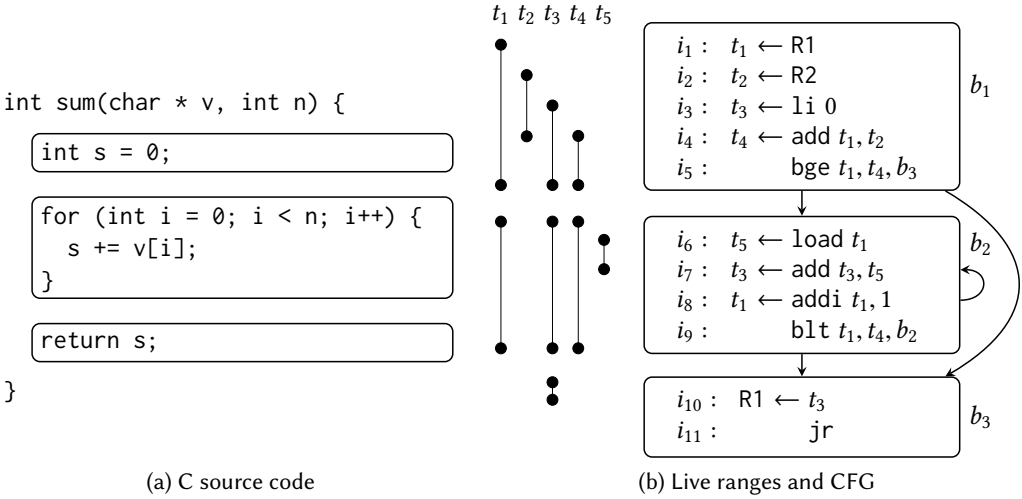

  \subfloat[C source code]{%
    \adjustbox{trim=0.85cm 0cm 0cm 0cm,clip=true}{\input{./figures/styles}%
\begin{tikzpicture}
    \node (bin) {};
    \node [below=0.4cm of bin] (b1) {};
    \node [below=1cm of b1]  (b2) {};
    \node [below=1cm of b2]  (b3) {};
    \node [below=0.4cm of b3]  (bout) {};
    \node [right=of bin, xshift=-0.5cm] (blockin) {\BUseVerbatim{sum-in}};
    \node [sum cfg block, right=of b1, minimum height=0.5cm] (block1) {};
    \node [below right] at (block1.north west) {\BUseVerbatim{sum-b1}};
    \node [sum cfg block, right=of b2, minimum height=1.33cm] (block2) {};
    \node [below right] at (block2.north west) {\BUseVerbatim{sum-b2}};
    \node [sum cfg block, right=of b3, minimum height=0.5cm] (block3) {};
    \node [below right] at (block3.north west) {\BUseVerbatim{sum-b3}};
    \node [right=of bout, xshift=-0.5cm] (blockout) {\BUseVerbatim{sum-out}};
    \node [below=0.5cm of bout] (foo) {};
\end{tikzpicture}}
    \label{fig:sum-c}
  }
  \hspace{0.14cm}
  \subfloat[Live ranges and CFG]{%
    \adjustbox{trim=0.1cm 0cm 0.35cm 0cm,clip=true}{\input{./figures/styles}%
\begin{tikzpicture}
    \node (b1-label) {};
    \node [below=0.7cm of b1-label] (b1) {};
    \node [below=0.7cm of b1] (b2-label) {};
    \node [below=1.15cm of b2-label] (b2) {};
    \node [below=0.7cm of b2] (b3-label) {};
    \node [below=0.54cm of b3-label] (b3) {};
    \node [cfg block, right=of b1, minimum height=2.2cm] (block1) {};
    \node [above right] at (block1.east) {$\B{1}$};
    \node [below right] at (block1.north west) {%
      \begin{labeledBasicBlock}
        \InstrFstMove{}: & \;\; \simpleCopyOperation{\Temp{1}}{\register{R1}}
        \InstrSndMove{}: & \simpleCopyOperation{\Temp{2}}{\register{R2}}
        \InstrLi{}: & \naturalOperation{\Temp{3}}{\loadImmediateInstruction{}}{0}
        \InstrAddAddr{}: & \naturalOperation{\Temp{4}}{\addInstruction{}}{\Temp{1}, \Temp{2}}
        \InstrBge{}: & \branchOperation{\jumpIfGEInstruction{}}{\Temp{1}, \Temp{4}, \B{3}}
      \end{labeledBasicBlock}
    };
    \node [cfg block, right=of b2, minimum height=1.8cm] (block2) {};
    \node [below right, yshift=-0.1cm] at (block2.north east) {$\B{2}$};
    \node [below right] at (block2.north west) {%
      \begin{labeledBasicBlock}
        \InstrLoad{}: & \;\; \naturalOperation{\Temp{5}}{\loadInstruction{}}{\Temp{1}}
        \InstrAdd{}: & \naturalOperation{\Temp{3}}{\addInstruction{}}{\Temp{3}, \Temp{5}}
        \InstrInc{}: & \naturalOperation{\Temp{1}}{\addiInstruction{}}{\Temp{1}, 1}
        \InstrBlt{}: & \branchOperation{\jumpIfLTInstruction{}}{\Temp{1}, \Temp{4}, \B{2}}
      \end{labeledBasicBlock}
    };
    \node [cfg block, right=of b3, minimum height=1.05cm] (block3) {};
    \node [right] at (block3.east) {$\B{3}$};
    \node [below right] at (block3.north west) {%
      \begin{labeledBasicBlock}
        \InstrRetMove{}: & \;\; \simpleCopyOperation{\register{R1}}{\Temp{3}}
        \InstrJr{}: & \branchOperation{\returnInstruction{}}{}
      \end{labeledBasicBlock}
    };
    \begin{scope}[flow]
      \draw (block1) -- (block2);
      \draw ([xshift=-0.2cm]block1.south east) to[bend left=60, distance=1.5cm] ([xshift=-0.2cm]block3.north east);
      \draw ([yshift=-0.2cm]block2.east) to[bend left=-90, distance=0.4cm] ([yshift=0.2cm]block2.east);
      \draw (block2) -- (block3);
    \end{scope}

    \node [left=1.8cm of block1.north west,yshift=0.1cm,xshift=0.15cm] (t1) {$\Temp{1}$};
    \coordinate [yshift=-0.22cm] (ls11) at (block1.north west -| t1);
    \coordinate (le11) at (block1.south west -| t1);
    \coordinate (ls12) at (block2.north west -| t1);
    \coordinate (le12) at (block2.south west -| t1);

    \begin{scope}[live range]
      \draw (ls11.center) -- (le11.center);
      \draw (ls12.center) -- (le12.center);
    \end{scope}

    \node [right=-0.15cm of t1] (t2) {$\Temp{2}$};
    \coordinate [yshift=-0.4cm] (ls21) at (ls11 -| t2);
    \coordinate [yshift=0.64cm] (le21) at (le11 -| t2);

    \begin{scope}[live range]
      \draw (ls21.center) -- (le21.center);
    \end{scope}

    \node [right=-0.15cm of t2] (t3) {$\Temp{3}$};
    \coordinate [yshift=-0.4cm] (ls31)  at (ls21 -| t3);
    \coordinate (le31) at (block1.south west -| t3);
    \coordinate (ls32) at (block2.north west -| t3);
    \coordinate (le32) at (block2.south west -| t3);
    \coordinate (ls33) at (block3.north west -| t3);
    \coordinate [yshift=0.685cm] (le33) at (block3.south west -| t3);

    \begin{scope}[live range]
      \draw (ls31.center) -- (le31.center);
      \draw (ls32.center) -- (le32.center);
      \draw (ls33.center) -- (le33.center);
    \end{scope}

    \node [right=-0.15cm of t3] (t4) {$\Temp{4}$};
    \coordinate [yshift=-0.4cm] (ls41) at (ls31 -| t4);
    \coordinate (le41) at (block1.south west -| t4);
    \coordinate (ls42) at (block2.north west -| t4);
    \coordinate (le42) at (block2.south west -| t4);

    \begin{scope}[live range]
      \draw (ls41.center) -- (le41.center);
      \draw (ls42.center) -- (le42.center);
    \end{scope}

    \node [right=-0.15cm of t4] (t5) {$\Temp{5}$};
    \coordinate [yshift=-0.24cm] (ls52) at (block2.north west -| t5);
    \coordinate [yshift=-0.78cm] (le52) at (block2.north west -| t5);

    \begin{scope}[live range]
      \draw (ls52.center) -- (le52.center);
    \end{scope}

\end{tikzpicture}}
    \label{fig:sum-cfg}
  }
  \caption{Running example: \code{sum} function.}
  \label{fig:sum-example}
\end{figure}

Figure~\ref{fig:sum-c} shows the C source code of a function returning
the sum of the \code{n} elements of an array~\code{v}.
Figure~\ref{fig:sum-cfg} shows its corresponding CFG in the form taken
as input to register allocation.
In this form, temporaries $\Temp{1}$, $\Temp{2}$, and $\Temp{3}$
correspond directly to the C variables \code{v}, \code{n}, and
\code{s}; $\Temp{4}$ corresponds to the end of the array (\code{v +
  n}); and $\Temp{5}$ holds the element loaded in each iteration.
$\Temp{1}$, $\Temp{3}$, $\Temp{4}$, and~$\Temp{5}$ interfere with each
other and $\Temp{2}$ interferes with $\Temp{1}$ and $\Temp{3}$, as
can be seen from the live ranges depicted to the left of the CFG.
The example uses the following MIPS32-like
instructions~\cite{Sweetman2006}: \code{li} (load immediate),
\code{add} (add), \code{addi} (add immediate), \code{bge} (branch if
greater or equal), \code{load} (load from memory), \code{blt} (branch
if lower than), and \code{jr} (jump and return).
The \code{sum} function is used as running example throughout the paper.

\begin{SaveVerbatim}{sum-in}
int sum(char * v, int n) {
\end{SaveVerbatim}
\begin{SaveVerbatim}{sum-b1}
int s = 0;
\end{SaveVerbatim}
\begin{SaveVerbatim}{sum-b2}
for (int i = 0; i < n; i++) {
  s += v[i];
}
\end{SaveVerbatim}
\begin{SaveVerbatim}{sum-b3}
return s;
\end{SaveVerbatim}
\begin{SaveVerbatim}{sum-out}
}
\end{SaveVerbatim}

\paragraph{Register allocation and assignment}

Register allocation maps temporaries to either processor registers or
memory.
The former are usually preferred as they have faster access times.
\emph{Multiple allocation} allows temporaries to be allocated to both
memory and processor registers simultaneously (at the same program
point), which can be advantageous in certain scenarios~\cite[Section
  2.2]{Colombet2015}.
\emph{Register assignment} gives specific registers to
register-allocated temporaries.
The same register can be assigned to multiple, non-interfering
temporaries to improve register utilization.

\paragraph{Spilling}

In general, the availability of enough processor registers is not
guaranteed and some temporaries must be \emph{spilled} (that is,
allocated to memory).
Spilling a temporary $t$ requires the insertion of store and load
instructions to move $t$'s value to and from memory.
The simplest strategy (known as \emph{spill-everywhere}) inserts store
and load instructions at each definition and use of~$t$.
\emph{Load-store optimization} allows $t$ to be spilled at a finer
granularity to reduce spill code overhead.

\paragraph{Coalescing}

The input program may contain temporaries related by \emph{copies}
(operations that replicate the value of a temporary into another).
Non-interfering copy-related temporaries can be \emph{coalesced}
(assigned to the same register) to discard the corresponding copies
and thereby improve efficiency and code size.
Likewise, copies of temporaries to or from registers (such as
$\simpleCopyOperation{\Temp{1}}{\register{R1}}$ and
$\simpleCopyOperation{\register{R1}}{\Temp{3}}$ in
Figure~\ref{fig:sum-cfg}) can be discarded by assigning the
temporaries to the corresponding registers whenever possible.

\paragraph{Live-range splitting}

Sometimes it is desirable to allocate a temporary $t$ to different
locations during different parts of its live range.
This is achieved by \emph{splitting} $t$ into a temporary for each
part of the live range that might be allocated to a different
location.

\paragraph{Packing}

Each temporary has a certain bit-width which is determined by its
source data type (for example, \code{char} versus \code{int} in C).
Many processors allow several temporaries of small widths to be
assigned to different parts of the same register of larger width.
This feature is known as \emph{register aliasing}.
For example, Intel's x86~\cite{x86} combines pairs of 8-bits
registers (\register{AH}, \register{AL}) into 16-bit registers
(\register{AX}).
Packing non-interfering temporaries into the same register is key to
improving register utilization.

\paragraph{Rematerialization}

In processors with a limited number of registers, it can sometimes be
beneficial to recompute (that is, \emph{rematerialize}) a value to be
reused rather than occupying a register until its later use or
spilling the value.

\paragraph{Multiple register banks}

Some processors include multiple register banks clustered around
different types of functional units, which often leads to alternative
temporary allocations.
To handle these architectures effectively, register allocation needs
to take into account the cost of allocating a temporary to different
register banks and moving its value across them.

\paragraph{Scope}

\emph{Local} register allocation deals with one basic block at a time,
spilling all temporaries that are live at basic block boundaries.
\emph{Global} register allocation considers entire functions, yielding
better code as temporaries can be kept in the same register across
basic blocks.
All approaches reviewed in this section are global.

\paragraph{Evaluation methods}

Combinatorial approaches to code generation tasks can be evaluated
statically (based on a cost estimation by the objective function),
dynamically (based on the actual cost from the execution of the
generated code), or by a mixture of the two (based on a static cost
model instantiated with execution measurements).
For runtime objectives such as speed, the accuracy of static
evaluations depends on how well they predict the behavior of the
processor and benchmarks.
For register allocation, dynamic evaluations are usually preferred
since they are most accurate and capture interactions with later tasks
such as instruction scheduling.
Mixed evaluations tend to be less accurate but can isolate the effect
of register allocation from other tasks.
Static evaluations require less implementation effort and are suitable
for static objectives (such as code size minimization) or when an
execution platform is not available.

\paragraph{Outline}

\begin{table}[b]%
  \newcommand{\registerAllocationApproach}[9]{#7 & #6 & #2 & #1 & #3 & #8 & #4 & #5 & #9}
  \renewcommand{\columnWidth}{0.58cm}
  \setlength{\tabcolsep}{2pt}
  \setlength{\tabulinesep}{0.03cm}
  \caption{Register allocation approaches: technique (TC),
    scope (SC), spilling (SP), register assignment (RA), coalescing
    (CO), load-store optimization (LO), register packing (RP),
    live-range splitting~(LS), rematerialization (RM), multiple
    register banks (MB), multiple allocation (MA), size of largest
    problem solved optimally (SZ) in number of
    instructions, and whether a dynamic evaluation is available (DE).\label{tab:register-allocation}}
    \begin{tabu}{%
        l%
        c%
        c%
        >{\centering\arraybackslash}p{\columnWidth}%
        >{\centering\arraybackslash}p{\columnWidth}%
        >{\centering\arraybackslash}p{\columnWidth}%
        >{\centering\arraybackslash}p{\columnWidth}%
        >{\centering\arraybackslash}p{\columnWidth}%
        >{\centering\arraybackslash}p{\columnWidth}%
        >{\centering\arraybackslash}p{\columnWidth}%
        >{\centering\arraybackslash}p{\columnWidth}%
        >{\centering\arraybackslash}p{\columnWidth}%
        c%
        >{\centering\arraybackslash}p{\columnWidth}}\hline
      \rowfont{\bfseries}
      approach & TC & SC & \registerAllocationApproach{LO}{CO}{RP}{RM}{MB}{RA}{SP}{LS}{MA} & SZ & DE \\
      \hline
      % The 2000 number is taken from Figures 15 and 16 in [Fu2002]
      ORA & IP & global & \registerAllocationApproach{\yes}{\yes}{\yes}{\yes}{\yes}{\yes}{\yes}{\yes}{\yes} & $\sim{}$2000 & \yes\\
      % The 200 number is taken from Table 2 in [Scholz2002] (assuming #temps ~= #instrs)
      \rowcolor{tblrow} Scholz~\etal{}~2002 & PBQP & global & \registerAllocationApproach{\no}{\yes}{\yes}{\yes}{\no}{\yes}{\yes}{\no}{\no} & $\sim{}$200 & \yes \\
      PRA & IP & global & \registerAllocationApproach{\yes}{\no}{\no}{\yes}{\yes}{\yes}{\yes}{\yes}{\no} & ? & \yes \\
      \rowcolor{tblrow} SARA & IP & global & \registerAllocationApproach{\yes}{\no}{\no}{\no}{\no}{\yes}{\yes}{\yes}{\no} & ? & \yes \\
      Barik~\etal{}~2007 & IP & global & \registerAllocationApproach{\yes}{\no}{\yes}{\yes}{\yes}{\yes}{\yes}{\yes}{\no} & 302 & \no \\
      \rowcolor{tblrow} Naik and Palsberg~2002 & IP & global & \registerAllocationApproach{\no}{\no}{\no}{\no}{\yes}{\no}{\no}{\no}{\no} & 850 & \no \\
      % The 1000 number assumes #temps ~= #instrs
      Falk~\etal{}~2011 & IP & global & \registerAllocationApproach{\yes}{\no}{\no}{\no}{\yes}{\yes}{\yes}{\yes}{\yes} & $\sim{}$1000 & \yes \\
      % The 2000 number is taken from Figure 7 in [Appel2001]
      \rowcolor{tblrow} Appel and George~2001 & IP & global & \registerAllocationApproach{\yes}{\no}{\no}{\no}{\no}{\no}{\yes}{\yes}{\no} & $\sim{}$2000 & \yes \\
      Ebner~\etal{}~2009 & IP & global & \registerAllocationApproach{\yes}{\no}{\no}{\no}{\no}{\no}{\yes}{\yes}{\no} & ? & \yes \\
      \rowcolor{tblrow} Colombet~\etal{}~2015 & IP & global & \registerAllocationApproach{\yes}{\no}{\no}{\yes}{\no}{\no}{\yes}{\yes}{\yes} & ? & \yes \\
  \end{tabu}
\end{table}

Table~\ref{tab:register-allocation} classifies combinatorial register
allocation approaches with information about their optimization
technique, scope, problem
coverage, approximate scalability, and evaluation method\footnote{%
For simplicity, Table~\ref{tab:register-allocation} classifies mixed evaluations as
  dynamic.}.
\emph{Problem coverage} refers to the subproblems that each approach
solves in integration with combinatorial optimization.
Approaches might exclude subproblems for scalability, modeling
purposes, or because they do not apply to their processor model.
The running text discusses the motivation behind each approach.
Scalability in this classification is approximated by the size of
largest problem solved optimally as reported by the original
publications.
Question marks are used when this figure could not be retrieved (no
reevaluation has been performed in the scope of this survey).
Improvements in combinatorial solving and increased computational
power should be taken into account when comparing approaches across
time.

Section~\ref{sec:basic-approaches} covers the first approaches that
include register assignment as part of their combinatorial models,
forming a baseline for all subsequent combinatorial register
allocation approaches.
Sections~\ref{sec:model-extensions}
and~\ref{sec:alternative-optimization-objectives} cover the study of
additional subproblems and alternative optimization objectives.
Section~\ref{sec:decomposed-approaches} discusses approaches that
decompose register allocation (including spilling) and register
assignment (including coalescing) for scalability.
Section~\ref{sec:register-allocation-discussion} closes with a summary
of developments and challenges in combinatorial register allocation.

\subsection{Basic Approaches}
\label{sec:basic-approaches}%

\paragraph{Optimal Register Allocation}
Goodwin and Wilken introduce the first widely-recognized approach to
combinatorial register allocation~\cite{Goodwin1996}, almost three
decades after some early work in the area~\cite{Day1970,Luque1984}.
The approach, called \emph{Optimal Register Allocation} (ORA), is
based on an IP model that captures the full range of register
allocation subproblems (see Table~\ref{tab:register-allocation}).
Goodwin and Wilken's ORA demonstrated, for the first time, that
combinatorial global register allocation is feasible -- although
slower than heuristic approaches.

The ORA allocator derives an IP model in several steps.
First, a \emph{temporary graph} (Goodwin and Wilken refer to
temporaries as \emph{symbolic registers}) is constructed for each
temporary $t$ and register $r$ where the nodes are the program points
$p_1, p_2, \ldots, p_n$ at which $t$ is live and the arcs correspond
to possible control transitions.
Then, the program points are annotated with register allocation
decisions that correspond to \mbox{0-1} variables in the IP model and
linear constraints involving groups of decisions.
Figure~\ref{fig:ora-example} shows the temporary graph corresponding
to $\Temp{1}$ and $\register{R1}$ in the running example.

\newcommand{\defvar}[1]{\noMathVar{def}{\Temp{1},\register{R1},#1}}
\newcommand{\storevar}[1]{\noMathVar{store}{\Temp{1},\register{R1},#1}}
\newcommand{\contvar}[1]{\noMathVar{cont}{\Temp{1},\register{R1},#1}}
\newcommand{\loadvar}[1]{\noMathVar{load}{\Temp{1},\register{R1},#1}}
\newcommand{\useendvar}[1]{\noMathVar{use-end}{\Temp{1},\register{R1},#1}}
\newcommand{\usecontvar}[1]{\noMathVar{use-cont}{\Temp{1},\register{R1},#1}}

\begin{figure}
  \centering
  \scalebox{1}{\input{./figures/styles}%

\newcommand{\lefttable}[1]{\begin{tabular}{l}#1\end{tabular}}
\newcommand{\righttable}[1]{\begin{tabular}{r}#1\end{tabular}}

\newcommand{\vertsep}{0.08cm}

\begin{tikzpicture}
    \node [program point] (entry) {};
    \node [statement, below=\vertsep of entry] (s1) {};
    \node [program point, below=\vertsep of s1] (p0) {};
    \node [statement, below=\vertsep of p0] (s2) {};
    \node [program point, below=\vertsep of s2] (p1) {};
    \node [statement, below=\vertsep of p1] (s3) {};
    \node [program point, below=\vertsep of s3] (p2) {};
    \node [statement, below=\vertsep of p2] (s4) {};
    \node [program point, below=\vertsep of s4] (p3) {};
    \node [statement, below=\vertsep of p3] (s5) {};
    \node [program point, below=\vertsep of s5] (end1) {};

    \node [program point, below=0.35cm of end1] (begin2) {};
    \node [statement, below=\vertsep of begin2] (s6) {};
    \node [program point, below=\vertsep of s6] (p4) {};
    \node [statement, below=\vertsep of p4] (s7) {};
    \node [program point, below=\vertsep of s7] (p5) {};
    \node [statement, below=\vertsep of p5] (s8) {};
    \node [program point, below=\vertsep of s8] (p6) {};
    \node [statement, below=\vertsep of p6] (s9) {};
    \node [program point, below=\vertsep of s9] (end2) {};

    \coordinate [left=3.3cm of end2] (end2left);
    \coordinate [left=3.3cm of begin2] (begin2left);

  \begin{scope}[flow]
    \draw (entry) -- (p0);
    \draw (p0) -- (p1);
    \draw (p1) -- (p2);
    \draw (p2) -- (p3);
    \draw (p3) -- (end1);
    \draw (end1) -- (begin2);
    \draw (begin2) -- (p4);
    \draw (p4) -- (p5);
    \draw (p5) -- (p6);
    \draw (p6) -- (end2);
    \draw [-] (end2) -- (end2left);
    \draw [-] (end2left) -- (begin2left);
    \draw (begin2left) -- (begin2);
  \end{scope}

    \node [left=-0.2cm of s1] {\righttable{$\InstrFstMove{}:\; \simpleCopyOperation{\Temp{1}}{\register{R1}}$}};
    \node [left=-0.2cm of s2] {\righttable{$\InstrSndMove{}:\; \simpleCopyOperation{\Temp{2}}{\register{R2}}$}};
    \node [left=-0.2cm of s3] {\righttable{$\InstrLi{}:\; \naturalOperation{\Temp{3}}{\loadImmediateInstruction{}}{0}$}};
    \node [left=-0.2cm of s4] {\righttable{$\InstrAddAddr{}:\; \naturalOperation{\Temp{4}}{\addInstruction{}}{\Temp{1}, \Temp{2}}$}};
    \node [left=-0.2cm of s5] {\righttable{$\InstrBge{}:\; \leftBranchOperation{\jumpIfGEInstruction{}}{\Temp{1}, \Temp{4}, \B{3}}$}};
    \node [left=-0.2cm of s6] {\righttable{$\InstrLoad{}:\; \naturalOperation{\Temp{5}}{\loadInstruction{}}{\Temp{1}}$}};
    \node [left=-0.2cm of s7] {\righttable{$\InstrAdd{}:\; \naturalOperation{\Temp{3}}{\addInstruction{}}{\Temp{3}, \Temp{5}}$}};
    \node [left=-0.2cm of s8] {\righttable{$\InstrInc{}:\; \naturalOperation{\Temp{1}}{\addiInstruction{}}{\Temp{1}, 1}$}};
    \node [left=-0.2cm of s9] {\righttable{$\InstrBlt{}:\; \leftBranchOperation{\jumpIfLTInstruction{}}{\Temp{1}, \Temp{4}, \B{2}}$}};

  \begin{scope}[ora variable]
    \node [right=-0.2cm of p0] {\lefttable{$\defvar{p_2};\,\storevar{p_2};\,\contvar{p_2}$}};
    \node [right=-0.2cm of p1] {\lefttable{$$}};
    \node [right=-0.2cm of p2] {\lefttable{$\loadvar{p_4}$}};
    \node [right=-0.2cm of p3] {\lefttable{$\useendvar{p_5};\,\usecontvar{p_5};\,\loadvar{p_5}$}};
    \node [right=-0.2cm of end1] {\lefttable{$\useendvar{p_6};\,\usecontvar{p_6}$}};
    \node [right=-0.2cm of begin2] {\lefttable{$\loadvar{p_7}$}};
    \node [right=-0.2cm of p4] {\lefttable{$\useendvar{p_8};\,\usecontvar{p_8}$}};
    \node [right=-0.2cm of p5] {\lefttable{$\loadvar{p_9}$}};
    \node [right=-0.2cm of p6] {\lefttable{$\defvar{p_{10}};\,\storevar{p_{10}};\,\contvar{p_{10}};\,\loadvar{p_{10}}$}};
    \node [right=-0.2cm of end2] {\lefttable{$\useendvar{p_{11}};\,\usecontvar{p_{11}}$}};
  \end{scope}

\end{tikzpicture}}
  \caption{Simplified ORA temporary graph for $\Temp{1}$ and
    $\register{R1}$.}
  \label{fig:ora-example}
\end{figure}

The model includes four main groups of variables to capture different
subproblems, where each variable is associated to a specific program
point $p$ in the temporary graph:
\begin{inparaitem}[]
\item \emph{register assignment variables} \var{def}{t,r,p},
  \var{use-cont}{t,r,p}, and \var{use-end}{t,r,p} indicate whether
  temporary $t$ is assigned to $r$ at each definition and use of $t$
  ($\operatorname{use-cont}$ and $\operatorname{use-end}$ reflect
  whether the assignment is effective at the use point and in that
  case whether it continues or ends afterwards);
\item \emph{spilling variables} \var{store}{t,r,p}, \var{cont}{t,r,p},
  and \var{load}{t,r,p} indicate whether temporary $t$ which is
  assigned to register $r$ is stored in memory, whether the assignment
  to $r$ continues after a possible store, and whether $t$ is loaded
  from memory to $r$;
\item \emph{coalescing variables} \var{elim}{t,t',r,p} indicate
  whether the copy from $t$ to $t'$ is eliminated by assigning $t$ and
  $t'$ to~$r$; and
\item \emph{rematerialization variables} \var{remat}{t,r,p} indicate
  whether~$t$ is rematerialized into~$r$.
\end{inparaitem}
In the original notation each variable is prefixed by $x$ and suffixed
and superscripted by its corresponding register and
temporary\footnote{The original variable and constraint names in the
  reviewed publications are sometimes altered for clarity,
  consistency, and comparability.
  A note is made whenever this is the case.}.
Figure~\ref{fig:ora-example} shows the variables for $\Temp{1}$
and~$\register{R1}$ at different program points.

The model includes linear constraints to enforce that:
\begin{inparaitem}[]
\item at each program point, each register holds at most one
  temporary;
\item each temporary $t$ is assigned to a register at $t$'s definition
  and uses;
\item each temporary is assigned the same register where its live
  ranges are merged at the join points of the CFG; and
\item an assignment of temporary $t$ to a register that holds right
  before a use is conserved until the program point where $t$ is used.
\end{inparaitem}
For example, the temporary graph shown in Figure~\ref{fig:ora-example}
induces the constraint $\usecontvar{p_5} + \useendvar{p_5} =
\contvar{p_2} + \loadvar{p_4}$ to enforce that the assignment of
  $\Temp{1}$ to $\register{R1}$ can only continue or end at program
  point $p_5$ (after $\InstrAddAddr{}$) if $\Temp{1}$ is actually
  assigned to $\register{R1}$ at that point.
Other constraints to capture spilling, coalescing, and
rematerialization are listed in the original paper~\cite{Goodwin1996}.

The objective function minimizes the total cost of decisions reflected
in the spilling, coalescing, and rematerialization variables.
In the running example, the $\storevar{p}$ and $\loadvar{p}$ variables
are associated with the estimated cost of spilling at each program
point $p$ where they are introduced (based on estimated execution
frequency and type of spill instructions) while $\defvar{p_2}$ is
associated with the estimated benefit of discarding the copy
$\InstrFstMove{}$ by coalescing $\Temp{1}$ and~$\register{R1}$.

Goodwin and Wilken use a commercial IP solver and compare the results
against those of GCC's~\cite{GCC2017} register allocator for a
Hewlett-Packard PA-RISC processor~\cite{Kane1996}.
Their experiments reveal that in practice register allocation problems
have a manageable average complexity, and functions of hundreds of
instructions can be solved optimally in a time scale of minutes.

The results of Goodwin and Wilken encouraged further research based on
the ORA approach.
Kong and Wilken present a set of extensions to the original ORA model,
including register packing and multiple register banks, to
deal with irregularities in register architectures~\cite{Kong1998}.
The extensions are complete enough to handle Intel's x86~\cite{x86}
architecture, which presents a fairly irregular register file.
Kong and Wilken estimate that their extended ORA approach reduces
GCC's execution time overhead due to register allocation by 61\% on
average.
The estimation is produced by a mixed static-dynamic evaluation that
instantiates the model's objective function with the actual execution
count of spill, coalescing, and rematerialization instructions.
While this estimation is more accurate than a purely static one, a
study of its relation to the actual execution time is not available.
Besides improving code quality, Kong and Wilken speed up the solving
time of Goodwin and Wilken by two orders of magnitude.
The reasons behind this speedup are a reduction of the search space
due to both the availability of fewer registers and the introduction
of irregularities, and the use of a faster machine with a newer
version of the IP solver.
The results illustrate an interesting aspect of combinatorial
optimization: factors that complicate the design of heuristic
approaches such as processor irregularities do not necessarily affect
combinatorial approaches negatively -- sometimes quite the opposite.

Fu and Wilken reduce the (still large) solving time gap between the
ORA and heuristic approaches~\cite{Fu2002}.
Their faster ORA approach identifies numerous conditions under which
decisions are dominated (that is, provably suboptimal).
For example, the fact that the ORA model assumes a constant cost for
any placement of spill code within the same basic block makes certain
spilling decisions dominated.
The variables corresponding to such decisions are guaranteed to be
zero in some optimal solution and can thus be discarded to reduce the
model's complexity.

Fu and Wilken find that the solving process is roughly sped up by four
orders of magnitude compared to the original ORA approach: two due to
increased computational power and algorithmic improvements in the IP
solver during the six-year gap between the publications, and two due
to the removal of dominated variables and their corresponding
constraints.
According to their results, the improvements make it possible to solve
98.5\% of the functions in the SPEC92 integer benchmarks~\cite{CPU}
optimally with a time limit of 1024 seconds.

\paragraph{Scholz et al}

Scholz and Eckstein propose an alternative combinatorial approach that
models register allocation as a partitioned Boolean quadratic
programming (PBQP) problem~\cite{Scholz2002}.
The simplicity with which register allocation can be reduced to a PBQP
problem and the availability since 2008 of a production-quality
implementation in the LLVM compiler~\cite{Lattner2004} have made PBQP
a popular technique for this purpose.
However, the simplicity of this approach comes with limitations ---
the range of subproblems that are captured is narrower than that of
the more general ORA approach (see
Table~\ref{tab:register-allocation}).
Although more subproblems could be in principle modeled with
additional variables and costs, it remains an open question whether
the resulting scalability would match that of IP-based approaches.

In contrast to the rather sophisticated ORA model, Scholz and
Eckstein's model features a single class of variables \var{a}{t}
giving the register to which temporary $t$ is assigned.
The decisions to spill a temporary $t$ to memory or to rematerialize
it~\cite[Chapter 4]{Hames2011} are captured by including special
\emph{spilling} and \emph{rematerialization registers} $\register{sp},
\register{rm}$ to the domain of its variable \var{a}{t}.
In the original notation each variable \var{a}{t} is defined as a
collection of alternative Boolean variables
$\set{\noMathVar{x}{t,\register{R0}}, \noMathVar{x}{t,\register{R1}},
  \ldots, \noMathVar{x}{t,\register{sp}},
  \noMathVar{x}{t,\register{rm}}}$ where each Boolean variable
captures exactly one value of \var{a}{t} and \var{a}{t} is referred to
as a vector $\noMathVar{\vv{x}}{t}$.
As is characteristic of PBQP models (see Table~\ref{tab:techniques}),
constraints are defined by giving conceptually infinite costs to
forbidden single assignments $c(\noMathVar{a}{t})$ and pairs of
assignments $C(\noMathVar{a}{t}, \noMathVar{a}{t'})$.
Individual costs $c(\noMathVar{a}{t})$ are used to
\begin{inparaitem}[]
\item forbid the assignment of temporary $t$ to registers that do not
  belong to its supported register classes,
\item account for the overhead of spilling or rematerializing $t$, and
\item account for the benefit of coalescing $t$ with preassigned
  registers.
\end{inparaitem}
Costs of pairs of assignments $C(\noMathVar{a}{t},
\noMathVar{a}{t'})$ are used to
\begin{inparaitem}[]
\item forbid assignments of interfering temporaries to the same (or
  aliased) registers, and
\item account for the benefit of coalescing $t$ and $t'$\!.
\end{inparaitem}
The objective function minimizes the cost given to single assignments
and pairs of assignments, thus avoiding solutions forbidden by
conceptually infinite costs.

Figure~\ref{fig:pbqp-example} shows the assignment costs for the
running example from Figure~\ref{fig:sum-example}, where $c$ is the
estimated benefit of discarding a copy by coalescing and $s$ is the
estimated cost of spilling a temporary (uniform benefits and costs are
assumed for simplicity).
None of the temporaries can be rematerialized since their values
cannot be recomputed from available values~\cite{Chaitin1981}, hence
the cost of assigning them to \register{rm} is infinite.
Since $\Temp{1}$ and $\Temp{2}$ interfere, assignments to the same
register incur an infinite cost.
This is the case for all pairs of temporaries in the example except
$\sequence{\Temp{2}, \Temp{4}}$ and $\sequence{\Temp{2}, \Temp{5}}$
which yield null matrices as they do not interfere.

\begin{figure}
  \setlength{\tabcolsep}{3.5pt}
  \hfill
  \subfloat[Costs of individual assignments]{%
    \adjustbox{}{%
      \begin{tabu}{rcccccc}
        \\[0.05cm]
        & \register{R1} & \register{R2} & $\cdots$ & \register{R31} & \register{sp} & \register{rm}\\[0.05cm]
        \begin{tabular}{r}$\noMathVar{a}{\Temp{1}}$\\$\noMathVar{a}{\Temp{3}}$\end{tabular} & $-c$ & $0$ & $\cdots$ & $0$ & $s$ & $\infty$ \\[0.15cm]
        \begin{tabular}{r}\var{a}{\Temp{2}}\end{tabular} & $0$ & $-c$ & $\cdots$ & $0$ & $s$ & $\infty$ \\[0.09cm]
        \begin{tabular}{r}$\noMathVar{a}{\Temp{4}}$\\$\noMathVar{a}{\Temp{5}}$\end{tabular} & $0$ & $0$ & $\cdots$ & $0$ & $s$ & $\infty$ \\
      \end{tabu}
    }
    \label{fig:pbqp-example-individual}
  }
  \hfill
  \subfloat[Costs of assignments for $\Temp{1}$ and $\Temp{2}$]{%
    \adjustbox{}{%
      \begin{tabu}{lrcccccc}
        & & \multicolumn{6}{c}{\var{a}{\Temp{1}}}\\[0.05cm]
        & & \register{R1} & \register{R2} & $\cdots$ & \register{R31} & \register{sp} & \register{rm}\\[0.05cm]
        \multirow{6}{*}{\var{a}{\Temp{2}}}
        & \register{R1} & $\infty$ & $0$ & $\cdots$ & $0$ & $0$ & $0$\\
        & \register{R2} & $0$ & $\infty$ & $\cdots$ & $0$ & $0$ & $0$\\
        & $\cdots$ & $\cdots$ & $\cdots$ & $\cdots$ & $\cdots$ & $\cdots$ & $\cdots$\\
        & \register{R31} & $0$ & $0$ & $\cdots$ & $\infty$ & $0$ & $0$\\
        & \register{sp} & $0$ & $0$ & $\cdots$ & $0$ & $0$ & $0$\\
        & \register{rm} & $0$ & $0$ & $\cdots$ & $0$ & $0$ & $0$\\
      \end{tabu}
    }
    \label{fig:pbqp-example-pairs}
  }
  \hfill
  \
  \caption{Assignment costs in Scholz and Eckstein's model for the
    running example.}
  \label{fig:pbqp-example}
\end{figure}

Scholz and Eckstein propose both a heuristic and an optimal PBQP
solver.
When no reduction rule applies, the former applies applies greedy
elimination rules while the latter resorts to exhaustive enumeration
(as opposed to branch-and-bound search).
As this survey is concerned with combinatorial approaches, only
results related to the optimal PBQP solver are discussed.
Scholz and Eckstein experiment with five signal processing benchmarks
on Infineon's Carmel 20xx~\cite{Carmel} Digital Signal Processor (DSP)
to demonstrate the ease with which its special register allocation
constraints (that dictate register combinations allowed for certain
pairs of temporaries) are modeled in PBQP.
Their dynamic evaluation shows that the optimal solver can deliver up
to 13.6\% faster programs than graph coloring-based
heuristics~\cite{Smith2004}.
A complementary static evaluation of the optimal PBQP solver by
Hirnschrott \etal{} supports the conclusions of Scholz and Eckstein
for different versions of an ideal DSP with different number of
registers and instruction operands~\cite{Hirnschrott2003}.

Hames and Scholz extend the PBQP solver, based originally in
exhaustive enumeration, with a branch-and-bound search mechanism to
reduce the amount of search needed to find optimal
solutions~\cite{Hames2006}.
Hames and Scholz's static evaluation on Intel's x86~\cite{x86} shows
that their branch-and-bound PBQP solver solves 97.4\% of the
SPEC2000~\cite{CPU} functions over 24 hours, yielding a slight
estimated spill cost reduction of 2\% over the same heuristic approach
as in the original experiments.
This suggests the improvement potential of Scholz~\etal{}'s approach
over heuristics is limited for general-purpose processors and larger
for more constrained processors such as DSPs.

\paragraph{Progressive Register Allocation}
Koes and Goldstein introduce a \emph{progressive} register allocation
(PRA) approach~\cite{Koes2005,Koes2006}.
An ideal progressive solver should
\begin{inparaitem}[]
\item deliver reasonable solutions quickly,
\item find improved solutions if more time is allowed, and
\item find an optimal solution if enough time is available.
\end{inparaitem}
Although both the ORA and Scholz~\etal{}'s approaches can also
potentially behave progressively, in 2005 none of them was able to
meet the three conditions.

Koes and Goldstein propose modeling register allocation as a
multi-commodity network flow (MCNF) problem~\cite{Ahuja1993}, which
can be seen as a special case of an IP problem.
The MCNF problem consists in finding a flow of multiple commodities
through a network such that the cost of flowing through all arcs is
minimized and the flow capacity of each arc is not exceeded.
The reduction of register allocation to MCNF is intuitive: each
commodity corresponds to a temporary which flows through storage
locations (registers and memory) at each program point, and the
network's structure forces interfering temporaries to flow through
different registers.
This model can express detailed allocations and accurately take into
account their cost.
Furthermore, the reduction to MCNF enables exploiting well-understood
techniques to solve network problems progressively.
On the other hand, the flow abstraction cannot cope with either
coalescing, register packing, or multiple allocations of the same
temporary, which makes the PRA model less general than that of the ORA
approach.

In the PRA approach, each commodity in the network flow corresponds to
a temporary.
For each program point and storage location, a node is added to the
network.
The flow of a temporary $t$ through the network determines how $t$ is
allocated.
As with Scholz~\etal{}'s model, a single class of variables is
defined:
\var{a}{t,i,j} ($x^{t}_{i,j}$ in the original notation) indicates
whether the temporary $t$ flows through the arc $\sequence{i,j}$ where
$i,j$ represent storage locations at a program point.
Compared to Scholz~\etal{}'s model, the additional program point
dimension makes it possible to express more detailed register
allocations capturing live-range splitting.
The model includes linear constraints to enforce that:
\begin{inparaitem}[]
\item only one temporary flows through each register node (called
  \emph{bundle constraints} by Koes and Goldstein),
  and
\item flow is conserved through the nodes (the same amount of flow
  that enters a node exits it).
\end{inparaitem}
The objective function minimizes the arc traversal cost for all
flows.
The cost of an arc \var{c}{i,j} reflects the cost of moving a
temporary from the source location $i$ to the destination location
$j$:
\begin{inparaitem}[]
\item if $i$ and $j$ correspond to the same locations, no cost is
  incurred;
\item if $i$ and $j$ correspond to different registers, \var{c}{i,j}
  is the cost of a register-to-register move instruction; and
\item if one of $i$ and $j$ corresponds to memory and the other to a
  register, \var{c}{i,j} is the cost of a memory access instruction.
\end{inparaitem}

\begin{figure}
  \centering
  \scalebox{1.0}{\input{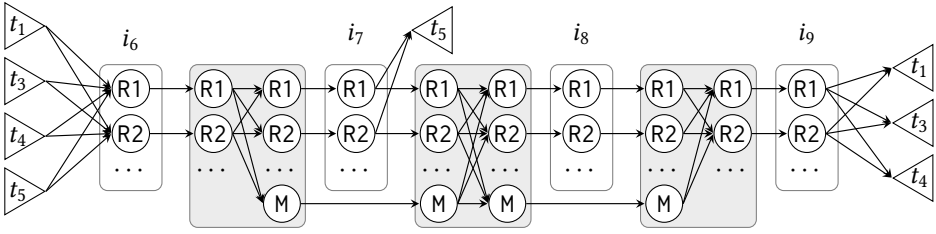}%
\pgfsetlayers{layer0,layer1}%
\newcommand{\righttable}[1]{\begin{tabular}{r}#1\end{tabular}}
\newcommand{\vertsep}{0.08cm}
\newcommand{\extravertsep}{-0.02cm}
\begin{tikzpicture}
  \begin{pgfonlayer}{layer1}
    \coordinate (source) {};
    \coordinate [right=1.5cm of source] (loadbar) {};
    \coordinate [right=1.1cm of loadbar] (crossbar1) {};
    \coordinate [right=1.9cm of crossbar1] (addbar) {};
    \coordinate [right=1.1cm of addbar] (crossbar2) {};
    \coordinate [right=1.9cm of crossbar2] (incbar) {};
    \coordinate [right=1.1cm of incbar] (crossbar3) {};
    \coordinate [right=1.9cm of crossbar3] (bltbar) {};
    \coordinate [right=1.5cm of bltbar] (sink) {};

  \begin{scope}
    \node [pra source right, below=-0.7cm of source] (source1) {$\Temp{1}$};
    \node [pra source right, below=0.3cm of source1] (source3) {$\Temp{3}$};
    \node [pra source right, below=0.3cm of source3] (source4) {$\Temp{4}$};
    \node [pra source right, below=0.3cm of source4] (source5) {$\Temp{5}$};

    \node [above=0.05cm of loadbar] {$\InstrLoad{}$};
    \node [pra storage, below=\vertsep of loadbar] (r1loadbar) {$\register{R1}$};
    \node [pra storage, below=\vertsep of r1loadbar] (r2loadbar) {$\register{R2}$};
    \node [pra dots, below=\vertsep of r2loadbar] (dotsloadbar) {$\cdots$};

    \node [pra storage, below=\vertsep of crossbar1] (r1crossbar1) {$\register{R1}$};
    \node [pra storage, below=\vertsep of r1crossbar1] (r2crossbar1) {$\register{R2}$};
    \node [pra dots, below=\vertsep of r2crossbar1] (dotscrossbar1) {$\cdots$};
    \node [pra storage, right=0.4cm of r1crossbar1] (r1crossbar1b) {$\register{R1}$};
    \node [pra storage, below=\vertsep of r1crossbar1b] (r2crossbar1b) {$\register{R2}$};
    \node [pra dots, below=\vertsep of r2crossbar1b] (dotscrossbar1b) {$\cdots$};
    \node [pra storage, below=\extravertsep of dotscrossbar1b] (memcrossbar1b) {$\register{M}$};

    \node [above=0.05cm of addbar] {\righttable{$\InstrAdd{}$}};
    \node [pra storage, below=\vertsep of addbar] (r1addbar) {$\register{R1}$};
    \node [pra storage, below=\vertsep of r1addbar] (r2addbar) {$\register{R2}$};
    \node [pra dots, below=\vertsep of r2addbar] (dotsaddbar) {$\cdots$};

    \node [pra storage, below=\vertsep of crossbar2] (r1crossbar2) {$\register{R1}$};
    \node [pra storage, below=\vertsep of r1crossbar2] (r2crossbar2) {$\register{R2}$};
    \node [pra dots, below=\vertsep of r2crossbar2] (dotscrossbar2) {$\cdots$};
    \node [pra storage, below=\extravertsep of dotscrossbar2] (memcrossbar2) {$\register{M}$};
    \node [pra storage, right=0.4cm of r1crossbar2] (r1crossbar2b) {$\register{R1}$};
    \node [pra storage, below=\vertsep of r1crossbar2b] (r2crossbar2b) {$\register{R2}$};
    \node [pra dots, below=\vertsep of r2crossbar2b] (dotscrossbar2b) {$\cdots$};
    \node [pra storage, below=\extravertsep of dotscrossbar2b] (memcrossbar2b) {$\register{M}$};

    \node [pra sink left, above=0.3cm of r1crossbar2] (sink5) {$\Temp{5}$};

    \node [above=0.05cm of incbar] {\righttable{$\InstrInc{}$}};
    \node [pra storage, below=\vertsep of incbar] (r1incbar) {$\register{R1}$};
    \node [pra storage, below=\vertsep of r1incbar] (r2incbar) {$\register{R2}$};
    \node [pra dots, below=\vertsep of r2incbar] (dotsincbar) {$\cdots$};

    \node [pra storage, below=\vertsep of crossbar3] (r1crossbar3) {$\register{R1}$};
    \node [pra storage, below=\vertsep of r1crossbar3] (r2crossbar3) {$\register{R2}$};
    \node [pra dots, below=\vertsep of r2crossbar3] (dotscrossbar3) {$\cdots$};
    \node [pra storage, below=\extravertsep of dotscrossbar3] (memcrossbar3) {$\register{M}$};
    \node [pra storage, right=0.4cm of r1crossbar3] (r1crossbar3b) {$\register{R1}$};
    \node [pra storage, below=\vertsep of r1crossbar3b] (r2crossbar3b) {$\register{R2}$};
    \node [pra dots, below=\vertsep of r2crossbar3b] (dotscrossbar3b) {$\cdots$};

    \node [above=0.05cm of bltbar] {\righttable{$\InstrBlt{}$}};
    \node [pra storage, below=\vertsep of bltbar] (r1bltbar) {$\register{R1}$};
    \node [pra storage, below=\vertsep of r1bltbar] (r2bltbar) {$\register{R2}$};
    \node [pra dots, below=\vertsep of r2bltbar] (dotsbltbar) {$\cdots$};

    \node [pra sink left, below=-0.15cm of sink] (sink1) {$\Temp{1}$};
    \node [pra sink left, below=0.3cm of sink1] (sink3) {$\Temp{3}$};
    \node [pra sink left, below=0.3cm of sink3] (sink4) {$\Temp{4}$};
  \end{scope}[]

  \begin{scope}[flow, line width=0.15mm, shorten >=0.1mm]

    \draw (source1.east) -- (r1loadbar.west);
    \draw (source1.east) -- (r2loadbar.west);
    \draw (source3.east) -- (r1loadbar.west);
    \draw (source3.east) -- (r2loadbar.west);
    \draw (source4.east) -- (r1loadbar.west);
    \draw (source4.east) -- (r2loadbar.west);
    \draw (source5.east) -- (r1loadbar.west);
    \draw (source5.east) -- (r2loadbar.west);

    \draw (r1loadbar.east) -- (r1crossbar1.west);
    \draw (r2loadbar.east) -- (r2crossbar1.west);

    \draw (r1crossbar1.east) -- (r1crossbar1b.west);
    \draw (r1crossbar1.east) -- (r2crossbar1b.west);
    \draw (r1crossbar1.east) -- (memcrossbar1b.west);
    \draw (r2crossbar1.east) -- (r1crossbar1b.west);
    \draw (r2crossbar1.east) -- (r2crossbar1b.west);
    \draw (r2crossbar1.east) -- (memcrossbar1b.west);

    \draw (memcrossbar1b.east) -- (memcrossbar2.west);

    \draw (r1crossbar1b.east) -- (r1addbar.west);
    \draw (r2crossbar1b.east) -- (r2addbar.west);

    \draw (r1addbar.east) -- (r1crossbar2.west);
    \draw (r2addbar.east) -- (r2crossbar2.west);

    \draw (r1addbar.east) -- (sink5.west);
    \draw (r2addbar.east) -- (sink5.west);

    \draw (r1crossbar2.east) -- (r1crossbar2b.west);
    \draw (r1crossbar2.east) -- (r2crossbar2b.west);
    \draw (r1crossbar2.east) -- (memcrossbar2b.west);
    \draw (r2crossbar2.east) -- (r1crossbar2b.west);
    \draw (r2crossbar2.east) -- (r2crossbar2b.west);
    \draw (r2crossbar2.east) -- (memcrossbar2b.west);
    \draw (memcrossbar2.east) -- (r1crossbar2b.west);
    \draw (memcrossbar2.east) -- (r2crossbar2b.west);
    \draw (memcrossbar2.east) -- (memcrossbar2b.west);

    \draw (memcrossbar2b.east) -- (memcrossbar3.west);

    \draw (r1crossbar2b.east) -- (r1incbar.west);
    \draw (r2crossbar2b.east) -- (r2incbar.west);

    \draw (r1incbar.east) -- (r1crossbar3.west);
    \draw (r2incbar.east) -- (r2crossbar3.west);

    \draw (r1crossbar3.east) -- (r1crossbar3b.west);
    \draw (r1crossbar3.east) -- (r2crossbar3b.west);
    \draw (r2crossbar3.east) -- (r1crossbar3b.west);
    \draw (r2crossbar3.east) -- (r2crossbar3b.west);
    \draw (memcrossbar3.east) -- (r1crossbar3b.west);
    \draw (memcrossbar3.east) -- (r2crossbar3b.west);

    \draw (r1crossbar3b.east) -- (r1bltbar.west);
    \draw (r2crossbar3b.east) -- (r2bltbar.west);

    \draw (r1bltbar.east) -- (sink1.west);
    \draw (r1bltbar.east) -- (sink3.west);
    \draw (r1bltbar.east) -- (sink4.west);
    \draw (r2bltbar.east) -- (sink1.west);
    \draw (r2bltbar.east) -- (sink3.west);
    \draw (r2bltbar.east) -- (sink4.west);

  \end{scope}
  \end{pgfonlayer}
  \begin{pgfonlayer}{layer0}
    \node [pra bar, fit=(r1loadbar) (dotsloadbar)] {};
    \node [pra crossbar, fit=(r1crossbar1) (memcrossbar1b)] {};
    \node [pra bar, fit=(r1addbar) (dotsaddbar)] {};
    \node [pra crossbar, fit=(r1crossbar2) (memcrossbar2) (r1crossbar2b)] {};
    \node [pra bar, fit=(r1incbar) (dotsincbar)] {};
    \node [pra crossbar, fit=(r1crossbar3) (memcrossbar3) (r1crossbar3b)] {};
    \node [pra bar, fit=(r1bltbar) (dotsbltbar)] {};
  \end{pgfonlayer}

\end{tikzpicture}}
  \caption{Simplified multi-commodity network flow for basic block
    $\B{2}$ in the PRA model.}
  \label{fig:pra-example}
\end{figure}

Figure~\ref{fig:pra-example} shows the MCNF corresponding to basic
block $\B{2}$ in the running example from
Figure~\ref{fig:sum-example}.
Temporary source and sink nodes are represented by triangles, while
storage locations (registers~$\register{R1}$, $\register{R2}$, \ldots,
and memory $\register{M}$) are represented by circles.
Each rectangle contains storage locations corresponding to either
instructions or program points.
The latter (colored in gray) allow temporaries to flow across
different storage locations between the execution of two instructions.
The MCNF is constructed to force temporaries used by an instruction to
flow through the storage locations supported by the instruction.
Unused temporaries can bypass instruction storage locations by
following additional arcs between gray rectangles (not depicted in
Figure~\ref{fig:pra-example} for clarity).
All arcs have capacity one except arcs between memory nodes which are
uncapacitated, allowing any number of temporaries to be simultaneously
spilled.

Although the PRA model is a valid IP model and can be thus solved by a
regular IP solver, Koes and Goldstein propose a dedicated solving
scheme to attain a more progressive behavior.
The scheme is based on a Lagrangian relaxation, a general IP technique
that, similarly to PBQP models, replaces hard constraints by terms in
the objective function that penalize their violation.
Relaxing the bundle constraints allows finding solutions heuristically
through shortest-path computations on the network.
The Lagrangian relaxation is used to guide the heuristics towards
improving solutions in an iterative process.
This solving scheme is not complete, but in practice it can often
prove optimality using bounds derived from the Lagrangian relaxation.

Koes and Goldstein compare their progressive solver with
GCC's~\cite{GCC2017} graph-coloring register allocator and a
commercial IP solver.
Their experiments with functions from different benchmarks show that
the PRA approach is indeed more progressive than standard IP: it
delivers first solutions in a fraction of the time taken by the IP
solver and solves 83.5\% of the functions optimally after~1000
iterations.
Koes and Goldstein report that their optimal solutions yield an
average code size reduction of 6.8\% compared to GCC's heuristic
approach.

\subsection{Model Extensions}\label{sec:model-extensions}

Further research on combinatorial register allocation has addressed
extensions of the baseline established by the basic approaches to cope
with different processor features.

Stack allocation is the problem of assigning specific stack locations
to spilled temporaries.
This problem is typically solved after register allocation, however
some processors provide features that can be best exploited if both
problems are solved in integration.
\emph{SARA}~\cite{Nandivada2006} is an IP approach that integrates
register allocation with stack allocation to exploit the double-store
and double-load instructions available in some ARM~\cite{ARM}
processors.
Such instructions can be used to spill pairs of temporaries but impose
additional constraints on the register assignment and stack allocation
of the spilled pairs, hence motivating the integrated SARA approach.
SARA's IP model is composed of a basic register allocation submodel
with load-store optimization and live-range splitting (see
Table~\ref{tab:register-allocation}) and a stack allocation submodel.
The latter includes
\begin{inparaitem}[]
\item \emph{location variables} \var{f}{t,l} indicating whether
  temporary $t$ is allocated to stack location $l$, and
\item explicit \emph{load and store pair variables}
  \var{load-pair}{i,t_1,t_2} and \var{store-pair}{i,t_1,t_2}
  indicating whether temporaries $t_1$, $t_2$ form a spill pair.
\end{inparaitem}
Linear constraints enforce that
\begin{inparaitem}[]
\item spilled temporaries are given a single location,
\item locations are not reused by multiple spilled temporaries, and
\item spill pairs satisfy the register assignment and stack allocation
  conditions.
\end{inparaitem}
The objective function minimizes estimated spill cost.
Nandivada and Palsberg's experiments for Intel's StrongARM processor
on 179 functions from different benchmarks show that the integrated
approach of SARA indeed generates faster code (4.1\%) than solving
each problem optimally but in isolation.

Bit-width aware register allocation extends register allocation to
handle processors that support referencing bit ranges within
registers, which is seen as a promising way to reduce register
pressure in multimedia and network processing
applications~\cite{Tallam2003,Nandivada2013}.
Handling such processors can be seen as a generalization of register
packing where register parts can be accessed with the finest
granularity and the bit-width of temporaries varies through the
program.
The only combinatorial approach to bit-width aware register allocation
is due to Barik~\etal{}~\cite{Barik2007}.
Their key contribution is an IP register allocation model that allows
multiple temporaries to be assigned to the same register $r$
simultaneously as long as the bit capacity of $r$ is not exceeded.
This is supported by generalizing the common constraints that ensure
that each register is not assigned more than one temporary
simultaneously into constraints that ensure that the sum of the
bit-width of the temporaries assigned to each register $r$
simultaneously does not exceed the capacity of~$r$.
The model does not capture the placement of temporaries within
registers and therefore it disregards the cost of \emph{defragmenting}
registers with register-to-register move instructions.
Barik~\etal{} perform a static evaluation with benchmarks from the
MediaBench~\cite{Lee1997} and Bitwise~\cite{Stephenson2000} suites on
an ideal processor with bitwise register addressing.
Their results show that, for such processors and applications,
extending combinatorial register allocation with bit-width awareness
can reduce the amount of spilling notably at the expense of
solving efficiency.

\subsection{Alternative Optimization Objectives}
\label{sec:alternative-optimization-objectives}

While most combinatorial register allocation approaches are concerned
with speeding up the average execution time of the generated code,
certain domains such as embedded systems show a high interest in
alternative optimization objectives such as minimizing code size,
energy, or worst-case execution time.

Naik and Palsberg introduce an IP approach~\cite{Naik2002} to minimize
code size for Zilog's Z86E30~\cite{Z86E30} processor.
This processor lacks stack memory and provides instead 16 register
banks of 16 registers each, which necessitates whole-program register
allocation without spilling.
Unlike other combinatorial register allocation approaches, Naik and
Palsberg assume that temporaries are always live and thus get
dedicated registers.
This assumption can increase register pressure significantly for
programs containing many short-lived temporaries, but this increase
might be acceptable for the Z86E30 processor due to its large amount
of registers and the expected modest size of its targeted
applications.
Both the lack of memory and the full-liveness assumption reduce the
amount of subproblems that need to be modeled significantly as seen in
Table~\ref{tab:register-allocation}.
The Z86E30 instructions can address a register by either specifying
its absolute address or an offset relative to a special register
pointer.
The latter mode is encoded with one byte less, creating an opportunity
to improve code size during register allocation.
For this purpose, Naik and Palsberg propose an IP model that
integrates register bank assignment with management of the register
pointer at different program points.
The model includes
\begin{inparaitem}[]
\item \emph{register bank assignment variables} \var{r}{t,b}
  indicating whether temporary~$t$ is assigned to register bank $b$,
\item \emph{current register pointer variables} \var{rp-val}{p,b}
  indicating whether the register pointer is set to register bank $b$
  at program point $p$, and
\item other variables to reflect the size of each instruction and the
  updates that are applied to the register pointer at different
  program points.
\end{inparaitem}
Linear constraints enforce that
\begin{inparaitem}[]
\item each temporary is assigned to one register bank,
\item the capacity of the register banks is not exceeded, and
\item the register pointer is updated accordingly to its intended
  register bank at each program point.
\end{inparaitem}
The objective function minimizes the total code size, including that
of the additional instructions needed to update the register pointer.
Naik and Palsberg' experiments show that their approach can match the
size of hand-optimized code for two typical control applications.

Real-time applications are usually associated with timing requirements
on their execution.
In that context, worst-case execution time (WCET) minimization is an
attractive optimization objective, since reducing the WCET of a
real-time application allows its computational cost to be reduced
without compromising the timing guarantees.
Falk~\etal{} present an IP approach for WCET
minimization~\cite{Falk2011} in the broader context of a WCET-aware
compiler.
The approach extends the basic ORA model with an alternative objective
function that minimizes the execution time of the longest execution
path through the function.
This is formulated by introducing \emph{WCET variables}~\var{w}{b} and
\emph{cost variables} \var{c}{b} corresponding to the execution time
of each basic block $b$ and of the longest execution path starting at
$b$, and letting the objective function minimize \var{w}{b_e}
where~$e$ is the entry basic block of the function.
The WCET and cost variables are related with linear constraints over
all paths of the function's CFG excluding arcs that form loops (called
\emph{back edges}).
For example, the CFG from Figure~\ref{fig:sum-cfg} yields the
constraints $\noMathVar{w}{\B{3}} = \noMathVar{c}{\B{3}}$;
$\noMathVar{w}{\B{2}} = \noMathVar{w}{\B{3}} + n \times
\noMathVar{c}{\B{2}}$; and $\noMathVar{w}{\B{1}} =
\max(\noMathVar{w}{\B{2}},\noMathVar{w}{\B{3}}) +
\noMathVar{c}{\B{1}}$, where $\max$ is in practice linearized with two
inequalities and $n$ is the given maximum iteration count of $\B{2}$.
The cost \var{c}{b} of a basic block $b$ corresponds to the cycles it
takes to perform all spills in $b$.
The cycles of a spill are modeled accurately by exploiting the
simplicity of the memory hierarchy of the target processor (Infineon
TriCore~\cite{TriCore}) and detailed pipeline knowledge.
Falk~\etal{}'s experiments on 55 embedded and real-time benchmarks
with high register pressure show that their approach reduces the WCET
by 14\% compared to a heuristic WCET-aware register allocator and, as
a side effect, speeds up code for the average case by 6.6\%.
Even though no problem sizes are reported, the authors find the
solving time acceptable in the context of WCET-aware compilation.

\subsection{Decomposed Approaches}
\label{sec:decomposed-approaches}

Register allocation includes a vast amount of subproblems, yielding
combinatorial problems of high average complexity as seen through this
section.
In an effort to improve scalability, a line of research has focused on
decomposing combinatorial register allocation into groups of
subproblems and solving each group optimality.
The decomposition that has received most attention is solving spilling
first (including strongly interdependent subproblems such as
load-store optimization, live-range splitting, and rematerialization)
followed by register assignment and coalescing.
The key assumption behind this decomposition scheme is that the impact
of spilling (in store and load instructions) is significantly higher
than the impact of coalescing (in register-to-register move
instructions), hence spilling is performed first.
The decomposition improves scalability because the size of the
spilling model becomes independent of the number of registers, and it
still delivers near-optimal solutions for processors where the
assumption holds.
For example, in an empirical investigation on the x86 and ARM
processors, Koes and Goldstein confirm that solving spilling optimally
and the remaining subproblems heuristically ``has no discernible
impact on performance''~\cite{Koes2009} compared to a non-decomposed,
optimal approach.
On the other hand, the solution quality can degrade for processors or
objectives such as code size minimization where the impact of spilling
does not necessarily dominate the cost~\cite{Koes2009}.
Also, the decomposition precludes a full integration with instruction
scheduling in a single combinatorial model as both spilling and the
remaining subproblems (register assignment and coalescing) have strong
interdependencies with instruction scheduling.
For decomposed approaches, Table~\ref{tab:register-allocation}
summarizes the first problem (spilling and associated subproblems)
where by construction the register assignment, coalescing, and
register packing subproblems do not apply.

Appel and George are the first to propose a decomposed scheme based on
IP~\cite{Appel2001}.
Their scheme solves spilling, load-store optimization, and live-range
splitting optimally first; and then solves register assignment and
coalescing either optimally or with heuristic algorithms.
The IP model for optimal spilling includes the variables \var{s}{t,p},
\var{l}{t,p} \var{r}{t,p}, \var{m}{t,p}, to indicate whether each
temporary $t$ at each program point $p$ is either stored to memory
($s$), loaded into a register ($l$), or kept in a register ($r$) or
memory ($m$).
The main model constraints ensure that the processor registers are not
overused at any program point.
The model also captures specific constraints of Intel's x86 to
demonstrate the flexibility of the approach.
The objective function minimizes the total cost of the spill code and
the x86 instruction versions required by the register allocation.
Appel and George's model is similar but not a strict subset of the ORA
model, as the latter limits live-range splitting to a set of
predetermined program points for scalability.
The same authors also propose a simple IP model for solving register
assignment and coalescing optimally, however they report that the
problems cannot be solved in reasonable time and resort to a heuristic
approach in the experiments.
Appel and George's experiments demonstrate that their approach indeed
scales better than the initial ORA solver (a comparison with the
improved ORA model is not available) and improves the speed of the
code generated by a heuristic register allocator by~9.5\%.
To encourage further research on optimal register assignment and
coalescing, the \emph{Optimal Coalescing Challenge}~\cite{OCC} is
proposed.
A few years later, Grund and Hack present an IP approach that
solves~90.7\% of the challenge's coalescing problems (some of them
with thousands of temporaries) optimally~\cite{Grund2007}.
The approach reduces the input problem by preassigning temporaries to
registers whenever it is safe and exploits the structure of the
interference graph to derive additional constraints that speed up
solving.
Grund and Hack's approach is not included in
Table~\ref{tab:register-allocation} since it does not address the core
register allocation problem.

Ebner~\etal{} recognize that the problem tackled by Appel and George
(spilling including live-range splitting and load-store optimization)
can be modeled as a minimum cut problem with capacity
constraints~\cite{Ebner2009}.
Such problems are well-researched in the IP literature and solvers can
typically handle large instances efficiently~\cite{Nemhauser1999}.
Ebner~\etal{} define a network where the nodes correspond to
temporaries at particular program points and the arcs correspond to
possible control transitions.
A solution corresponds to a cut in the network where one partition is
allocated to memory and the other one to registers.
The cost of a solution is the total cost of the arcs crossed by the
cut, where the cost of an arc is the spill cost at its corresponding
program point.
The capacity constraints ensure that the processor registers are not
overused at any program point.
Ebner~\etal{} solve the problem both with a dedicated solver based on
a Lagrangian relaxation (where the capacity constraints are relaxed)
and a commercial IP solver.
Interestingly, their experiments show that the IP solver delivers
optimal solutions in less solving time than all but the simplest
configuration of the dedicated solver.

Finally, Colombet~\etal{} introduce an alternative IP
approach~\cite{Colombet2015} that additionally captures
rematerialization and multiple allocation, and can handle programs in
Static Single Assignment (SSA) form~\cite{Cytron1991}.
SSA is a program form that defines temporaries only once and
explicates control-dependent definitions at join points of the CFG
with special $\phi$-instructions.
SSA has proven itself useful for register allocation as it enables
register assignment in isolation to be solved optimally in polynomial
time~\cite{Hack2006}.
The basic IP model of Colombet~\etal{} resembles that of Appel and
George but has the key difference that the variables corresponding to
\var{r}{t,p} and \var{m}{t,p} are not mutually exclusive, allowing
each temporary $t$ to be allocated to both memory and a register at
the same program point $p$.
The basic model is extended with variables and constraints to handle
rematerialization and some specifics of the SSA form.
Colombet~\etal{}'s compare experimentally their approach with that of
Appel and George and the corresponding subset of the PRA model using
the EEMBC~\cite{Poovey2009} and SPEC2000 benchmarks for
ST231~\cite{Faraboschi2000}, a Very Long Instruction Word (VLIW)
processor.
The results estimate statically that the introduced approach yields
significantly better results than Appel and George's (around~40\%) and
slightly better results (around~5\%) than the adapted PRA approach.
Around half of the estimated improvement over Appel and George's
approach is due to supporting rematerialization, while the other half
is mostly due to avoiding spurious store instructions by allocating
temporaries to memory and registers simultaneously (the adapted PRA
approach is not as penalized since it assigns spurious store
instructions a cost of zero).
An interesting finding is that these substantial estimated
improvements only correspond to modest runtime improvements, due to
spill cost model inaccuracies and interactions with later compiler
stages.
Colombet~\etal{} identify two stages that alter the cost estimated by
their objective function significantly: \emph{immediate folding} and
instruction scheduling.
The former removes immediate loads whenever the immediate can be
directly encoded as an operand in its user instruction (which is
common for rematerialization) while the latter tends to hide the cost
of spill code, in particular for a VLIW processor such as ST231 where
up to four instructions can be scheduled simultaneously.
While the effect of immediate folding can be directly captured in a
register allocation model, Colombet~\etal{} leave it as an open
question whether the effect of scheduling could be modeled without
resorting to an integrated approach as in
Section~\ref{sec:integrated}.

\subsection{Discussion}
\label{sec:register-allocation-discussion}

This section has reviewed combinatorial approaches to register
allocation proposed within the last~20 years.
Since the first proposed approach (ORA), combinatorial register
allocation has been able to handle a wide range of subproblems for
functions with up to a few thousands of instructions, and to
demonstrate actual code quality improvements on dynamic evaluations.
Subsequent extensions for different processor features and
optimization objectives illustrate the flexibility of the
combinatorial approach, for which IP remains essentially unchallenged
as the technique of choice.
Further scalability with virtually no performance degradation can be
attained by decomposing register allocation and focusing in solving
spilling and its closest subproblems (load-store optimization,
live-range splitting, \dots) optimally, as pioneered by Appel and
George's approach.
These benefits come at the expense of flexibility: as observed by Koes
and Goldstein, the approach is less effective when the remaining
subproblems have a high impact on the quality of the solution, such as
in code size optimization.

Despite its relative scalability and wide applicability, combinatorial
register allocation is rarely applied in general-purpose production
compilers.
The single-digit average speedup demonstrated by most reviewed
register allocation approaches is most likely not compelling enough
for compilers aiming at striking a balance with compilation time.
This raises two challenges for combinatorial register allocation:
improving the quality of its generated code and reducing solving time.
The first challenge calls for spill cost models that faithfully
capture the effect of complex memory hierarchies (common in modern
processors).
Falk~\etal{}'s approach takes a step in this direction.
For VLIW processors, an open question is whether their effect can be
captured by a combinatorial register allocation model without
resorting to an integrated approach as in
Section~\ref{sec:integrated}.
The second challenge could be addressed by multiple lines of research.
The study of the structure of IP models for register allocation could
result in significant scalability gains like in instruction scheduling
(see Section~\ref{sec:instruction-scheduling}).
Alternative techniques such as CP have proven successful for other
compiler problems but remain unexplored for register allocation.
Hybrid combinatorial approaches such as PBQP that can resort to
heuristics for large problems could allow compilers to benefit from
combinatorial optimization and scale up with lower maintenance costs.
Finally, combinatorial register allocation has potential to contribute
to the areas of compiler validation, security, and energy efficiency
due to its flexible yet formal nature.

\section{Instruction Scheduling}\label{sec:instruction-scheduling}

Instruction scheduling maps program instructions to basic blocks and
issue cycles within the blocks.
A valid instruction schedule must satisfy the dependencies among
instructions and cannot exceed the capacity of processor resources
such as data buses and functional units.
Typically, instruction scheduling aims at minimizing the
\emph{makespan} of the computed schedule -- the number of cycles it
takes to execute all its instructions.
All approaches presented in this section have makespan minimization as
their objective function unless otherwise stated.

\begin{wrapfigure}{r}{0.19\textwidth}
  \vspace{-0.35cm}%
  \centering
  \hspace{-0.2cm}%
  \adjustbox{trim=0.05cm 0cm 0.18cm 0cm,clip=true}{%
    \scalebox{1.0}{\input{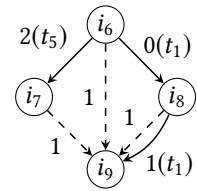}%

\begin{tikzpicture}

    \coordinate (center) {};
    \node [pra storage, above=0.7cm of center] (load) {$\InstrLoad{}$};
    \node [pra storage, left=0.7cm of center] (add) {$\InstrAdd{}$};
    \node [pra storage, right=0.7cm of center] (inc) {$\InstrInc{}$};
    \node [pra storage, below=0.7cm of center] (blt) {$\InstrBlt{}$};

  \begin{scope}[flow, auto, pra label]
    \draw (load) to node [xshift=-0.8cm,yshift=0.6cm] {$2 (\Temp{5})$} (add);
    \draw (load) to node [xshift=-0.1cm,yshift=-0.1cm] {$0 (\Temp{1})$} (inc);
    \draw [dashed] (load) to node [xshift=-0.43cm] {$1$} (blt);
    \draw [dashed] (add) to node [xshift=-0.4cm,yshift=-0.4cm] {$1$} (blt);
    \draw [dashed] (inc) to node [xshift=-0.37cm,yshift=0.45cm] {$1$} (blt);
    \draw (inc) to [bend left=20] node [xshift=-0.2cm,yshift=0cm] {$1 (\Temp{1})$} (blt);
  \end{scope}

\end{tikzpicture}}}
  \vspace{-0.35cm}%
  \caption{DG of $\B{2}$}
  \label{fig:dg}
  \vspace{-0.35cm}%
\end{wrapfigure}
\paragraph{Dependencies}

Data and control flow cause dependencies among instructions.
The dependencies in a basic block form a \emph{dependency graph} (DG)
where nodes represent instructions and an arc $\sequence{i,j}$
indicates that instruction $j$ depends on instruction $i$.
Each arc $\sequence{i,j}$ in a DG is labeled with the latency
\var{l}{i,j} between $i$ and $j$.
The latency dictates the minimum amount of cycles that must elapse
between the issue of the two instructions, and is usually (but not
necessarily) positive.
For modeling convenience, DGs often have an \emph{entry} (\emph{exit})
instruction which precedes (succeeds) all other instructions.

Figure~\ref{fig:dg} shows the DG of the basic block $\B{2}$ in the
running example from Figure~\ref{fig:sum-example}, where
$\InstrLoad{}$ is the entry instruction and $\InstrBlt{}$ is the exit
instruction.
Data dependencies are represented with solid arcs labeled with their
latency (and corresponding temporary, for clarity).
All instructions are assumed to have unit latency, except
$\InstrLoad{}$ ($\loadInstruction{}$) which is assumed to have a
latency of two cycles.
$\sequence{\InstrLoad{},\InstrInc{}}$ is called an
\emph{anti-dependency}: $\InstrLoad{}$ uses $\Temp{1}$ which is then
redefined by $\InstrInc{}$.
Due to the structure of instruction pipelines, instructions related by
anti-dependencies can typically run simultaneously in multiple-issue
processors (see the \emph{Instruction bundling} paragraph in this
section).
Hence, such dependencies are often labeled with a latency of zero
cycles.
Branch dependencies maintain the original control flow (for example
forcing $\InstrAdd{}$ to be scheduled before $\InstrBlt{}$) and are
represented with dashed arcs.
Branch dependencies that are redundant from a latency perspective such
as $\sequence{\InstrLoad{},\InstrBlt{}}$ are kept in the DG for
uniformity.

\paragraph{Resources}

Instructions share limited processor resources such as functional
units and data buses.
The organization of hardware resources varies widely among different
processors and affects the complexity of instruction scheduling
profoundly.
Resources are classified as \emph{single-} or \emph{multi-capacity}
depending on the number of instructions that can access them at the same
time.
Instructions can use either one resource each (\emph{single usage}) or
multiple resources (\emph{multiple usage}).
Resource usage can be \emph{uni-} or \emph{two-dimensional} depending on
whether its duration is one or longer than one.
Finally, resource usage is either \emph{synchronous} if usage by an
instruction $i$ starts at the issue cycle of $i$ or \emph{asynchronous}
otherwise.
Asynchronous resource usage typically arises from irregularities in
processor pipelines, where two instructions might contend for the same
resource (for example a pipeline stage) at a cycle different than
their issues.

\paragraph{Instruction bundling}

Classic RISC processors are \emph{single-issue}: they can issue only
one instruction at each clock cycle.
Modern processors are usually \emph{multiple-issue}: they can issue
several instructions every clock cycle.
To exploit this capability, in-order multiple-issue processors such as
VLIW processors require the compiler to perform \emph{instruction
  bundling}, combining instructions into bundles that are compatible
with the issue restrictions of the processor.
Such restrictions are often caused by constraints imposed by processor
resources.

\paragraph{Scope}

Unlike register allocation which is typically approached globally, the
instruction scheduling literature covers different problem scope
levels, where each level can potentially deliver better code at the
expense of increased complexity:
\begin{inparaitem}[]
\item \emph{local instruction scheduling} schedules instructions
  within basic blocks in isolation, under the assumption that each
  instruction is already placed in a certain basic block;
\item \emph{regional instruction scheduling} considers collections of
  basic blocks with a certain CFG structure; and
\item \emph{global instruction scheduling} considers entire functions.
\end{inparaitem}
Besides assigning each instruction $i$ to an issue cycle within its
block, regional and global instruction scheduling place $i$ into a
basic block.

\paragraph{Evaluation methods}

In general, dynamic evaluations provide the highest accuracy for
combinatorial instruction scheduling approaches that optimize runtime
objectives such as speed.
Static evaluations suit approaches that can capture the behavior of
the program and processor statically (such as local approaches on
in-order processors with constant instruction latencies).

\paragraph{Outline}

\begin{table}[b]%
  \newcommand{\instructionSchedulingApproach}[7]{#1 & #2 & #4 & #5 & #6 & #7}
  \renewcommand{\columnWidth}{0.6cm}
  \setlength{\tabcolsep}{2pt}
  \setlength{\tabulinesep}{0.03cm}
  \caption{Instruction scheduling approaches: technique
    (TC), scope (SC where EN stands for \emph{enumeration}), bundling (BD), multiple usage (MU),
    two-dimensional usage (2D), asynchronous usage (AS),
    size of largest problem
    solved optimally (SZ) in number of instructions, and whether a dynamic evaluation is available (DE).
    \label{tab:instruction-scheduling}}
    \begin{tabu}{%
        l%
        c%
        c%
        >{\centering\arraybackslash}p{\columnWidth}%
        >{\centering\arraybackslash}p{\columnWidth}%
        >{\centering\arraybackslash}p{\columnWidth}%
        >{\centering\arraybackslash}p{\columnWidth}%
        c%
        >{\centering\arraybackslash}p{\columnWidth}}\hline
      \rowfont{\bfseries}
      approach & TC & \instructionSchedulingApproach{SC}{BD}{MC}{MU}{2D}{AS}{SZ} & DE \\
      \hline
      Arya 1985 & IP & \instructionSchedulingApproach{local}{\no}{\no}{\yes}{\yes}{\yes}{36} & \no \\
      \rowcolor{tblrow} Ertl and Krall 1991 & CP & \instructionSchedulingApproach{local}{\no}{\no}{\yes}{\yes}{\yes}{24} & \yes \\
      Chou and Chung 1995 & EN & \instructionSchedulingApproach{local}{\yes}{\no}{\no}{\no}{\no}{20} & \no \\
      \rowcolor{tblrow} Leupers and Marwedel 1997 & IP & \instructionSchedulingApproach{local}{\yes}{\no}{\yes}{\no}{\no}{45} & \no \\
      Wilken~\etal{}~2000 & IP & \instructionSchedulingApproach{local}{\no}{\no}{\no}{\no}{\no}{1000} & \no \\
      \rowcolor{tblrow} van Beek and Wilken~2001 & CP & \instructionSchedulingApproach{local}{\no}{\no}{\no}{\no}{\no}{1000} & \no \\
      Malik~\etal{}~2008 & CP & \instructionSchedulingApproach{local}{\yes}{\yes}{\yes}{\no}{\no}{2600} & \no \\
      \rowcolor{tblrow} Shobaki and Wilken~2004 & EN & \instructionSchedulingApproach{superblock}{\yes}{\no}{\no}{\no}{\no}{$<$ 1236} & \no \\
      Malik~\etal{}~2008 & CP & \instructionSchedulingApproach{superblock}{\yes}{\yes}{\yes}{\yes}{\yes}{2600} & \no \\
      \rowcolor{tblrow} Beg and van Beek~2013 & CP & \instructionSchedulingApproach{superblock}{\yes}{\yes}{\yes}{\yes}{\yes}{100} & \no \\
      Shobaki~\etal{}~2009 & EN & \instructionSchedulingApproach{trace}{\yes}{\no}{\no}{\no}{\no}{$<$ 424} & \no \\
      \rowcolor{tblrow} Govindarajan \etal{}~1994a & IP & \instructionSchedulingApproach{sw.~pipelining}{\yes}{?}{\yes}{\yes}{\no}{?} & \no \\
      Altman \etal{}~1995 & IP & \instructionSchedulingApproach{sw.~pipelining}{\yes}{?}{\yes}{\yes}{\yes}{$\sim{}$20} & \no \\
      \rowcolor{tblrow} Winkel \etal{}~2007 & IP & \instructionSchedulingApproach{global}{\yes}{\no}{\yes}{\yes}{\no}{600} & \yes \\
  \end{tabu}
\end{table}

Table~\ref{tab:instruction-scheduling} classifies combinatorial
instruction scheduling approaches with information about their
optimization technique, scope, problem coverage, approximate
scalability, and evaluation method.
The discussion is structured by scope:
Sections~\ref{sec:local-instruction-scheduling},~\ref{sec:regional-instruction-scheduling},
and~\ref{sec:global-instruction-scheduling} cover local, regional, and
global combinatorial instruction scheduling.
Section~\ref{sec:instruction-scheduling-discussion} closes with a
summary of developments and challenges in combinatorial instruction
scheduling.
Appendix~\ref{app:rp-instruction-scheduling} complements this section
with a review of register pressure-aware instruction scheduling
approaches.

\subsection{Local Instruction Scheduling}~\label{sec:local-instruction-scheduling}

Local instruction scheduling is closely related to the
resource-constrained project scheduling problem
(RCPSP)~\cite{Artigues2008}, which aims at finding the
minimal-makespan schedule of a set of precedence-related activities
with resource demands such that the precedence and resource capacity
constraints are satisfied.
RCPSP solving is a well-researched area, and basic IP approaches are
proposed as early as in the
1960s~\cite{Bowman1959,Wagner1959,Manne1960,Pritsker1969}.
Local instruction scheduling can be seen as an application of the
general RCPSP that is concerned with specific code generation aspects
such as complex processor resources and non-uniform instruction
latencies~\cite{DeDinechin2004}.

\paragraph{Early approaches}

An early combinatorial approach to local instruction scheduling is
introduced by Arya~\cite{Arya1985}.
This approach uses IP to compute optimal schedules for single-issue
vector processors such as the early Cray-1~\cite{Russell1978}.
Arya's model can be seen as an application of Manne's RCPSP
model~\cite{Manne1960} with general (that is, non 0-1) integer
variables \var{s}{i} ($T_i$ in the original notation) representing the
issue cycle of each instruction $i$.
This model allows dependency constraints to be formulated with simple
inequalities on scheduling variables.
For example, the dependency $\sequence{\InstrLoad{}, \InstrAdd{}}$ in
Figure~\ref{fig:dg} yields $\noMathVar{s}{\InstrAdd{}} \ge
\noMathVar{s}{\InstrLoad{}} + 2$.
However, the use of general integer variables precludes multi-capacity
resources and requires auxiliary 0-1 variables \var{p}{i,j} to
indicate whether instruction~$i$ precedes instruction $j$ or vice
versa.
A constraint is added for each pair of instructions $i$, $j$ to relate
the auxiliary and the scheduling variables.
For example, the corresponding constraint for $\InstrAdd{}$
and~$\InstrInc{}$ is $1 \le M \times
\noMathVar{p}{\InstrAdd{},\InstrInc{}} + \noMathVar{s}{\InstrAdd{}} -
\noMathVar{s}{\InstrInc{}} \le M - 1$, where $M$ is a large positive
constant.
Each value of $\noMathVar{p}{\InstrAdd{},\InstrInc{}}$ activates one
of the sides of the compound inequality and enforces a precedence
among $\InstrAdd{}$ and $\InstrInc{}$ in one or the opposite
direction.
Arya shows on three basic blocks of up to 36 instructions that the
approach is feasible and improves the makespan of hand-optimized code
significantly.

Ertl and Krall introduce the first local instruction scheduling
approach based on constraint programming~\cite{Ertl1991}.
In particular, the approach applies \emph{constraint logic
  programming}~\cite{Jaffar1987} which embeds constraint propagation
into logic programming.
Their model targets single-issue, pipelined RISC processors where the
resource consumption of an instruction must be considered at each
stage of the pipeline.
The model includes, for each instruction $i$ and pipeline stage $k$, a
finite integer domain variable \var{s}{i,k} representing the cycle in
which $i$ resides in $k$ (the original notation uses a set of
variables $\set{D_i, W_i, \dots}$ for each instruction $i$ where each
variable corresponds to a pipeline stage such as \emph{decode} and
\emph{writeback}).
Dependencies among instructions are modeled as in Arya's approach.
Equality constraints link the different stage scheduling variables of
an instruction.
An \emph{all-different} global constraint~\cite{CPH:global} ensures
that only one instruction resides in each pipeline stage~$k$ at a time
by forcing all variables $\set{\noMathVar{s}{i_1,k},
  \noMathVar{s}{i_2,k}, \ldots, \noMathVar{s}{i_n,k}}$ to take
different values.
The use of \emph{all-different} yields a compact model as it removes
the need for a quadratic number of auxiliary variables present in
Arya's approach.
For example, enforcing that only one of the four instructions in the
DG of Figure~\ref{fig:dg} can be executed at a time (at the pipeline
stage called $\exStage{}$) is captured by the constraint
$\alldifferent{\set{\noMathVar{s}{\InstrLoad{},\exStage{}},
    \noMathVar{s}{\InstrAdd{},\exStage{}},
    \noMathVar{s}{\InstrInc{},\exStage{}},
    \noMathVar{s}{\InstrBlt{},\exStage{}}}}$.
Ertl and Krall's experiments on a few programs for the Motorola 88100
processor~\cite{Alsup1990} show that their approach improves~8.5\%
and~19\% of the basic blocks scheduled by the Harris C compiler and
GCC~1.3~\cite{GCC2017} respectively.
However, the richness of the resource model comes at the cost of low
scalability: their experiments show that the approach cannot handle
larger problems than Arya's scheduler despite the six-year gap between
them.

Chou and Chung introduce an early combinatorial approach to
instruction scheduling and bundling for multiple-issue
processors~\cite{Chou1995}.
The approach proposes a special-purpose enumeration technique based on
the depth-first search exploration of a search tree where nodes
correspond to bundles of instructions and edges correspond to bundle
issues.
The search tree is pruned by detecting and exploiting domination and
equivalent relations among instructions.
The detection analysis is an extension of early work on multiprocessor
scheduling~\cite{Ramamoorthy1972} where latencies are taken into
account.
Chou and Chung present a limited set of experiments where they
optimally schedule random basic blocks of up to 20 instructions on
processors with a bundle width of four instructions.
The basic blocks are generated with different dependency densities (a
density $d$ means that $d\%$ of all instruction pairs are connected by
a dependency).
They find that the hardest basic blocks have a dependency density of
around 12\%: less (more) dependencies reduce the search by
increasing the number of equivalent (dominated) instructions.

Inspired by early work on IP-based scheduling for high-level
synthesis~\cite{Hwang1991,Gebotys1991}, Leupers and Marwedel introduce
an IP model that includes \emph{alternative instruction versions} and
\emph{side effect handling} for irregular, multiple-issue
DSPs~\cite{Leupers1997}.
Alternative instruction versions (called \emph{alternative encodings}
by Leupers and Marwedel) supports the selection of different
instruction implementations, each using different resources, for each
input instruction.
Side effect handling
\begin{inparaitem}[]
\item supports inserting and scheduling \emph{no-operation
  instructions} that inhibit writes to certain registers, and
\item deals with pairs of instructions that must be scheduled together
  if a particular version of one of them is chosen.
\end{inparaitem}
To schedule instructions in a basic block and take into account the
particularities of DSPs, Leupers and Marwedel define two types of 0-1
variables:
\begin{inparaitem}[]
\item \emph{scheduling variables} \var{s}{v,i,k} ($v_{i,v,k}$ in the
  original notation) indicate whether version $v$ of instruction $i$
  is issued in cycle $k$, and
\item \emph{no-operation variables} \var{n}{r,k} indicate whether
  register $r$ is inhibited by a no-operation in cycle $k$.
\end{inparaitem}
The model includes constraints to ensure that
\begin{inparaitem}[]
\item each instruction is implemented and issued exactly once,
\item dependencies are satisfied,
\item instruction versions that use the same resource are not issued
  in the same cycle, and
\item no-operations are inserted to prevent destroying live data
  stored in registers.
\end{inparaitem}
This model can be seen as a generalization of Bowman's RCPSP
approach~\cite{Bowman1959}.
The use of 0-1 scheduling variables for each cycle allows resource
constraints to be expressed in a more direct and general form than in
Arya's approach.
On the other hand, the number of variables becomes dependent on an
upper bound of the makespan and modeling the dependency constraints
becomes more complex: for each dependency $\sequence{i,j}$ and cycle
$k$ in which $i$ can be issued, a constraint is needed to enforce that
if \var{s}{v,i,k} holds then some \var{s}{v',j,l} must hold for $l$
greater than $k$ (such implications can be expressed in IP models with
inequalities).
For example (assuming that only one version $v_1$ is available for
each instruction), the dependency $\sequence{\InstrLoad{},
  \InstrAdd{}}$ in Figure~\ref{fig:dg} yields the constraint for each
cycle $k$ that if $\noMathVar{s}{v_1,\InstrLoad{},k}$ holds then one
of $\noMathVar{s}{v_1,\InstrAdd{},k + 2},
\noMathVar{s}{v_1,\InstrAdd{},k + 3}, \dots$ must hold.
Leupers and Marwedel present some experimental results on a few
signal-processing basic blocks for Texas Instrument's TMS320C2x~\cite{C2x} and
Motorola DSP56k~\cite{Kloker1987} processors, reporting optimal
solutions for basic blocks of up to 45 instructions.

\paragraph{Modern approaches}

Wilken~\etal{} triggered a new research phase in combinatorial
instruction scheduling~\cite{Wilken2000}.
This phase is characterized by a higher ambition on scalability where
the goal is to solve the largest basic blocks (containing thousands of
instructions) from established benchmark suites such as SPEC95 and
SPEC2000~\cite{CPU}.
The focus shifts from being able to deal with highly irregular
architectures to
\begin{inparaitem}[]
\item understanding and exploiting the structure of the DG (a key
  structure of instruction scheduling), and
\item improving the underlying solving methods by applying
  problem-specific knowledge.
\end{inparaitem}

Wilken~\etal{} use a simple IP model that can be seen as an
application of Pritsker~\etal{}'s RCPSP approach~\cite{Pritsker1969}
targeting an ideal, single-issue processor as seen in
Table~\ref{tab:instruction-scheduling}.
The model contains scheduling variables \var{s}{i,k} ($x^k_i$ in the
original notation) indicating whether instruction $i$ is issued in
cycle $k$, and single-scheduling constraints as in Leupers and
Marwedel's approach.
Following the RCPSP approach of Pritsker~\etal{} yields more compact
dependency constraints than in Leupers and Marwedel's model: if $M$
is the maximum makespan of the schedule, the issue cycle of an
instruction~$i$ can be expressed as $\sum_{k = 0}^{M} k \times
\noMathVar{s}{i,k}$, and a dependency $\sequence{i,j}$ with latency
$l$ can be expressed as $\sum_{k = 0}^{M} k \times \noMathVar{s}{i,k}
+ l \le \sum_{k = 0}^{M} k \times \noMathVar{s}{j,k}$.
For example, the dependency $\sequence{\InstrLoad{}, \InstrAdd{}}$ in
Figure~\ref{fig:dg} yields the single constraint $\sum_{k = 0}^{M} k
\times \noMathVar{s}{\InstrLoad{},k} + 2 \le \sum_{k = 0}^{M} k \times
\noMathVar{s}{\InstrAdd{},k}$.

The main contributions of Wilken~\etal{} are DG transformations and
methods to ease solving the IP problem.
Wilken~\etal{} assume single-issue processors although the model
itself can be generalized for multiple-issue processors.
The DG transformations include:
\begin{inparaitem}[]
\item \emph{partitioning} to decompose DGs while preserving
  optimality,
\item \emph{redundant dependency elimination} to discard dependencies
  that do not affect the solution but incur overhead in the IP model
  (such as the branch dependency $\sequence{\InstrLoad{},
    \InstrBlt{}}$ in Figure~\ref{fig:dg}), and
\item \emph{region linearization} to determine the order in which
  instructions from specific components of the DG called
  \emph{regions} must appear in the optimal solution.
\end{inparaitem}
Before solving, CP-style constraint propagation is applied to discard
scheduling variables and reduce the size of the dependency
constraints.
During solving, violations of the dependency constraints are detected
in the linear relaxations and corrected for all subsequent relaxations
by adding additional constraints that are logically implied.
Wilken~\etal{}'s experiments on an ideal processor with latencies of
up to three cycles show that their improvements enable scheduling all
basic blocks (with up to 1000 instructions) from the SPEC95~\cite{CPU}
floating-point benchmarks with optimal makespan.

Shortly after the paper of Wilken~\etal{}, van Beek and Wilken
introduce a CP approach targeting the same processor
model~\cite{VanBeek2001}.
The model contains finite integer scheduling variables \var{s}{i}
(just~$i$ in the original notation) representing the issue cycle of
each instruction $i$.
Similarly to Ertl and Krall, dependency constraints are directly
expressed as inequalities on the scheduling variables and an
\emph{all-different} global constraint is used to force a single issue
in each cycle.
The model also contains two types of \emph{implied constraints}
(logically redundant constraints that increase the amount of
constraint propagation and thus reduce the search space):
\begin{inparaitem}[]
\item \emph{distance constraints}, which impose a minimum issue
  distance among the boundary instructions of the regions as defined
  by Wilken~\etal{}; and
\item \emph{predecessor/successor constraints}, which enforce
  lower/upper bounds on the scheduling variable domains of
  instructions with multiple predecessors/successors (which can be
  seen as an application of \emph{edge-finding}, a propagation
  algorithm for global scheduling constraints~\cite{CPH:global}).
\end{inparaitem}
For example, instruction $\InstrBlt{}$ in the DG of
Figure~\ref{fig:dg} yields the predecessor constraint
$\noMathVar{s}{\InstrBlt{}} \ge \min \set{\noMathVar{s}{\InstrAdd{}},
  \noMathVar{s}{\InstrInc{}}} + 2$ which may improve propagation on
the earliest possible issue cycle of $\InstrBlt{}$.
Van Beek and Wilken demonstrate the effect of combining their implied
constraints with a custom, lightweight CP solver: in their
experiments, the CP approach is significantly faster (22 times) than
the original IP approach for the same basic blocks from the SPEC95
floating-point benchmarks.

Heffernan and Wilken contribute two further DG transformations to ease
solving: \emph{superior nodes} (an extension to Chou and Chung's
dominance relation) and \emph{superior
  subgraphs}~\cite{Heffernan2006}.
The first transformation identifies pairs of independent instructions
with equal resource usage that can be ordered with artificial
dependencies while preserving the optimality of the original
scheduling problem.
For example, $\InstrInc{}$ in the DG from Figure~\ref{fig:dg} can
always be scheduled before $\InstrAdd{}$ without loss of optimality,
since $\InstrInc{}$ is \emph{closer} to its predecessor $\InstrLoad{}$
and $\InstrAdd{}$ is not \emph{further away} from its successor
$\InstrBlt{}$ in terms of shortest-path distance.
Hence, an artificial dependency $\sequence{\InstrInc{}, \InstrAdd{}}$
with latency 0 can be safely added to the DG.
The second transformation is a computationally heavier generalization
that identifies pairs of isomorphic subgraphs in the DG such that
their instructions can also be ordered while preserving optimality.
The transformations are applied iteratively as their outcome might
expose further transformation opportunities.
Heffernan and Wilken show that their transformations reduce
significantly the upper and lower makespan bounds of the basic blocks
in the SPEC2000 floating-point benchmarks, particularly for
single-issue processors.

Malik~\etal{} (including van Beek) introduce a CP approach that
combines and generalizes the ideas of van Beek, Wilken, and Heffernan
to multiple-issue processors while remaining
scalable~\cite{Malik2008}.
The model is an extension of van Beek and Wilken's model where
single-issue \emph{all-different} constraints are replaced by the more
general \emph{global-cardinality} constraints~\cite{CPH:global}.
A \emph{global-cardinality} constraint for each processor resource
limits the amount of instructions of the same type that can be issued
in the same cycle.
For instance, in Figure~\ref{fig:dg} a processor with a bundle width
of two instructions yields the constraint
$\globalcardinality{\set{\noMathVar{s}{\InstrLoad{}},
    \noMathVar{s}{\InstrAdd{}}, \noMathVar{s}{\InstrInc{}},
    \noMathVar{s}{\InstrBlt{}}}, 2}$, limiting the number of
scheduling variables that can be assigned the same cycle to two.
These constraints reduce the search space using efficient propagation
based on network flow algorithms and reduce solving overhead by
eliminating the need to track the resource usage of each instruction
with additional variables and constraints.
Furthermore, the model contains four types of implied constraints:
\begin{inparaitem}[]
\item \emph{distance} and \emph{predecessor/successor constraints}
  generalized to multiple-issue processors,
\item \emph{safe pruning constraints} which discard suboptimal
  schedules with idle cycles, and
\item \emph{superior subgraph constraints} (called \emph{dominance
  constraints} by Malik \etal{}) which apply Heffernan and Wilken's
  transformations as constraint propagation.
\end{inparaitem}
Unlike earlier modern approaches, Malik \etal{} experiment on a
multiple-issue processor with a bundle width ranging from one to six
and latencies similar to those of the PowerPC
architecture~\cite{Diefendorff1994}.
Their results for the SPEC2000 benchmarks using a custom CP solver
show that most basic blocks (up to 2600 instructions) can be scheduled
with provably optimal makespan.
Interestingly, the experiments show that the number of basic blocks
that are solved optimally in time does not decrease for bundle widths
beyond two instructions.

\subsection{Regional Instruction Scheduling}
\label{sec:regional-instruction-scheduling}

\begin{figure}
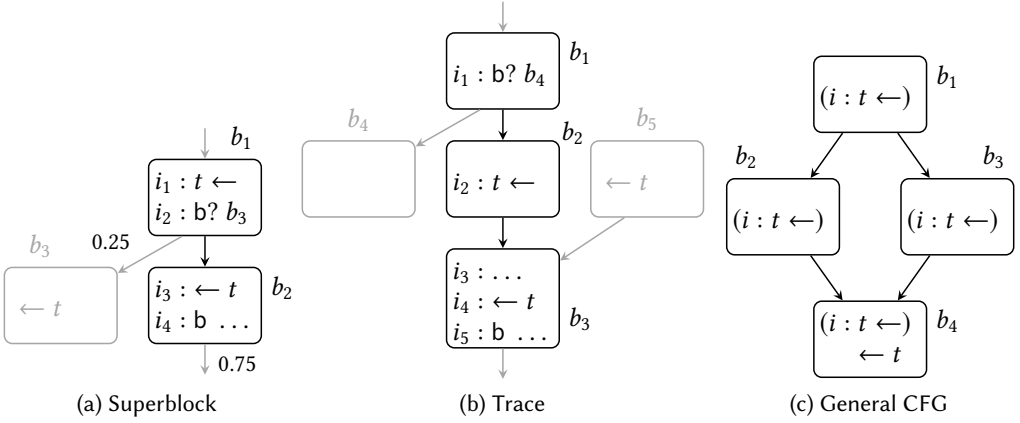

  \hfill
  \subfloat[Superblock]{%
    \adjustbox{trim=0.3cm 0cm 0.2cm 0cm,clip=true}{%
    \scalebox{1.0}{\input{./figures/styles}%
\begin{tikzpicture}
    \coordinate [] (in) {};
    \node [mini cfg block, below=0.4cm of in] (block1) {};
    \node [above left] at (block1.north east) {$\B{1}$};
    \node [mini cfg code] at (block1.north west) {%
      \begin{miniBasicBlock}
        i_1: & t \leftarrow\\
        i_2: & \instruction{b?} \; \B{3}\\
      \end{miniBasicBlock}
    };
    \node [mini cfg block, below=of block1] (block2) {};
    \node [below right] at (block2.north east) {$\B{2}$};
    \node [mini cfg code] at (block2.north west) {%
      \begin{miniBasicBlock}
        i_3: & \leftarrow t\\
        i_4: & \instruction{b} \; \dots\\
      \end{miniBasicBlock}
    };
    \coordinate [below=0.4cm of block2] (out) {};

    \begin{scope}[draw=background, text=background]
    \node [mini cfg block, left=of block2] (block3) {};
    \node [above left] at (block3.north) {$\B{3}$};
    \node [mini cfg code, single instruction] at (block3.north west) {%
      \begin{miniBasicBlock}
        \simpleCopyOperation{}{t}
      \end{miniBasicBlock}
    };
    \end{scope}

    \begin{scope}[flow, auto, font=\small]
      \draw (block1) -- (block2);
      \begin{scope}[draw=background]
        \draw (in) -- (block1);
        \draw [draw=background] ($(block1.south west) !.6! (block1.south)$) to node [xshift=-0.9cm, yshift=0.4cm] {0.25} (block3);
        \draw [draw=background] (block2) to node [xshift=0.05cm, yshift=-0.05cm] {0.75} (out);
      \end{scope}
    \end{scope}

\end{tikzpicture}}}
    \label{fig:superblock-cfg}
  }
  \hspace{-0.25cm}
  \hfill
  \subfloat[Trace]{%
    \adjustbox{trim=0.3cm 0cm 0.05cm 0cm,clip=true}{%
      \scalebox{1.0}{\input{./figures/styles}%
\begin{tikzpicture}
    \coordinate [] (in) {};
    \node [mini cfg block, below=0.4cm of in] (block1) {};
    \node [below right] at (block1.north east) {$\B{1}$};
    \node [mini cfg code, single instruction] at (block1.north west) {%
      \begin{miniBasicBlock}
        i_1: & \instruction{b?} \; \B{4}\\
      \end{miniBasicBlock}
    };
    \node [mini cfg block, below=of block1] (block2) {};
    \node [right, xshift=-0.1cm, yshift=0.1cm] at (block2.north east) {$\B{2}$};
    \node [mini cfg code, single instruction] at (block2.north west) {%
      \begin{miniBasicBlock}
        i_2: & t \leftarrow\\
      \end{miniBasicBlock}
    };
    \node [mini cfg block, below=of block2, minimum height=1.3cm] (block3) {};
    \node [below right] at (block3.east) {$\B{3}$};
    \node [mini cfg code] at (block3.north west) {%
      \begin{miniBasicBlock}
        i_3: & \dots \\
        i_4: & \leftarrow t\\
        i_5: & \instruction{b} \; \dots\\
      \end{miniBasicBlock}
    };
    \coordinate [below=0.4cm of block3] (out) {};

    \begin{scope}[draw=background, text=background]
    \node [mini cfg block, left=of block2] (block4) {};
    \node [above] at (block4.north) {$\B{4}$};
    \node [mini cfg code] at (block4.north west) {%
      \begin{miniBasicBlock}
      \end{miniBasicBlock}
    };
    \node [mini cfg block, right=of block2] (block5) {};
    \node [above] at (block5.north) {$\B{5}$};
    \node [mini cfg code, single instruction] at (block5.north west) {%
      \begin{miniBasicBlock}
        & \leftarrow t\\
      \end{miniBasicBlock}
    };
    \end{scope}

    \begin{scope}[flow, auto]
      \draw (block1) -- (block2);
      \draw (block2) -- (block3);
      \begin{scope}[draw=background]
      \draw (in) -- (block1);
      \draw ($(block1.south west) !.6! (block1.south)$) -- (block4);
      \draw ($(block5.south west) !.6! (block5.south)$) -- (block3);
      \draw (block3) -- (out);
      \end{scope}
    \end{scope}

\end{tikzpicture}}}
    \label{fig:trace-cfg}
  }
  \hfill
  \subfloat[General CFG]{%
    \adjustbox{trim=0.4cm 0cm 0.3cm 0cm,clip=true}{%
      \scalebox{1.0}{\input{./figures/styles}%
\newcommand{\blockShift}{}
\begin{tikzpicture}
    \node [mini cfg block] (block1) {};
    \node [below right] at (block1.north east) {$\B{1}$};
    \node [mini cfg code, single instruction] at (block1.north west) {%
      \begin{miniBasicBlock}
        (i: & t \leftarrow)\\
      \end{miniBasicBlock}
    };
    \coordinate [below=1.1cm of block1] (center) {};

    \node [mini cfg block, left=of center] (block2) {};
    \node [above right] at (block2.north west) {$\B{2}$};
    \node [mini cfg code, single instruction] at (block2.north west) {%
      \begin{miniBasicBlock}
        (i: & t \leftarrow)\\
      \end{miniBasicBlock}
    };
    \node [mini cfg block, right=of center] (block3) {};
    \node [above left] at (block3.north east) {$\B{3}$};
    \node [mini cfg code, single instruction] at (block3.north west) {%
      \begin{miniBasicBlock}
        (i: & t \leftarrow)\\
      \end{miniBasicBlock}
    };

    \node [mini cfg block, below=1.1cm of center] (block4) {};
    \node [below right] at (block4.north east) {$\B{4}$};
    \node [mini cfg code] at (block4.north west) {%
      \begin{miniBasicBlock}
         (i: & t \leftarrow)\\
         & \leftarrow t\\
       \end{miniBasicBlock}
    };

    \begin{scope}[flow, auto]
      \draw (block1) -- (block2);
      \draw (block1) -- (block3);
      \draw (block2) -- (block4);
      \draw (block3) -- (block4);
    \end{scope}

\end{tikzpicture}}}
    \label{fig:general-cfg}
  }
  \hspace{0.1cm}
  \hfill
  \caption{Example CFGs (unrelated to the running example).}
\end{figure}

Local instruction scheduling is well understood and, despite its
theoretical complexity, has been shown in practice to be feasible with
combinatorial techniques.
Unfortunately though, the scope of local instruction scheduling
severely limits the amount of instruction-level parallelism that can
be exploited, particularly for control-intensive programs.
To overcome this limitation, regional instruction scheduling
approaches have been proposed that operate on multiple basic blocks
with a certain CFG structure.
A popular structure in the context of combinatorial instruction
scheduling is the \emph{superblock}.
A superblock is a collection of consecutive basic blocks with a single
entry point and possibly multiple exit points~\cite{Hwu1993}.
Figure~\ref{fig:superblock-cfg} shows an example superblock with two
basic blocks $\B{1}$ and $\B{2}$ annotated with their exit
likelihoods.

\paragraph{Handling compensation code}

Moving an instruction from one basic block to another often requires
the insertion of \emph{compensation code} into yet other basic blocks
to preserve program semantics.
In the case of superblocks, such a situation only arises when an
instruction is moved after an exit point where the result of the
instruction must be available.
For example, temporary $t$ in Figure~\ref{fig:superblock-cfg} must be
available at the exit point of $\B{1}$ since it is used in $\B{3}$.
Therefore, moving the instruction~$i_1$ that defines $t$ to $\B{2}$
must be compensated by the insertion of a duplicate of instruction
$i_1$ in $\B{3}$.
Most regional instruction scheduling approaches based on combinatorial
optimization avoid handling compensation code by either disregarding
the additional cost of placements of instructions that require
compensation code or just disallowing such placements.
The latter can be achieved by adding dependencies from such
instructions to their corresponding exit points, at the expense of
limiting the scope for optimization.

\paragraph{Superblock scheduling}

Shobaki and Wilken present the first combinatorial approach to
superblock scheduling~\cite{Shobaki2004}.
The approach introduces the foundations of a special-purpose
enumeration technique that is later extended in multiple
directions~\cite{Shobaki2009,Shobaki2013}.
The processor model is similar to that of Heffernan and
Wilken~\cite{Heffernan2006} and Malik \etal{}~\cite{Malik2008} for
local instruction scheduling.
The key difference to local scheduling lies in the objective function:
while local scheduling aims at minimizing the makespan of a basic
block, the objective function in superblock scheduling minimizes the
\emph{weighted makespan}.
The weighted makespan of a superblock is the sum of the cycles in
which each exit instruction is scheduled weighted by their exit
likelihood:
$\textSummation{b \in B}{\exitWeight{b} \times \noMathVar{s}{\exitInstruction{b}}}$
where $B$ is the set of basic blocks in the superblock,
$\exitInstruction{b}$ gives the exit instruction of basic block $b$,
and \var{s}{i} gives the issue cycle of instruction $i$.
For example, the objective function of the superblock from
Figure~\ref{fig:superblock-cfg} is $0.25 \times \noMathVar{s}{i_2} +
0.75 \times \noMathVar{s}{i_4}$.
This objective function is common to all available combinatorial
optimization approaches for superblock scheduling.

The enumeration technique follows a solving scheme akin to CP: it
combines depth-first search with constraint propagation (called
\emph{pruning} by the authors) to reduce the search space.
The technique searches for schedules with cost smaller than a certain
upper bound and incrementally increases the upper bound until a
schedule is found.
The incrementality of the process and the completeness of the search
guarantee that the found schedule is optimal.
Each node in the search tree corresponds to an instruction to be
scheduled next, and each root-to-leaf path corresponds to a complete
schedule.
At each non-leaf node, four propagation algorithms (called
\emph{tests} by Shobaki and Wilken) are applied to discard subtrees
that cannot yield a schedule with cost smaller or equal than the upper
bound:
\begin{inparaitem}[]
\item testing whether the partial schedule up to the node is dominated
  according to the relation defined by Chou and Chung
  (Section~\ref{sec:local-instruction-scheduling}),
\item checking whether all instructions can be scheduled after
  simulating the propagation of dependency constraints,
\item comparing the partial schedule to \emph{similar} schedules seen
  before during search likewise Kessler~\cite{Kessler1998}
  (see Appendix~\ref{app:rp-instruction-scheduling}), and
\item applying branch-and-bound with the cost lower bound obtained
  from a relaxation of the problem~\cite{Rim1994,Langevin1996}.
\end{inparaitem}
If any of the algorithms finds that the partial schedule cannot be
completed, backtracking is applied.

Shobaki and Wilken experiment with superblocks from SPEC2000 on
different versions of an ideal processor with bundle width ranging
from one to six instructions.
Their results show that the approach improves the weighted makespan of
the solutions generated by the best available heuristics at the time
by 2.7\% in average using a modest time limit of one second per
superblock.

Malik \etal{} present a CP approach that extends their local scheduler
(discussed in Section~\ref{sec:local-instruction-scheduling}) to
superblocks and a richer resource model~\cite{Malik2008b}.
The basic model adds two elements to the local scheduling one:
\begin{inparaitem}[]
\item the objective function for superblocks as in Shobaki and
  Wilken's approach, and
\item additional variables and constraints to model two-dimensional
  resource usage and instructions that use all resources in their
  issue cycle.
\end{inparaitem}
To scale to larger problem sizes, the model also incorporates
superblock adaptations of
\begin{inparaitem}[]
\item \emph{distance} and
\item \emph{superior subgraph constraints}.
\end{inparaitem}
Malik \etal{} employ a more sophisticated solving procedure than in
previous approaches, involving:
\begin{inparaitem}[]
\item \emph{portfolio search}, where the level and complexity of
  constraint propagation is increased as solving proceeds;
\item \emph{impact-based search}, where the solver selects first the
  variable that causes most propagation to try different search
  alternatives on; and
\item \emph{decomposition}, where the independent basic blocks that
  belong to larger superblocks are solved separately.
\end{inparaitem}
An overview of these techniques is provided by van
Beek~\cite{CPH:search} and Gomes and Walsh~\cite{CPH:random}.
Malik \etal{} present a thorough (though static) experimental
evaluation on the SPEC2000~\cite{CPU} benchmarks for different
variations of the PowerPC~\cite{Diefendorff1994} processor.
Their results show that the approach improves the makespan obtained
from heuristic algorithms and conserves the scalability demonstrated
for basic blocks, despite the generalization to superblocks and the
incorporation of a richer resource model.

Beg and van Beek integrate the superblock approach of Malik \etal{}
with \emph{spatial scheduling}~\cite{Beg2013} using CP.
Spatial scheduling assigns instructions to clusters of functional
units and register banks in clustered processors which are common in
embedded systems.
Spatial and instruction scheduling are interdependent as assigning
data-dependent instructions to different clusters requires scheduling
additional intercluster copy instructions.
The problem is solved by decomposition where clusters are first
assigned and for each given assignment an optimal schedule is computed
as in Malik \etal{}
Experiments by Beg and van Beek on SPEC2000 benchmarks and different
configurations of an ideal clustered processor based on
PowerPC~\cite{Diefendorff1994} show that integrating superblock and
spatial scheduling can improve significantly the weighted makespan of
heuristic approaches for clustered processors (by up to 26\% on
average, depending on the configuration) at the expense of compilation
scalability (the largest superblock solved optimally contains 100
instructions compared to the 2600 instructions of Malik \etal{}).

\paragraph{Trace scheduling}

Superblocks can be generalized into \emph{traces} to further increase
the scope for scheduling, at the expense of higher complexity in
handling compensation code.
A trace is a collection of consecutive basic blocks with possibly
multiple entry and exit points~\cite{Fisher1981}.
Figure~\ref{fig:trace-cfg} shows a trace with three basic blocks
$\B{1}$, $\B{2}$, and $\B{3}$.
Trace scheduling considers all possible execution paths through a
trace.
The example trace consists of three paths: $p_1 = \sequence{\B{1}}$,
$p_2 = \sequence{\B{1}, \B{2}, \B{3}}$, and $p_3 = \sequence{\B{3}}$.
Besides requiring compensation code due to exits (as with superblock
scheduling), trace scheduling must insert compensation code when
moving instructions before entry points.
For example, $i_4$ in Figure~\ref{fig:trace-cfg} is executed by $p_2$
and $p_3$.
Therefore, moving $i_4$ to $b_2$ (above the entry point of $p_3$) must
be compensated by the insertion of a duplicate of $i_4$ in $\B{5}$.

The only available combinatorial approach to trace scheduling is an
extension~\cite{Shobaki2009} of Shobaki and Wilken's superblock
scheduler.
The main difference with the superblock approach lies in the objective
function, which minimizes a combination of the weighted makespan of
all possible execution paths and the cost of inserting compensation
code outside the trace:
$k I + \textSummation{p \in P}{\exitWeight{p} \times \p{\noMathVar{s}{\exitInstruction{p}} - \noMathVar{s}{\entryInstruction{p}}}}$
where $I$ is a variable indicating the number of compensation
instructions introduced in external blocks, $k$ is a constant factor
to control the trade-off between trace speed and size of the rest of
the function, and $P$ is the set of paths from some entry to some exit
in the trace.
For example, assuming that the weights of $p_1$, $p_2$, and $p_3$ are
0.3, 0.5, and 0.2, the objective function of the trace from
Figure~\ref{fig:trace-cfg} is $k I + 0.3 \times \p{\noMathVar{s}{i_1}
  - 0} + 0.5 \times \p{\noMathVar{s}{i_5} - 0} + 0.2 \times
\p{\noMathVar{s}{i_5} - \noMathVar{s}{i_3}}$.
Shobaki \etal{}'s experimental results for the SPEC2006 integer
benchmarks and a UltraSPARC~IV~\cite{Greenley1995} processor (with an
instruction bundle width of four instructions) show that~91\% of the
traces (of up to 424~instructions) can be solved optimally with a
one-second timeout, improving the weighted makespan slightly (2.7\% on
average) and reducing of the compensation code noticeably (15\% on
average) compared to heuristic solutions.

\paragraph{Software pipelining}

\begin{figure}
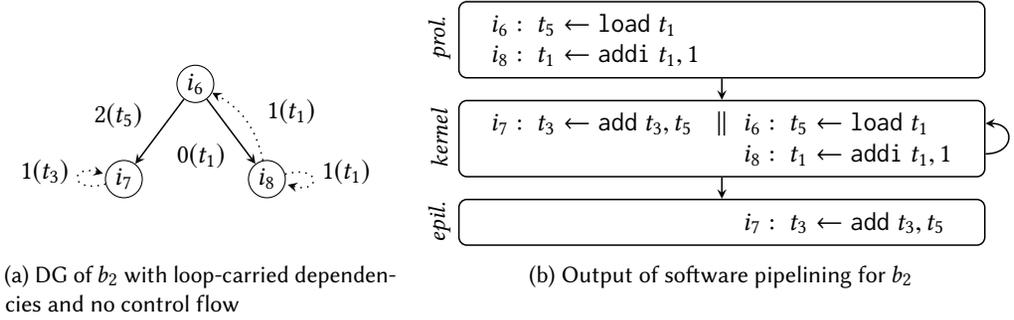

  \hfill
  \subfloat[DG of $\B{2}$ with loop-carried dependencies and no control flow]{%
    \adjustbox{trim=0cm 0.5cm 0cm 0cm,clip=true}{%
      \scalebox{1.0}{\input{./figures/styles}%

\begin{tikzpicture}

    \coordinate (center);
    \node [pra storage, above=1cm of center] (load) {$\InstrLoad{}$};
    \node [pra storage, left=0.7cm of center] (add) {$\InstrAdd{}$};
    \node [pra storage, right=0.7cm of center] (inc) {$\InstrInc{}$};
    \node [below=1.2cm of center] (bottom) {};

  \begin{scope}[flow, auto, pra label]
    \draw (load) to node [xshift=-1cm,yshift=0.5cm] {$2 (\Temp{5})$} (add);
    \draw (load) to node [xshift=-0.85cm,yshift=-0.6cm] {$0 (\Temp{1})$} (inc);
    \draw [dotted] (inc) to [bend right=25] node [xshift=1.05cm, yshift=0.4cm] {$1 (\Temp{1})$} (load);
    \draw [dotted] (inc) edge [loop right] node [yshift=0.1cm] {$1 (\Temp{1})$} (inc);
    \draw [dotted] (add) edge [loop left] node [yshift=0.1cm] {$1 (\Temp{3})$} (add);
  \end{scope}

\end{tikzpicture}}}
    \label{fig:dg-swp}
  }
  \hfill
  \subfloat[Output of software pipelining for $\B{2}$]{%
    \adjustbox{trim=0cm 0cm 0cm 0cm,clip=true}{%
      \scalebox{1.0}{\input{./figures/styles}%
\begin{tikzpicture}
    \coordinate (pro) {};
    \coordinate [below=1.3cm of pro] (ker) {};
    \coordinate [below=1.1cm of ker] (epi) {};
    \node [big cfg block, right=of pro, minimum height=1cm] (prologue) {};
    \node [left, xshift=-0.25cm, yshift=0.43cm, rotate=90] at (prologue.west) {\small{\emph{prol.}}};
    \node [below right] at (prologue.north west) {%
      \begin{basicBlock}
        \InstrLoad{}:\; \naturalOperation{\Temp{5}}{\loadInstruction{}}{\Temp{1}}
        \InstrInc{}:\; \naturalOperation{\Temp{1}}{\addiInstruction{}}{\Temp{1}, 1}
      \end{basicBlock}
    };
    \node [big cfg block, right=of ker, minimum height=1cm] (kernel) {};
    \node [left, xshift=-0.25cm, yshift=0.53cm, rotate=90] at (kernel.west) {\small{\emph{kernel}}};
    \node [below right] at (kernel.north west) {%
      \begin{basicBlock}
         \InstrAdd{}:\; \naturalOperation{\Temp{3}}{\addInstruction{}}{\Temp{3}, \Temp{5}}
      \end{basicBlock}
    };
    \node [below] at (kernel.north) {%
      \begin{basicBlock}
        \parallel\\
      \end{basicBlock}
    };
    \node [below right, xshift=-0.15cm] at (kernel.north) {%
      \begin{basicBlock}
        \InstrLoad{}:\; \naturalOperation{\Temp{5}}{\loadInstruction{}}{\Temp{1}}
        \InstrInc{}:\; \naturalOperation{\Temp{1}}{\addiInstruction{}}{\Temp{1}, 1}
      \end{basicBlock}
    };
    \node [big cfg block, right=of epi, minimum height=0.6cm] (epilogue) {};
    \node [left, xshift=-0.25cm, yshift=0.39cm, rotate=90] at (epilogue.west) {\small{\emph{epil.}}};
    \node [below right, xshift=-0.15cm] at (epilogue.north) {%
      \begin{basicBlock}
        \InstrAdd{}:\; \naturalOperation{\Temp{3}}{\addInstruction{}}{\Temp{3}, \Temp{5}}
      \end{basicBlock}
    };
    \begin{scope}[flow]
      \draw (prologue) -- (kernel);
      \draw ([yshift=-0.2cm]kernel.east) to[bend left=-90, distance=0.4cm] ([yshift=0.2cm]kernel.east);
      \draw (kernel) -- (epilogue);
    \end{scope}

\end{tikzpicture}}}
    \label{fig:swp-output}
  }
  \hfill
  \caption{Software pipelining applied to the running example.}
  \label{fig:software-pipelining}
\end{figure}

A fundamentally different form of regional scheduling is
\emph{software pipelining}~\cite{Rau1981,Lam1988}.
Software pipelining schedules the instructions of multiple iterations
of a loop body simultaneously to increase the amount of available
instruction-level parallelism.
Typically, software pipelining is applied to inner-most loops
consisting of single basic blocks although extensions are proposed to
deal with larger structures~\cite[Section 20.4]{Rong2007}.
Similarly to other instruction scheduling approaches, the input is
represented by a DG, which in the case of software pipelining is
extended with dependencies across iterations (called
\emph{loop-carried dependencies}).
Figure~\ref{fig:dg-swp} shows the DG of Figure~\ref{fig:dg} extended
with loop-carried dependencies as dotted arcs.
Instruction $\InstrBlt{}$ is omitted for simplicity (correct handling
of control instructions needs extra care in software pipelining).

The output of software pipelining is a loop kernel with as short a
makespan as possible (called \emph{initiation interval} in this
context), together with a prologue and epilogue to the kernel where
the instructions of the initial and final iterations of the loop are
scheduled.
Common to all reviewed software pipelining approaches is the
construction of schedules by successively increasing the length of the
initiation interval and invoking the underlying optimization
technique.
The kernel of the first schedule found is thus guaranteed to be of
minimum makespan.
Most approaches define secondary objective functions to complement
initiation interval minimization, usual candidates are register
pressure (see Appendix~\ref{app:rp-instruction-scheduling}) and
resource usage minimization.
Figure~\ref{fig:swp-output} shows the output of software pipelining
for the DG from Figure~\ref{fig:dg-swp} and a multiple-issue
processor.
The kernel is formed with the overlap of two iterations so that the
sum of the array element ($\InstrAdd{}$) can be issued in parallel
($\parallel$) with the load of the next element ($\InstrLoad{}$).
This improves the makespan of the example loop kernel by one cycle
compared to the non-pipelined version.
A comprehensive review of the problem is given by Allan
\etal{}~\cite{Allan1995}.

The basic combinatorial approach to software pipelining, based on IP,
is due to Govindarajan \etal{}~\cite{Govindarajan1994}.
The model includes two main groups of variables:
\begin{inparaitem}[]
\item \emph{kernel scheduling variables} \var{a}{i,k} ($a_{k,i}$ in
  the original notation) indicate whether instruction $i$ is scheduled
  in cycle $k$ within the loop kernel, and
\item \emph{offset scheduling variables} \var{t}{i} give the cycle
  in which the first instance of instruction $i$ is issued.
\end{inparaitem}
The example from Figure~\ref{fig:swp-output} corresponds to a solution
where: \var{a}{\InstrLoad{},0}, \var{a}{\InstrAdd{},0}, and
\var{a}{\InstrInc{},1} are set to one; the remaining \var{a}{i,k} are
set to zero; and $\noMathVar{t}{\InstrLoad{}} = 0$,
$\noMathVar{t}{\InstrAdd{}} = 2$, and $\noMathVar{t}{\InstrInc{}} =
1$.
These variables are subject to linear constraints to enforce that:
\begin{inparaitem}[]
\item only one instance of each instruction is scheduled within the
  loop kernel,
\item the capacity of the resources is not exceeded in any cycle,
\item the data dependencies among instructions are satisfied, and
\item the cycle of an instruction $i$ in the kernel given by
  \var{a}{i,k} corresponds to its offset cycle \var{t}{i}.
\end{inparaitem}
Govindarajan \etal{}'s approach is processor resource-centric: the
model minimizes resource usage as a secondary objective function and
deals with rich resource models including two-dimensional resources
and multiple resource usage.
Their experiments on 22 loops from Livermore~\cite{McMahon1986},
SPEC92~\cite{CPU}, and Whetstone~\cite{Curnow1976} benchmarks for a
multiple-issue processor with a bundle width varying from one to six
show that the optimal solution can be found in reasonable time (less
than 30 seconds) for 91\% of the loops.
Although the maximum size of a loop is not indicated, the reported
initiation intervals suggest that the approach can handle loops in
the order of tens of instructions.

Altman \etal{} (including Govindarajan) extend the resource usage
minimization model~\cite{Altman1995} to handle asynchronous resource
usage by tracking each individual use of each instruction with 0-1
variables and relating the use to its instruction's issue.
Altman \etal{}'s experiments on 1066 loops from Livermore, SPEC92,
Linpack~\cite{Dongarra2003}, NAS~\cite{Bailey1985}, and Whetstone
benchmarks for a variant of the PowerPC-604~\cite{Diefendorff1994}
show that, despite a significant increase in the complexity of the
resource model, their extended approach can solve 75\% of the loops
optimally within 18 minutes.

Most subsequent work on combinatorial software pipelining focuses on
register pressure minimization (see
Appendix~\ref{app:rp-instruction-scheduling}).

\subsection{Global Instruction Scheduling}
\label{sec:global-instruction-scheduling}

The ultimate scope of instruction scheduling is to consider whole
functions simultaneously, without making any assumption on the CFG
structure.
This is referred to as \emph{global instruction scheduling} in this
survey.
Global instruction scheduling allows an instruction to be moved from
its original basic block (called \emph{home basic block}) to any of
its predecessor or successor basic blocks (called \emph{candidate
  basic blocks}).
An exception is made for \emph{non-speculative} instructions (for
example, those which can trigger exceptions), whose candidate basic
blocks are limited to avoid altering the semantics of the original
program.
Handling compensation code in global scheduling becomes a general
problem in which multiple copies of a moved instruction might need to
be placed in different blocks to compensate for its motion.

\paragraph{Winkel}

The first combinatorial approach to global instruction scheduling that
handles compensation code and does not make any significant assumption
on the CFG structure is due to Winkel~\cite{Winkel2004,Winkel2007}.
Winkel's IP approach is highly ambitious in that it captures 15
different types of code motion proposed in previous literature.
However, the basic model is relatively compact, although the complete
model comprises eleven types of variables and 35 types of constraints
to deal with the idiosyncrasies of the targeted processor and advanced
model extensions~\cite{Winkel2007}.
The key idea is to consider for each basic block $b$ and each original
instruction $i$ an optional copy of the instruction, and to enforce
with linear constraints that one such copy is scheduled in every CFG
path through the home basic block of $i$.
The basic types of variables are:
\begin{inparaitem}[]
\item \emph{scheduling variables} \var{s}{i,k,b} ($x^{b,k}_i$ in the
  original notation) to indicate whether a copy of instruction $i$ is
  scheduled in cycle $k$ in basic block $b$,
\item \emph{global motion variables} \var{c}{i,b} ($a^{\uparrow b}_i$
  in the original notation) to indicate whether a copy of instruction
  $i$ is scheduled on all program paths preceding basic block $b$, and
\item \emph{block makespan variables} \var{m}{b,k} ($B^b_k$ in the
  original notation) to indicate whether basic block $b$ has a
  makespan of~$k$.
\end{inparaitem}
The basic types of constraints enforce that:
\begin{inparaitem}[]
\item every path leading to the home basic block of each
  instruction~$i$ contains a scheduled copy of~$i$,
\item the capacity of the resources is not exceeded in any cycle,
\item the data dependencies among instructions are satisfied, and
\item the makespan of each basic block~$b$ is equal to the latest
  cycle of the instructions scheduled in~$b$.
\end{inparaitem}
Additional variables and constraints are proposed, for example, to
discard bundles containing incompatible groups of
instructions~\cite{Winkel2004} and to handle two-dimensional resource
usage across basic block boundaries~\cite{Winkel2007}.
A detailed account of the extensions is given in Winkel's doctoral
dissertation~\cite[Chapter 6]{Winkel2004b}.
The objective function is to minimize the weighted makespan of the
function, that is the sum of the makespan of each basic block weighed
by its estimated execution frequency.
A second solving step is proposed where the makespan of each basic
block is fixed and the amount of code motion is minimized with the
intention to reduce register pressure.
Figure~\ref{fig:general-cfg} shows an example CFG with four basic
blocks $\B{1}$, $\B{2}$, $\B{3}$, and $\B{4}$.
$\B{4}$~is the home basic block of the original instruction $i$ which
has been expanded into one optional copy (parenthesized) per basic
block.
The result $t$ is later used in $\B{4}$.
The motion of $i$ to $\B{2}$ is modeled by scheduling $i$ in $\B{2}$
(that is, setting \var{s}{i,k,\B{2}} to one for some cycle $k$).
Given that decision, the model constraints force the scheduling of $i$
also in $\B{3}$ to compensate the motion.
If $i$ is instead scheduled in $\B{1}$ or in $\B{4}$, every path
leading to $\B{4}$ contains a copy of $i$ and no other copy of $i$ is
scheduled.

Winkel's approach solves the generated IP problems with a commercial
IP solver.
Winkel demonstrates analytically that the model yields problems which
can be solved efficiently with IP techniques because their LP
relaxations provide very tight bounds.
This is achieved by proving that different relaxations of the model
(for example disregarding resource capacity constraints) have LP
relaxations with guaranteed integral solutions~\cite[Chapter
  5]{Winkel2004b}.
Winkel confirms the analytic results with a dynamic experimental
evaluation on the 104 functions of up to 600 instructions that account
for 90\% of the execution time in the SPEC2006~\cite{CPU}
benchmarks.
His experiments for the Itanium~2~\cite{McNairy2003} processor (with a
bundle width of six instructions) show that the \emph{efficiency} of
the IP model makes it feasible to solve functions with hundreds of
instructions, yielding a runtime speedup of 10\% compared to a
heuristic global scheduler and 91\% compared to optimal local
instruction scheduling.
Although Winkel's model does not directly capture the notion of
register pressure, the second solving step is shown to be an effective
replacement as it reduces the percentage of spill code from the
heuristic global scheduler's 0.8\% to 0.2\%.

\subsection{Discussion}
\label{sec:instruction-scheduling-discussion}

This section has reviewed combinatorial approaches to instruction
scheduling proposed within the last 30 years.
For local and superblock instruction scheduling, cross-fertilization
from different combinatorial techniques has resulted in CP approaches
that can optimally solve problems of thousands of instructions in
seconds.
The focus has been on handling in-order processors, and combinatorial
models are available that capture many of the features of complex
pipelines for such processors.
For larger scopes, state-of-the-art approaches based on IP and
special-purpose enumeration are able to deliver optimal solutions for
medium-sized problems of hundreds of instructions.

Similarly to register allocation, combinatorial instruction scheduling
has not seen much adoption in general-purpose production compilers
yet.
The solution quality improvements reported for local and superblock
approaches tend to be modest and, as a consequence of targeting
in-order processors, only statically evaluated.
A challenge for these approaches is to increase the currently limited
range of supported processors by refining and evaluating dynamically
the underlying combinatorial models.
The remaining regional and global approaches show higher potential to
improve code quality, but they do not scale beyond medium-sized
problems yet, which limits their applicability.
These approaches are almost exclusively based on IP and
special-purpose enumeration and might benefit from cross-fertilization
with other techniques as in the local case.
Also, as the processor-memory gap increases, register pressure-aware
approaches (reviewed in Appendix~\ref{app:rp-instruction-scheduling})
could become increasingly relevant.
A particular challenge in this context is to capture the cost incurred
by high register pressure more accurately without suffering from the
low scalability that characterizes the integrated approaches from
Section~\ref{sec:integrated}.

\section{Integrated Register Allocation and Instruction Scheduling}\label{sec:integrated}

As many approaches reviewed in Sections~\ref{sec:register-allocation}
and~\ref{sec:instruction-scheduling} illustrate, the tight
interdependencies between register allocation and instruction
scheduling call for integrated approaches where the trade-offs between
these tasks can be accurately reflected.
Combinatorial optimization eases this effort (at the expense of some
scalability) by allowing the composition of models of different tasks
into a single combinatorial model that reflects the interdependencies
among the tasks and can be solved with the same methods as for each
task in isolation.
Consequently, multiple combinatorial approaches integrating register
allocation and instruction scheduling have been proposed.

\paragraph{Outline}

\begin{table}[b]%
  \newcommand{\integratedApproachRA}[9]{#7 & #6 & #2 & #1 & #3 & #8 & #4 & #5 & #9}
  \newcommand{\integratedApproachIS}[5]{#1 & #3 & #4 & #5}
  \renewcommand{\columnWidth}{0.51cm}
  \setlength{\tabcolsep}{2pt}
  \setlength{\tabulinesep}{0.03cm}
  \caption{Integrated register allocation and instruction scheduling approaches: technique (TC), scope
    (SC), spilling (SP), register assignment~(RA), coalescing (CO),
    load-store optimization (LO), register packing (RP), live-range
    splitting (LS), rematerialization (RM), multiple register banks
    (MB), multiple allocation (MA), bundling (BD), multiple usage~(MU),
    two-dimensional usage (2D), asynchronous usage (AS),
    size of largest problem solved
    optimally (SZ) in number of instructions, and whether a dynamic evaluation is available (DE).\label{tab:integrated}}
    \begin{tabu}{%
        l%
        c%
        c%
        >{\centering\arraybackslash}p{\columnWidth}%
        >{\centering\arraybackslash}p{\columnWidth}%
        >{\centering\arraybackslash}p{\columnWidth}%
        >{\centering\arraybackslash}p{\columnWidth}%
        >{\centering\arraybackslash}p{\columnWidth}%
        >{\centering\arraybackslash}p{\columnWidth}%
        >{\centering\arraybackslash}p{\columnWidth}%
        >{\centering\arraybackslash}p{\columnWidth}%
        >{\centering\arraybackslash}p{\columnWidth}%
        >{\centering\arraybackslash}p{\columnWidth}%
        >{\centering\arraybackslash}p{\columnWidth}%
        >{\centering\arraybackslash}p{\columnWidth}%
        >{\centering\arraybackslash}p{\columnWidth}%
        c%
        >{\centering\arraybackslash}p{\columnWidth}}\hline
      \rowfont{\bfseries}
      approach & TC & SC & \integratedApproachRA{LO}{CO}{RP}{RM}{MB}{RA}{SP}{LS}{MA} & \integratedApproachIS{BD}{MC}{MU}{2D}{AS} & SZ & DE \\
      \hline
      PROPAN & IP & superbl. & \integratedApproachRA{\no}{\no}{\no}{\no}{\yes}{\yes}{\no}{\no}{\no} & \integratedApproachIS{\yes}{\no}{\yes}{\yes}{\no} & 39 & \no \\
      \rowcolor{tblrow} UNISON & CP & global   & \integratedApproachRA{\yes}{\yes}{\yes}{\no}{\yes}{\yes}{\yes}{\yes}{\yes} & \integratedApproachIS{\yes}{\yes}{\yes}{\yes}{\no} & 605 & \no \\
      Chang~1997 & IP & local & \integratedApproachRA{\yes}{\no}{\no}{\no}{\no}{\no}{\yes}{\no}{\no} & \integratedApproachIS{\yes}{\no}{\yes}{\no}{\no} & $\sim{}$10 & \no \\
      \rowcolor{tblrow} Nagar.~2007 & IP & sw.~pip. & \integratedApproachRA{\yes}{\no}{\no}{\no}{\no}{\yes}{\yes}{\yes}{\yes} & \integratedApproachIS{\yes}{\yes}{\no}{\no}{\no} & ? & \no \\
    \end{tabu}%
\end{table}

Table~\ref{tab:integrated} classifies integrated register allocation
and instruction scheduling approaches with information about their
optimization technique, scope, problem coverage, approximate
scalability, and evaluation method.
Sections~\ref{sec:propan}, \ref{sec:unison},
and~\ref{sec:other-integrated} discuss these approaches in further
detail.
Section~\ref{sec:integrated-discussion} closes with a summary of
developments and challenges in integrated combinatorial approaches.
Appendix~\ref{app:full-integration} complements this section with a
review of approaches that also incorporate instruction selection into
the combinatorial model.

\subsection{PROPAN}\label{sec:propan}

K\"{a}stner proposes an IP approach to integrated instruction
scheduling and register assignment in \emph{PROPAN}, a system to
reoptimize assembly code~\cite{Kastner2001}.
Spilling and its associated subproblems are not handled by such
systems (see Table~\ref{tab:integrated}) as the input programs
originate from assembly code and have thus acceptable register
pressure.
PROPAN provides two alternative IP models: a \emph{time-based model}
and an \emph{order-based model}.
The \emph{time-based model} captures only instruction scheduling and
has a structure similar to other reviewed IP models such as that of
Leupers and Marwedel in
Section~\ref{sec:local-instruction-scheduling}, hence the discussion
focuses here on the \emph{order-based model}.

\begin{figure}
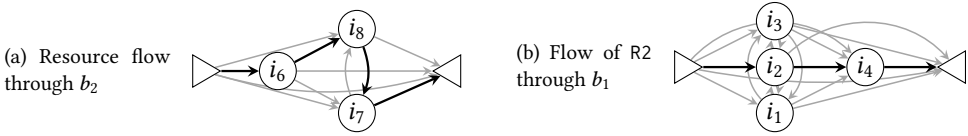

  \hfill
  \subfloat{%
    \scalebox{1.0}{\input{./figures/styles}%
\pgfsetlayers{main,layer0,layer1}%
\begin{tikzpicture}

\begin{pgfonlayer}{layer1}
    \node [pra source right] (source) {};
    \node [pra storage, right=0.5cm of source] (load) {$\InstrLoad{}$};
    \coordinate [right=0.8cm of load] (center) {};
    \node [pra storage, below=0.3cm of center] (add) {$\InstrAdd{}$};
    \node [pra storage, above=0.3cm of center] (inc) {$\InstrInc{}$};
    \node [pra sink left, right=1cm of center] (sink) {};
    \node [font=\footnotesize, left=0cm of source] {\begin{tabular}{p{2.2cm}}(a) Resource flow through $\B{2}$\end{tabular}};
\end{pgfonlayer}

\begin{pgfonlayer}{layer0}
  \begin{scope}[flow, draw=background]
    \draw (source) -- (add);
    \draw (source) -- (inc);
    \draw (source) to [bend right=15] (sink);
    \draw (load) -- (add);
    \draw (load) -- (sink);
    \draw (add) to [bend left=20] (inc);
    \draw (inc) -- (sink);
  \end{scope}

  \begin{scope}[flow, line width=0.3mm]
    \draw (source) -- (load);
    \draw (load) -- (inc);
    \draw (inc) to [bend left=20] (add);
    \draw (add) -- (sink);
  \end{scope}
\end{pgfonlayer}

\end{tikzpicture}}
    \label{fig:propan-resource-fn}
  }
  \hfill
  \subfloat{%
   \scalebox{1.0}{\input{./figures/styles}%
\pgfsetlayers{main,layer0,layer1}%
\begin{tikzpicture}

\begin{pgfonlayer}{layer1}
    \node [pra source right] (source) {};
    \node [pra storage, right=0.7cm of source] (sndmove) {$\InstrSndMove{}$};
    \node [pra storage, below=0.1cm of sndmove] (fstmove) {$\InstrFstMove{}$};
    \node [pra storage, above=0.1cm of sndmove] (li) {$\InstrLi{}$};
    \node [pra storage, right=0.7cm of sndmove] (addaddr) {$\InstrAddAddr{}$};
    \node [pra sink left, right=0.7cm of addaddr] (sink) {};
    \node [font=\footnotesize, left=0cm of source] {\begin{tabular}{p{1.8cm}}(b) Flow of $\code{R2}$ through $\B{1}$\end{tabular}};
\end{pgfonlayer}

\begin{pgfonlayer}{layer0}
  \begin{scope}[flow, draw=background]
    \draw (source) -- (fstmove);
    \draw (source) -- (li);
    \draw (source) to [bend left=45] (addaddr);
    \draw (source) to [bend right=10] (sink);
    \draw (fstmove) to [bend left=10] (sndmove);
    \draw (fstmove) to [bend left=45] (li);
    \draw (fstmove) -- (addaddr);
    \draw (fstmove) -- (sink);
    \draw (sndmove) to [bend left=10] (fstmove);
    \draw (sndmove) to [bend left=10] (li);
    \draw (sndmove) to [bend left=45] (sink);
    \draw (li) to [bend left=15] (sndmove);
    \draw (li) to [bend left=45] (fstmove);
    \draw (li) -- (addaddr);
    \draw (li) -- (sink);
  \end{scope}

  \begin{scope}[flow, line width=0.3mm]
    \draw (source) -- (sndmove);
    \draw (sndmove) -- (addaddr);
    \draw (addaddr) -- (sink);
  \end{scope}
\end{pgfonlayer}

\end{tikzpicture}}
   \label{fig:propan-register-fn}
  }
  \hfill\hspace{0.2cm}
  \caption{Running example in PROPAN.}
  \label{fig:propan}
\end{figure}

The \emph{order-based model} uses general scheduling variables
\var{s}{i} ($t_i$ in the original notation) for each instruction $i$
and inequalities on pairs of scheduling variables to model dependency
constraints as in Arya's approach in
Section~\ref{sec:local-instruction-scheduling}.
Unlike earlier IP-based instruction scheduling models, it captures
resource constraints as a compact resource allocation subproblem based
on flow networks~\cite{Zhang1996,Kastner1999,Kastner2000b}.
A \emph{resource flow graph} is defined for each resource $l$ where
nodes represent instructions (including source and sink nodes), and
each arc $\sequence{i,j}$ represents a potential \emph{flow} of $l$
through instructions $i$ and $j$.
Hence, a flow of~$l$ through a sequence of instructions represents
successive use of $l$ by these instructions and imposes a partial
order among them.
The flow of resource $l$ through instructions $i$ and $j$ is reflected
on the 0-1 variables \var{u}{l,i,j} ($x^l_{i,j}$ in the original
notation).
The partial order given by the \var{u}{l,i,j} variables is connected
to the scheduling variables with additional constraints.
Figure~\ref{fig:propan-resource-fn} shows the flow of a
single-capacity resource through basic block $\B{2}$ from
Figure~\ref{fig:sum-cfg}, where instruction $\InstrBlt{}$ is omitted
for simplicity.
The resource flows through all instructions, enforcing a sequential
schedule with the order $\sequence{\InstrLoad{}, \InstrInc{},
  \InstrAdd{}}$.

Register assignment is handled similarly to resource allocation: a
\emph{register flow graph} (similar to \emph{reuse
  graphs}~\cite{Berson1996,Touati2002}) is constructed where nodes
correspond to instructions and each arc $\sequence{i,j}$ corresponds
to a potential \emph{flow} of a register through instructions $i$ and
$j$.
In this graph, a flow through a sequence of instructions implies that
the temporaries defined by these instructions reuse a register.
The flow of a register through instructions $i$ and $j$ is reflected
on the 0-1 variables \var{r}{i,j} ($x^r_{i,j}$ in the original
notation).
Additional constraints ensure that two instructions $i$, $j$ can only
reuse a register if $j$ is scheduled after the live range of the
temporaries defined by~$i$ (similarly to how instruction-level
parallelism is restricted by Malik~\cite{Malik2008c} and Barany and
Krall's~\cite{Barany2013} approaches discussed in
Appendix~\ref{app:rp-instruction-scheduling}).
Figure~\ref{fig:propan-register-fn} shows a flow of register
$\code{R2}$ through $\B{1}$ from Figure~\ref{fig:sum-cfg}, where
instruction $\InstrBge{}$ is omitted for simplicity: first,
$\code{R2}$ is assigned to $\Temp{2}$ by $\InstrSndMove{}$; then, it
is assigned to $\Temp{4}$ by $\InstrAddAddr{}$, which is possible
since $\Temp{2}$ is last used at that point.
The \emph{register flow graph} can be seen as a simplified, dual
version of the PRA's \emph{multi-commodity network flow} from
Section~\ref{sec:basic-approaches} where temporaries flow through
registers.

The scope of PROPAN is the superblock~\cite[Chapter 7]{Kastner2000b},
and the objective function minimizes its makespan~\cite[Chapter
  5]{Kastner2000b}.
K\"{a}stner evaluates the order-based model with integrated
instruction scheduling and register assignment on five superblocks of
up to 39 instructions from DSP applications for the ADSP-2106x
processor~\cite{SH}.
In this setup, PROPAN generates code with almost 20\% shorter
makespan than list scheduling and optimal register assignment in
isolation, but it does not seem to scale beyond small problems of up
to 39 instructions~\cite{Kastner2001}.

\subsection{UNISON}\label{sec:unison}

Casta\~{n}eda Lozano \etal{} present \emph{UNISON}, a CP approach that
integrates multiple subproblems of register allocation with
instruction scheduling~\cite{Castaneda2012}.
Unlike most integrated combinatorial approaches, UNISON focuses on
capturing a wide array of register allocation subproblems, including
register assignment and packing, coalescing, and spilling for multiple
register banks.
The scope of UNISON's register allocation is global while instructions
are only scheduled locally.

\begin{figure}
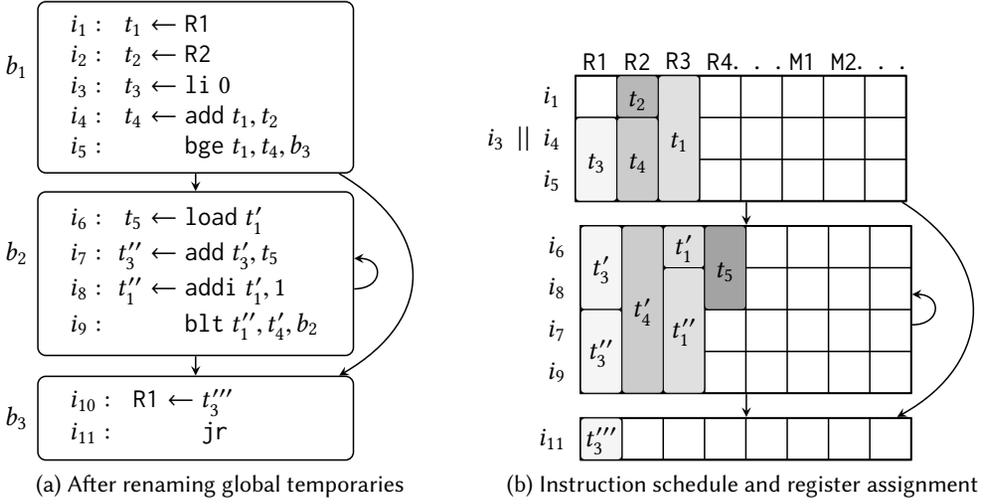

  \hfill
  \subfloat[After renaming global temporaries]{%
    \adjustbox{}{\input{./figures/styles}%
\begin{tikzpicture}
    \coordinate (b1);
    \coordinate [below=2.45cm of b1] (b2);
    \coordinate [below=1.9cm of b2] (b3);
    \node [cfg block, right=of b1, minimum height=2.25cm] (block1) {};
    \node [above left] at (block1.west) {$\B{1}$};
    \node [below right] at (block1.north west) {%
      \begin{labeledBasicBlock}
        \InstrFstMove{}: & \;\; \simpleCopyOperation{\Temp{1}}{\register{R1}}
        \InstrSndMove{}: & \simpleCopyOperation{\Temp{2}}{\register{R2}}
        \InstrLi{}: & \naturalOperation{\Temp{3}}{\loadImmediateInstruction{}}{0}
        \InstrAddAddr{}: & \naturalOperation{\Temp{4}}{\addInstruction{}}{\Temp{1}, \Temp{2}}
        \InstrBge{}: & \branchOperation{\jumpIfGEInstruction{}}{\Temp{1}, \Temp{4}, \B{3}}
      \end{labeledBasicBlock}
    };
    \node [cfg block, right=of b2, minimum height=2.15cm] (block2) {};
    \node [above left] at (block2.west) {$\B{2}$};
    \node [below right] at (block2.north west) {%
      \begin{labeledBasicBlock}[1.1]
        \InstrLoad{}: & \;\; \naturalOperation{\Temp{5}}{\loadInstruction{}}{\Temp{1}'}
        \InstrAdd{}: & \naturalOperation{\Temp{3}''}{\addInstruction{}}{\Temp{3}', \Temp{5}}
        \InstrInc{}: & \naturalOperation{\Temp{1}''}{\addiInstruction{}}{\Temp{1}', 1}
        \InstrBlt{}: & \branchOperation{\jumpIfLTInstruction{}}{\Temp{1}'', \Temp{4}', \B{2}}
      \end{labeledBasicBlock}
    };
    \node [cfg block, right=of b3, minimum height=1.1cm] (block3) {};
    \node [left] at (block3.west) {$\B{3}$};
    \node [below right] at (block3.north west) {%
      \begin{labeledBasicBlock}[1]
        \InstrRetMove{}: & \;\; \simpleCopyOperation{\register{R1}}{\Temp{3}'''}
        \InstrJr{}: & \branchOperation{\returnInstruction{}}{}
      \end{labeledBasicBlock}
    };
    \begin{scope}[flow]
      \draw (block1) -- (block2);
      \draw ([xshift=-0.2cm]block1.south east) to[bend left=60, distance=1.5cm] ([xshift=-0.2cm]block3.north east);
      \draw ([yshift=-0.2cm]block2.east) to[bend left=-90, distance=0.4cm] ([yshift=0.2cm]block2.east);
      \draw (block2) -- (block3);
    \end{scope}

\end{tikzpicture}}
    \label{fig:unison-lssa}
  }
  \hfill
  \subfloat[Instruction schedule and register assignment]{%
   \adjustbox{trim=1cm 0cm 0.5cm 0cm, clip=true}{\input{./figures/unison-sol}}
   \label{fig:unison-sol}
  }
  \hfill
  \caption{Running example in UNISON.}
  \label{fig:unison}
\end{figure}

Casta\~{n}eda Lozano \etal{} apply two program transformations before
formulating a CP problem.
First, the global register allocation problem is decomposed by
renaming \emph{global temporaries} (temporaries whose live ranges span
several basic blocks) into one temporary per basic block, and relating
the renamed temporaries across basic blocks similarly to SSA (see
Section~\ref{sec:decomposed-approaches}).
Figure~\ref{fig:unison-lssa} shows the running example after the
renaming.
Renamed temporaries that originate from the same global temporary (for
example $\Temp{1}$, $\Temp{1}'$, and $\Temp{1}''$) are defined and
used at basic block boundaries (for example,~$\Temp{1}$ is used at the
exit of $\B{1}$ and $\Temp{1}'$ is defined at the entry of $\B{2}$),
although such definitions and uses are omitted from the example for
simplicity.
Then, optional copy instructions are inserted at each definition and
use of a temporary.
These copies can be implemented by alternative instructions such as
register-to-register moves, loads, and stores to support live-range
splitting, spilling, and handling multiple register banks; or
discarded to coalesce their corresponding temporaries.
Memory locations are modeled as individual registers, yielding a
compact model where the register assigned to a temporary determines
its allocation (processor registers or memory), and spill code is
reduced to simple copies between registers.

The CP model includes three main types of finite integer variables:
\begin{inparaitem}[]
\item \emph{register variables} \var{r}{t} give the register to
  which temporary $t$ is assigned,
\item \emph{instruction variables} \var{i}{i} give the instruction
  that implements copy $i$, and
\item \emph{scheduling variables} \var{s}{i} ($c_i$ in the original
  notation) give the issue cycle of instruction $i$.
\end{inparaitem}
Register assignment is modeled as a rectangle packing problem for each
basic block using the \emph{no-overlap} global
constraint~\cite{Beldiceanu2001}, which makes the number of variables
in the model independent of the number of registers.
In this subproblem, the live range of each temporary yields a
rectangle to be packed into a grid where columns correspond to
registers, similarly to the approach of Pereira and
Palsberg~\cite{Pereira2008}.
Overlapping rectangles are forbidden, which models precisely that
interfering temporaries cannot share registers.
The integration between register allocation and instruction scheduling
is reflected on the vertical dimension, where rows correspond to issue
cycles and the top and bottom of temporary $t$'s rectangle are
determined by the issue cycles of $t$'s definer and last user
instruction.
Figure~\ref{fig:unison-sol} shows a solution to the rectangle packing
problems (and hence a schedule and register assignment) for the
example in Figure~\ref{fig:unison-lssa}, assuming a multiple-issue
processor with latencies as given by the DG of Figure~\ref{fig:dg}.
In this example, the rectangle corresponding to $t_5$ is packed in the
column $\noMathVar{r}{t_5} = \register{R4}$ and stretches from row
$\noMathVar{s}{i_6} = 1$ to $\noMathVar{s}{i_7} = 3$ ($i_6$ defines
$t_5$ while $i_7$ uses it).
The \emph{registers} \texttt{M1}, \texttt{M2}, \dots correspond to
memory locations, which are unused in the example since there is no
spilling (all copies are discarded).
The packing subproblems from different basic blocks are connected by
constraints that assign renamed temporaries that originate from the
same global temporary (such as $t_1$, $t_1'$, and $t_1''$) to the same
register.
To ensure that the capacity of processor resources is not exceeded,
UNISON uses \emph{cumulative} global constraints~\cite{CPH:global}.
Compared to Malik \etal{}'s approach based on
\emph{global-cardinality} constraints (see
Section~\ref{sec:local-instruction-scheduling}), the \emph{cumulative}
constraints enable a more compact model as two-dimensional usage can
be modeled without resorting to additional variables and constraints.
Finally, additional constraints are used to model the behavior of the
copies, allocating the temporaries of a copy $c$ to processor or
memory \emph{registers} depending on the instruction \var{i}{c}
implementing $c$, or forcing them to the same register if $c$ is
discarded.
The objective is to minimize the \emph{weighted cost} of the input
function
$\textSummation{b \in B}{\exitWeight{b} \times \blockCost{b}}$
where $B$ is the set of basic blocks in the function and
$\blockCost{b}$ is a generic cost function for basic block $b$.
By combining different definitions of $\exitWeight{b}$ and
$\blockCost{b}$ the objective function can be adapted to optimize for
different criteria.
For example setting $\exitWeight{b}$ to the estimated execution
frequency of $b$ and $\blockCost{b}$ to $b$'s makespan results in
speed optimization, while setting $\exitWeight{b}$ to one and
$\blockCost{b}$ to the total size of the instructions in $b$ results
in code size optimization.

Casta\~{n}eda Lozano \etal{} propose a model extension to capture
load-store optimization and multiple allocation, and two methods to
improve the robustness and scalability of UNISON~\cite{Castaneda2014}:
\begin{inparaitem}[]
\item \emph{presolving} reformulates the CP problem by adding implied
  constraints and bounds; and
\item \emph{decomposition} splits the CP problem into a global problem
  (where the decisions affecting multiple basic blocks are taken), and
  a local problem per basic block $b$ (where the internal decisions of
  $b$ are taken independently).
\end{inparaitem}

Casta\~{n}eda Lozano \etal{}'s evaluate UNISON statically with 55
functions of up to 1000 instructions from MediaBench
benchmarks~\cite{Lee1997} for Hexagon~\cite{Codrescu2014}, a VLIW
processor with a bundle width of four instructions and 32
general-purpose registers.
The results show that UNISON reduces the weighted makespan of a
heuristic approach (LLVM) by 7\% on average, the approach can be
easily adapted for code size optimization (resulting code size is on
par with LLVM), and the extension with load-store optimization and
multiple allocation indeed improves solution quality.
The presolving and decomposition methods allow UNISON to scale up to
functions of hundreds of instructions.
An open question is whether instruction scheduling in UNISON could be
extended to superblocks or beyond while preserving this scalability.

\subsection{Other Approaches}\label{sec:other-integrated}
Other IP approaches have been proposed that integrate multiple-issue
instruction scheduling with subproblems of register allocation for
different program scopes.
Chang \etal{} propose an early IP model for local instruction
scheduling and spilling that includes load-store
optimization~\cite{Chang1997}.
Similarly to UNISON, input programs are extended with optional store
and load instructions after temporary definitions and before temporary
uses.
The model includes
\begin{inparaitem}[]
\item \emph{scheduling variables} \var{s}{i,k} to indicate whether
  instruction $i$ is scheduled in cycle $k$,
\item \emph{live range variables} \var{l}{t,k} to indicate whether
  temporary $t$ is live in cycle~$k$ (similarly to
  Govindarajan~\etal{}'s register pressure-aware scheduling approach
  reviewed in Appendix~\ref{app:rp-instruction-scheduling}), and
\item \emph{activeness variables} \var{a}{i,k} to indicate whether the
  store or load instruction $i$ is active in cycle~$k$.
\end{inparaitem}
These variables correspond to $x_{i,k}$, $U_{t,k}$, and $f_{i,k}$ in
the original notation.
Single-scheduling, dependency, and resource constraints are defined
similarly to other time-based models in
Section~\ref{sec:local-instruction-scheduling}.
The number of simultaneous live temporaries in each cycle is
constrained to be less than or equal to the number of processor
registers.
The model only considers a store (load) instruction as active if it is
not scheduled in the same cycle as its predecessor (successor)
instruction in the DG.
Two alternative objective functions are proposed:
\begin{inparaitem}[]
\item to minimize the used registers given a fixed makespan, and
\item to minimize the makespan given a fixed number of processor
  registers.
\end{inparaitem}
Chang \etal{} compare two manually-specified models of a basic block
with ten instructions for an ideal processor, with and without
spilling, and find that the model with spilling takes one order of
magnitude more time to be solved optimally.

Nagarakatte and Govindarajan propose an IP model for register
allocation and spill-code scheduling in software
pipelining~\cite{Nagarakatte2007}.
The approach assumes a software-pipelined loop kernel with a given
initiation interval and computes (if feasible) a register allocation
(including assignment, spilling, load-store optimization, and multiple
allocation) and a schedule of the spill code.
The IP model includes three types of variables:
\begin{inparaitem}[]
\item \emph{register assignment variables} \var{a}{t,r,k}
  indicate whether temporary $t$ is assigned to register
  $r$ in cycle $k$;
\item \emph{store variables} \var{s}{t,r,k} indicate
  whether temporary $t$ is stored from register $r$ to
  memory in cycle $k$; and
\item \emph{load variables} \var{l}{t,r,k} indicate
  whether temporary $t$ is loaded from memory to register
  $r$ in cycle $k$.
\end{inparaitem}
These variables correspond to $TN_{t,r,k}$, $STN_{t,r,k}$, and
$LTN_{t,r,k}$ in the original notation.
The model has linear constraints to ensure that each temporary:
\begin{inparaitem}[]
\item is assigned to registers in the cycles in which it is defined
  and used,
\item is assigned to a register in a certain cycle as a continuation
  of an assignment in previous cycles or if it is just loaded from
  memory,
\item is assigned to a register before being stored to memory, and
\item is stored in memory before being loaded into a register.
\end{inparaitem}
Other constraints enforce that multiple temporaries are not assigned
to the same register, and that the capacity of the memory functional
units is not exceeded.
The model also includes dominance constraints, for instance to enforce
that each temporary is stored at most once, and that temporaries whose
values are already contained in registers are not loaded from memory.
The objective function minimizes spill code.
Nagarakatte and Govindarajan's experiments for an ideal processor with
a bundle width of ten instructions on 268 loops from the
SPEC~\cite{CPU} and Perfect Club~\cite{Berry1988} benchmarks show that
their approach yields 18\% (13\%) less spill code than its
state-of-the-art heuristic counterpart~\cite{Zalamea2000}, avoids
increasing the initiation interval in~11\% (12\%) of the non-trivial
loops, and takes on average 78 (18) seconds to solve each loop
optimally for 32 (16) registers.

\subsection{Discussion}
\label{sec:integrated-discussion}

This section has reviewed combinatorial approaches to integrated
register allocation and instruction scheduling proposed within the
last 20 years.
As a consequence of their intrinsic complexity, the proposed IP and CP
approaches only scale to small problems, except UNISON which handles
medium-sized problems.
The improvement potential of the integrated approaches should
theoretically be greater than the combined effect of solving each
problem optimally in isolation, however it is not fully determined yet
due to scalability limitations (it has been observed that the
improvement over heuristic approaches grows with the problem size) and
the unavailability of dynamic evaluations.

The main challenge for integrated approaches is to scale up to larger
problems: only then a thorough dynamic evaluation could reveal their
improvement potential.
Promising directions are identifying and exploiting problem
decompositions as in UNISON and employing hybrid combinatorial
optimization techniques~\cite{CPH:local,CPH:or}.
At the same time, improving on state-of-the-art heuristic approaches
presupposes combinatorial models that cover the same subproblems at
the same scope.
None of the reviewed approaches has yet reached this level of coverage
(for example, rematerialization is not addressed), and many of them
are limited in scope (for example, performing register allocation
locally).
Overcoming these limitations while scaling to at least medium-sized
problems remains a major challenge.

\section{Conclusion}\label{sec:conclusion}

This paper reviews and classifies the current body of literature on
combinatorial register allocation and instruction scheduling, two
central compiler back-end problems, thus allowing researchers and
practitioners in the field to identify developments, trends,
challenges, and opportunities.

Significant progress has been made for each of the two problems in
isolation since the application of combinatorial optimization was
first proposed in the 1980s.
Today, the vast majority of register allocation and local instruction
scheduling problems found in practice can be solved optimally in the
order of seconds.
Yet, combinatorial optimization is rarely applied in production
compilers.
Making combinatorial register allocation or local instruction
scheduling more compelling involves three challenges: reducing solving
time, improving the quality of the generated code by capturing the
cost of each optimization decision accurately in the combinatorial
model, and demonstrating the improvement with thorough dynamic
evaluations.
Less fundamental factors such as licensing mismatches and software
integration obstacles have to be dealt with as well.

More sophisticated instruction scheduling and integrated approaches
have the potential to bring code quality improvements to motivate a
paradigm shift, however leveraging this potential presupposes
combinatorial models that are accurate and complete enough to deliver
substantial improvements in code quality.
Devising such models and the corresponding solving methods to scale up
to problem sizes of practical interest remains a major challenge for
researchers in compilers and combinatorial optimization.

\begin{acks}
The authors are grateful for helpful comments from Magnus Boman, David
Broman, Mats Carlsson, Mattias Eriksson, Lang Hames, Gabriel Hjort
Blindell, Sven Mallach, Saranya Natarajan, Bernhard Scholz, and the
anonymous reviewers.
This work has been partially funded by
\grantsponsor{Ericsson}{Ericsson AB}{https://www.ericsson.com/} and
the \grantsponsor{VR}{Swedish Research Council
  (VR)}{https://www.vr.se/} under grant~\grantnum{VR}{621-2011-6229}.
\end{acks}

\bibliographystyle{ACM-Reference-Format}
\bibliography{paper}

\newpage

\appendix

\section{Register Pressure-Aware Instruction Scheduling}\label{app:rp-instruction-scheduling}

This appendix reviews register pressure-aware (\emph{RP}) instruction
scheduling approaches.
This task is performed before register allocation (see
Figure~\ref{fig:backend}) and aims at minimizing register pressure
(number of temporaries that need to be stored simultaneously) or
striking a balance with makespan minimization.
The importance of each objective is typically dictated by the
characteristics of the target: out-of-order processors tend to benefit
from spill reduction by minimizing register pressure while VLIW
processors tend to benefit from higher instruction-level parallelism
by minimizing makespan.
When register pressure minimization is the only concern, instruction
scheduling is typically reduced to its simplest configuration
(\emph{instruction ordering} which assumes a single-issue processor
with a trivial resource model and unit latency instructions as
reflected in Table~\ref{tab:rp-instruction-scheduling} for the
corresponding approaches).
Addressing also instruction-level parallelism necessitates more
advanced processor configurations.
Register pressure-aware approaches have shown moderate code quality
improvements over heuristics on dynamic evaluations for out-of-order
processors.

\begin{figure}
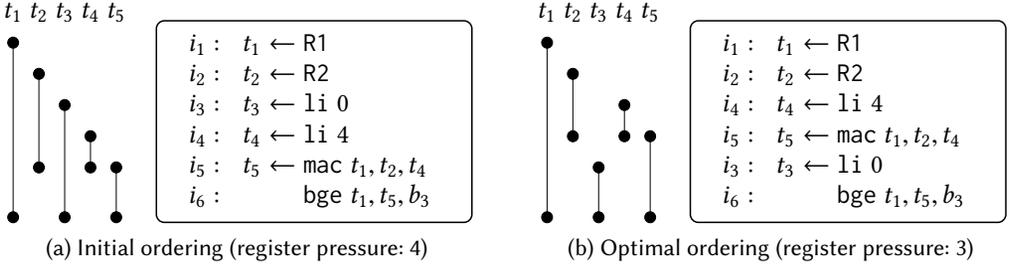

  \hspace{0.05cm}
  \subfloat[Initial ordering (register pressure: 4)]{%
    \scalebox{1.0}{\input{./figures/styles}%
\begin{tikzpicture}
    \node (b1-label) {};
    \node [below=0.7cm of b1-label] (b1) {};
    \node [cfg block, right=of b1, minimum height=2.65cm] (block1) {};
    \node [below right] at (block1.north west) {%
      \begin{labeledBasicBlock}
        \InstrPRFstMove{}: & \;\; \simpleCopyOperation{\Temp{1}}{\register{R1}}
        \InstrPRSndMove{}: & \simpleCopyOperation{\Temp{2}}{\register{R2}}
        \InstrPRFstLi{}: & \naturalOperation{\Temp{3}}{\loadImmediateInstruction{}}{0}
        \InstrPRSndLi{}: & \naturalOperation{\Temp{4}}{\loadImmediateInstruction{}}{4}
        \InstrPRMac{}: & \naturalOperation{\Temp{5}}{\macInstruction{}}{\Temp{1}, \Temp{2}, \Temp{4}}
        \InstrPRBge{}: & \branchOperation{\jumpIfGEInstruction{}}{\Temp{1}, \Temp{5}, \B{3}}
      \end{labeledBasicBlock}
    };

    \node [left=1.8cm of block1.north west,yshift=0.1cm,xshift=0.15cm] (t1) {$\Temp{1}$};
    \coordinate [yshift=-0.22cm] (ls1) at (block1.north west -| t1);
    \coordinate (le1) at (block1.south west -| t1);

    \node [right=-0.15cm of t1] (t2) {$\Temp{2}$};
    \coordinate [yshift=-0.41cm] (ls2) at (ls1 -| t2);
    \coordinate [yshift=0.655cm] (le2) at (le1 -| t2);

    \node [right=-0.15cm of t2] (t3) {$\Temp{3}$};
    \coordinate [yshift=-0.41cm] (ls3)  at (ls2 -| t3);
    \coordinate (le3) at (block1.south west -| t3);

    \node [right=-0.15cm of t3] (t4) {$\Temp{4}$};
    \coordinate [yshift=-0.41cm] (ls4)  at (ls3 -| t4);
    \coordinate (le4) at (le2 -| t4);

    \node [right=-0.15cm of t4] (t5) {$\Temp{5}$};
    \coordinate [yshift=-0.41cm] (ls5) at (ls4 -| t5);
    \coordinate (le5) at (block1.south west -| t5);

    \begin{scope}[live range]
      \draw (ls1.center) -- (le1.center);
      \draw (ls2.center) -- (le2.center);
      \draw (ls3.center) -- (le3.center);
      \draw (ls4.center) -- (le4.center);
      \draw (ls5.center) -- (le5.center);
    \end{scope}

\end{tikzpicture}}
    \label{fig:sum-int-cfg}
  }
  \hspace{0.4cm}
  \subfloat[Optimal ordering (register pressure: 3)]{%
    \scalebox{1.0}{\input{./figures/styles}%
\begin{tikzpicture}
    \node (b1-label) {};
    \node [below=0.7cm of b1-label] (b1) {};
    \node [cfg block, right=of b1, minimum height=2.65cm] (block1) {};
    \node [below right] at (block1.north west) {%
      \begin{labeledBasicBlock}
        \InstrPRFstMove{}: & \;\; \simpleCopyOperation{\Temp{1}}{\register{R1}}
        \InstrPRSndMove{}: & \simpleCopyOperation{\Temp{2}}{\register{R2}}
        \InstrPRSndLi{}: & \naturalOperation{\Temp{4}}{\loadImmediateInstruction{}}{4}
        \InstrPRMac{}: & \naturalOperation{\Temp{5}}{\macInstruction{}}{\Temp{1}, \Temp{2}, \Temp{4}}
        \InstrPRFstLi{}: & \naturalOperation{\Temp{3}}{\loadImmediateInstruction{}}{0}
        \InstrPRBge{}: & \branchOperation{\jumpIfGEInstruction{}}{\Temp{1}, \Temp{5}, \B{3}}
      \end{labeledBasicBlock}
    };

    \node [left=1.8cm of block1.north west,yshift=0.1cm,xshift=0.15cm] (t1) {$\Temp{1}$};
    \coordinate [yshift=-0.22cm] (ls1) at (block1.north west -| t1);
    \coordinate (le1) at (block1.south west -| t1);

    \node [right=-0.15cm of t1] (t2) {$\Temp{2}$};
    \coordinate [yshift=-0.41cm] (ls2) at (ls1 -| t2);
    \coordinate [yshift=1.065cm] (le2) at (le1 -| t2);

    \node [right=-0.15cm of t2] (t3) {$\Temp{3}$};

    \node [right=-0.15cm of t3] (t4) {$\Temp{4}$};
    \coordinate [yshift=-0.41cm] (ls4)  at (ls2 -| t4);
    \coordinate (le4) at (le2 -| t4);

    \node [right=-0.15cm of t4] (t5) {$\Temp{5}$};
    \coordinate [yshift=-0.41cm] (ls5) at (ls4 -| t5);
    \coordinate (le5) at (block1.south west -| t5);

    % coordinates of t3's live range
    \coordinate [yshift=-0.41cm] (ls3)  at (ls5 -| t3);
    \coordinate (le3) at (block1.south west -| t3);

    \begin{scope}[live range]
      \draw (ls1.center) -- (le1.center);
      \draw (ls2.center) -- (le2.center);
      \draw (ls3.center) -- (le3.center);
      \draw (ls4.center) -- (le4.center);
      \draw (ls5.center) -- (le5.center);
    \end{scope}

\end{tikzpicture}}
    \label{fig:sum-int-rp-cfg}
  }
  \caption{Register pressure-aware ordering for an extended version
    of $\B{1}$.}
  \label{fig:sum-int-rp-example}
\end{figure}

Figure~\ref{fig:sum-int-rp-example} shows two instruction orderings of
an extended version of $\B{1}$ from Figure~\ref{fig:sum-cfg} where the
value held by $\Temp{5}$ is computed by a multiply-and-accumulate
instruction ($\macInstruction$) that adds the base address of an array
($\Temp{1}$) with the product of its length ($\Temp{2}$) and its
storage size ($\Temp{4}$).
In Figure~\ref{fig:sum-int-cfg}, four temporaries ($\set{\Temp{1},
  \Temp{2}, \Temp{3}, \Temp{4}}$) need to be stored simultaneously in
different registers between $\InstrPRSndLi{}$ and $\InstrPRMac{}$.
In Figure~\ref{fig:sum-int-rp-cfg}, register pressure-aware scheduling
delays the issue of $\InstrPRFstLi{}$ to achieve optimal register
pressure with at most three temporaries simultaneously live (first
$\set{\Temp{1}, \Temp{2}, \Temp{4}}$ and then $\set{\Temp{1},
  \Temp{3}, \Temp{5}}$).

\paragraph{Outline}

Table~\ref{tab:rp-instruction-scheduling} classifies register
pressure-aware instruction scheduling approaches with information
about their optimization technique, scope, problem coverage,
approximate scalability, and evaluation method.
As in Section~\ref{sec:instruction-scheduling}, the discussion is
structured by scope:
Sections~\ref{sec:local-rp-instruction-scheduling},~\ref{sec:regional-rp-instruction-scheduling},
and~\ref{sec:global-rp-instruction-scheduling} cover local, regional,
and global approaches.

\begin{table}[b]%
  \newcommand{\instructionSchedulingApproach}[7]{#1 & #2 & #4 & #5 & #6 & #7}
  \renewcommand{\columnWidth}{0.6cm}
  \setlength{\tabcolsep}{2pt}
  \setlength{\tabulinesep}{0.03cm}
  \caption{Register pressure-aware instruction scheduling approaches: technique
    (TC), scope (SC where EN stands for \emph{enumeration}), bundling (BD), multiple usage (MU),
    two-dimensional usage (2D), asynchronous usage (AS),
    size of largest problem
    solved optimally (SZ) in number of instructions, and whether a dynamic evaluation is available (DE).
    \label{tab:rp-instruction-scheduling}}
    \begin{tabu}{%
        l%
        c%
        c%
        >{\centering\arraybackslash}p{\columnWidth}%
        >{\centering\arraybackslash}p{\columnWidth}%
        >{\centering\arraybackslash}p{\columnWidth}%
        >{\centering\arraybackslash}p{\columnWidth}%
        c%
        >{\centering\arraybackslash}p{\columnWidth}}\hline
      \rowfont{\bfseries}
      approach & TC & \instructionSchedulingApproach{SC}{BD}{MC}{MU}{2D}{AS}{SZ} & DE \\
      \hline
      Kessler~1998 & EN & \instructionSchedulingApproach{local}{\no}{\no}{\no}{\no}{\no}{25} & \no \\
      \rowcolor{tblrow} Govindarajan~\etal{}~2003 & IP & \instructionSchedulingApproach{local}{\no}{\no}{\no}{\no}{\no}{$\sim{}$20} & \yes \\
      Malik~2008 & CP & \instructionSchedulingApproach{local}{\yes}{\yes}{\yes}{\no}{\no}{50} & \no \\
      \rowcolor{tblrow} Shobaki~\etal{}~2013 & EN & \instructionSchedulingApproach{local}{\no}{\no}{\no}{\no}{\no}{664} & \yes \\
      Govindarajan \etal{}~1994b & IP & \instructionSchedulingApproach{sw.~pipelining}{\yes}{?}{\yes}{\no}{\no}{?} & \no \\
      \rowcolor{tblrow} Eichenberger \etal{}~1997 & IP & \instructionSchedulingApproach{sw.~pipelining}{\yes}{?}{\yes}{\yes}{\yes}{41} & \no \\
      Dupont De Dinechin~2004 & IP & \instructionSchedulingApproach{sw.~pipelining}{\yes}{?}{\yes}{\yes}{\yes}{?} & \no \\
      \rowcolor{tblrow} Domaga\l{}a \etal{}~2016 & CP & \instructionSchedulingApproach{loop unrolling}{\no}{?}{\no}{\no}{\no}{?} & \yes \\
      Barany~\etal{}~2013 & IP & \instructionSchedulingApproach{global}{\no}{\no}{\no}{\no}{\no}{$\sim{}$1000} & \yes \\
  \end{tabu}
\end{table}

\subsection{Local RP Instruction Scheduling}~\label{sec:local-rp-instruction-scheduling}

The first combinatorial approach to local register pressure-aware
instruction scheduling is due to Kessler~\cite{Kessler1998}.
Kessler proposes a special-purpose enumeration technique for ordering
and a scheduling extension with long instruction latencies.
The technique explores a search tree where nodes correspond to sets of
instructions that can be issued next (because their DG predecessors
are already issued) and edges correspond to instruction issues.
For example, the root node of the search tree for
Figure~\ref{fig:sum-int-rp-example} is
$\set{\InstrPRFstMove,\InstrPRSndMove,\InstrPRFstLi,\InstrPRSndLi}$,
and issuing one of $\InstrPRFstMove$, $\InstrPRSndMove$,
$\InstrPRFstLi$, and $\InstrPRSndLi$ induces its children
$\set{\InstrPRSndMove,\InstrPRFstLi,\InstrPRSndLi}$,
$\set{\InstrPRFstMove,\InstrPRFstLi,\InstrPRSndLi}$,
$\set{\InstrPRFstMove,\InstrPRSndMove,\InstrPRSndLi}$ and
$\set{\InstrPRFstMove,\InstrPRSndMove,\InstrPRFstLi}$.
The register pressure (called \emph{register need} by Kessler) up to
each node is kept during the search.
A key improvement comes from the realization that nodes with the same
sets of instructions can be merged in a dynamic programming fashion by
simply selecting the one with lowest register pressure.
This is possible since the particular order in which earlier
instructions are scheduled to arrive at a certain search node does not
need to be known to compute the optimal solution.
For example, the only information that must be kept at search node
$\set{\InstrPRFstMove,\InstrPRSndMove,\InstrPRSndLi}$ is that the
lowest register pressure in any schedule including the predecessors of
$\set{\InstrPRFstMove,\InstrPRSndMove,\InstrPRSndLi}$ is three
regardless of their particular order.
Kessler experiments with both random and real basic blocks from
Livermore~\cite{McMahon1986} and Perfect Club~\cite{Berry1988}
benchmarks.
The instruction ordering approach manages to produce optimal solutions
with a minimum register pressure of up to 13 simultaneously live
temporaries for basic blocks of up to 51 instructions.
An extension with long instruction latencies and any combination of
makespan and register pressure minimization objectives is also
proposed.
Kessler's experiments show that handling multiple objectives (where
makespan minimization is set to be the primary one) limits the
scalability of the approach to basic blocks of up to 25 instructions.

Govindarajan~\etal{} introduce the first IP model for register
pressure-aware instruction ordering~\cite{Govindarajan2003}.
The model is conceived to study the quality of a heuristic approach
for spill reduction in out-of-order processors, hence sophisticated
processor models for instruction-level parallelism are not considered.
The IP model can be seen as a register pressure-aware extension of
Arya's with \var{s}{i} variables ($f_i$ in the original notation)
representing the issue cycle of each instruction~$i$.
Besides variables and constraints as in Arya's approach (in its turn
an application of Manne's RCPSP model~\cite{Manne1960}), the model
includes \emph{live range variables} \var{l}{t,k} ($s_{t,k}$ in the
original notation) to indicate whether temporary $t$ is live in cycle
$k$ and constraints to enforce this if $\noMathVar{s}{d} \le k$ and
$\noMathVar{s}{u} \ge k$ for the definer $d$ and some user $u$ of $t$.
The objective function minimizes register pressure, that is the
maximum sum of live temporaries among all cycles.
Govindarajan~\etal{} evaluate the quality of their heuristic approach
by comparing with the results of the IP approach on 675 basic blocks
from SPEC92, Linpack~\cite{Dongarra2003},
Livermore~\cite{McMahon1986}, and NAS~\cite{Bailey1985} benchmarks.
Given the low scalability of their IP approach, only basic blocks of
limited size (averaging 19 instructions) are considered.
The comparison shows that Govindarajan~\etal{}'s heuristic approach is
near-optimal for these basic blocks: the IP approach only manages to
slightly improve six of them.
An open question is whether this conclusion can be extended to larger
basic blocks: related
studies~\cite{Wilken2000,Malik2008,Eriksson2011,Shobaki2013} suggest
that the quality gap between heuristic and combinatorial approaches
grows with the size of the problems.
Govindarajan~\etal{} also present a thorough evaluation of the impact
of their heuristic approach on the out-of-order MIPS R10000
processor~\cite{Yeager1996} for the SPEC95 floating-point benchmarks.
Their experiments show that register pressure-aware scheduling reduces
significantly the number of executed stores and loads (by 7\% and 11\%
on a processor configuration with 16 and~32 floating-point registers),
and that this reduction yields a more modest speedup of around 3\% and
4\%.
An open question is whether a higher speedup could be achieved by
balancing register pressure minimization with instruction-level
parallelism on a more refined processor model.

\begin{table}[b]%
  \centering
  \newcommand{\minimizeGoal}{min.}
  \setlength{\tabcolsep}{3.5pt}
  \caption{Objectives of local register pressure-aware instruction scheduling.
  \label{tab:rp-vs-makespan}}
  \begin{tabu}{lcc}
    \hline%
    \textbf{approach} & \textbf{reg.~pressure} & \textbf{makespan}\\\hline
    Kessler~1998 & \minimizeGoal{} & \minimizeGoal{}\\
    Govindarajan~\etal{}~2003 & \minimizeGoal{} & - \\
    Malik~2008 & no excess & \minimizeGoal{}\\
    Shobaki~\etal{}~2013 & min.~excess & \minimizeGoal{}\\
  \end{tabu}
\end{table}

Govindarajan~\etal{} introduce and exploit the notion of
\emph{lineages} (similar to \emph{reuse
  graphs}~\cite{Berson1996,Touati2002}) in their heuristic approach
for register pressure-aware instruction scheduling.
A \emph{lineage} is a path of data-dependent instructions in a basic
block (for example
$\sequence{\InstrPRSndMove,\InstrPRMac,\InstrPRBge}$ in
Figure~\ref{fig:sum-int-rp-example}).
Lineages have the interesting property that all temporaries that
connect them can reuse the same register.
Multiple lineages can be fused using artificial data dependencies to
control the register pressure of a basic block, at the expense of
limiting instruction-level parallelism.
Malik's doctoral dissertation~\cite[Chapter 5]{Malik2008c} explores
the use of lineages in a combinatorial approach to find
minimum-makespan schedules without \emph{register pressure excess}
(that is, not exceeding a given register pressure), as opposed to
earlier approaches where register pressure is minimized as shown in
Table~\ref{tab:rp-vs-makespan}.
%,
Just eliminating register pressure excess assumes that, once the
register pressure is equal to the number of processor registers,
minimizing it further might only restrict instruction-level
parallelism without providing any additional benefit.
Since Malik's approach forbids register pressure excess, by
construction it cannot handle basic blocks whose register pressure
necessarily exceeds the number of processor registers.
The dissertation shows that the DG transformations of Heffernan and
Wilken~\cite{Heffernan2006} do not increase register pressure and that
finding an instruction ordering with minimum register pressure is
equivalent to finding a minimum cover of the DG by fusing lineages
(where only the last instruction of a lineage is allowed to overlap
with another lineage).
Based on these insights, Malik proposes an approach that enumerates
candidate fusions of lineages covering the DG and solves a
minimum-makespan instruction scheduling problem for each candidate (as
in earlier work by Malik~\etal{}), iterating until the optimal
solution is found.
Malik's experiments with the same basic blocks and processor model as
in his earlier work show that the overhead of addressing register
pressure is significant: the approach can scale up to basic blocks of
up to 50 instructions, two orders of magnitude less than the
underlying CP approach for regular instruction scheduling.
Comparing to Govindarajan~\etal{}'s heuristic, Malik shows that
reformulating the register pressure-minimization problem as a
makespan-minimization problem under register pressure constraints can
improve the makespan of up to 8\% of all basic blocks.

The latest register pressure-aware instruction scheduling approach is
due to Shobaki~\etal{}~\cite{Shobaki2013}.
The approach is based on a special-purpose enumeration technique
developed previously by Shobaki and Wilken for regional instruction
scheduling (see Section~\ref{sec:regional-instruction-scheduling})
that is extended to handle register pressure.
Its major contribution is to improve the scalability of its related
approaches by one order of magnitude (see
Table~\ref{tab:rp-instruction-scheduling}).
However, this comparison is complicated since each approach optimizes
according to different objectives: for Shobaki~\etal{}, the objective
is to minimize a combination of register pressure excess (that is,
likewise Malik's approach only excessive register pressure is seen as
undesirable) and makespan (see Table~\ref{tab:rp-vs-makespan}).
As in previous approaches targeting out-of-order processors, the
processor model is relatively simple, although long instruction
latencies are considered for makespan minimization.
The enumeration technique extends Shobaki and Wilken's depth-first
search, constraint propagation-based technique (see
Section~\ref{sec:regional-instruction-scheduling}) with register
pressure-awareness.
In particular, an additional propagation algorithm is introduced that
evaluates the register pressure of a partial schedule to discard
subtrees that cannot improve the best schedule found so far.
Shobaki~\etal{} are the first to experiment with a combinatorial
instruction scheduling approach for Intel's x86, the most prominent
out-of-order processor in the last two decades.
Two configurations are considered: x86-32 which has very few registers
and x86-64 which includes an additional set of registers.
Their experiments use the functions that take most execution time for
each SPEC2006 benchmark, and use the register pressure-aware
instruction scheduling heuristics in LLVM~2.9~\cite{Lattner2004} as a
baseline.
The results show that optimal solutions can be obtained for
medium-sized basic blocks of hundreds of instructions.
The runtime speedup results are consistent with those of
Govindarajan~\etal{}~who also compare to a heuristic approach on an
out-out-order processor: a modest speedup of 2.4\% is obtained for
the floating-point benchmarks on x86-64 while the speedup for the
integer benchmarks is nearly negligible.
Additionally, Shobaki~\etal{}~find that balancing register pressure
with makespan minimization has indeed some impact.
In particular, improving the x86 processor model with more precise
latencies~\cite{Fog2017} gives an additional~0.4\% speedup.

\subsection{Regional RP Instruction Scheduling}
\label{sec:regional-rp-instruction-scheduling}

\paragraph{Software pipelining}

Govindarajan \etal{} present an alternative version of their basic
software pipelining approach
(Section~\ref{sec:regional-instruction-scheduling}) that minimizes
register pressure as a secondary goal~\cite{Govindarajan1994b} instead
of focusing on rich resource modeling.
The approach approximates register pressure by the amount of register
\emph{buffers} needed to store the values defined by each instruction
in multiple iterations.
Their experiments on 27 loops from Livermore, SPEC,
Linpack~\cite{Dongarra2003}, and Whetstone benchmarks for different
configurations of a multiple-issue processor show that the optimal
solution can be found in less than 5 seconds for 87\% of the loops,
and that in most cases the solutions found improve those of
alternative heuristic approaches in either initiation interval, amount
of register buffers, or both.
Although a direct comparison is not available, the reported solving
times suggest that the register pressure-aware approach is slightly
less scalable than its resource usage minimization counterpart.

Eichenberger and Davidson reformulate the dependency constraints in
the basic IP model of Govindarajan \etal{} to ease
solving~\cite{Eichenberger1997}.
The reformulation, based on earlier insights on the structure of IP
scheduling models~\cite{Chaudhuri1994}, makes the dependency
constraints \mbox{\emph{0-1 structured}}: each variable appears at
most once and has only unit coefficients (see
Table~\ref{tab:techniques}).
This property improves the linear relaxation of the problem and thus
reduces the amount of branch-and-bound search needed to solve it.
Additionally, Eichenberger \etal{} propose a secondary objective
function~\cite{Eichenberger1995} that captures register pressure
minimization more accurately than Govindarajan \etal{}
Eichenberger and Davidson experiment with 1327 loops from the Perfect
Club~\cite{Berry1988}, SPEC89, and Livermore benchmarks scheduled for
the Cydra 5 processor~\cite{Beck1993} under different secondary
objective functions.
The results show that the reformulation can reduce solving time by
almost an order of magnitude and almost double the number of
instructions that can be scheduled optimally in a given time budget,
and that using the more accurate register pressure objective function
instead of the register buffer minimization by Govindarajan \etal{}
does not increase solving time.

Dupont De Dinechin introduces an alternative IP model that minimizes
register pressure as accurately as Eichenberger \etal{} with a simpler
structure but a potentially larger number of
variables~\cite{DeDinechin2004}.
Unlike earlier IP approaches to software pipelining, Dupont De
Dinechin's model has a single type of 0-1 \emph{scheduling variables}
\var{x}{i,k} indicating whether instruction $i$ is scheduled in cycle
$k$ relative to the initial loop iteration.
For example, Figure~\ref{fig:swp-output} corresponds to a solution
where \var{x}{\InstrLoad{},0}, \var{x}{\InstrAdd{},2}, and
\var{x}{\InstrInc{},1} are set to one and the remaining \var{x}{i,k}
are set to zero.
Compared to the earlier approaches, Dupont De Dinechin's model
features a better linear relaxation of the dependency constraints and
is more amenable to heuristic
approaches~\cite{Blachot2006,DeDinechin2007} but suffers from a
potentially larger number of variables, since the time horizon for the
$x$ variables is longer than for the former $a$ variables.
An implementation of Dupont De Dinechin's approach is available in the
LAO code generator~\cite{DeDinechin2000} which is a part of
STMicroelectronics' ST200~\cite{Faraboschi2000} production compiler.

\paragraph{Loop unrolling}

A simpler alternative with a similar goal to software pipelining is
\emph{loop unrolling}.
Loop unrolling duplicates $n$ iterations of a loop body ($n$ is often
called the \emph{unrolling factor}) to increase the available
instruction-level parallelism and reduce the overhead imposed by
control instructions.
Similarly to software pipelining, loop unrolling tends to increase
register pressure and processor resource usage.

Domaga\l{}a \etal{} present a register pressure-aware CP approach to
integrated loop unrolling and instruction ordering where the loop body
can be partitioned into \emph{tiles} with different unrolling
factors~\cite{Domagala2016}.
The approach proposes a sophisticated CP model to determine the order
of each instruction, the partitioning of instructions into tiles, and
the unrolling factor of each tile.
The model follows Malik's local approach~\cite{Malik2008c} in that it
disallows excessive register pressure rather than minimizing it.
The objective in Domaga\l{}a \etal{}'s approach is to minimize the
number of spills per iteration due to the structure of the resulting
unrolling and tiling.
Domaga\l{}a \etal{}'s experiments show that about half of the
high-register-pressure loops from the SPEC benchmarks can be solved
optimally and that the overall number of spills is indeed reduced even
though loop unrolling is performed.

\subsection{Global RP Instruction Scheduling}
\label{sec:global-rp-instruction-scheduling}

\paragraph{Barany and Krall}

While multiple combinatorial approaches to local and regional register
pressure-aware instruction scheduling have been proposed, Barany and
Krall are the first to propose a combinatorial approach that addresses
the problem globally~\cite{Barany2013}.
The strategy differs significantly from previous approaches: rather
than directly finding a schedule that minimizes register pressure (or
a combination with makespan minimization), Barany and Krall use IP to
find code motion restrictions that might reduce register pressure by
allowing pairs of temporaries to reuse registers.
A code motion restriction is a set of conditions that allows two
temporaries to reuse the same register, and it can be seen as a
generalization of Govindarajan~\etal{}'s notion of \emph{lineage
  fusion} explained in
Section~\ref{sec:local-rp-instruction-scheduling}.
Pairs of temporaries that might reuse the same register and their
associated code motion restrictions are called \emph{reuse candidates}
and \emph{blames} by Barany and Krall.
Because the IP model does not have a global view of register pressure,
the choice of restrictions that indeed reduce register pressure is
delegated to a register allocator.
After enforcing such restrictions, aggressive code motion can finally
run with a reduced risk of increasing register pressure.

The IP model defining the set of code motion restrictions has three
main types of variables:
\begin{inparaitem}[]
\item \emph{restriction selection variables} \var{select}{r} to
  indicate whether code motion restriction $r$ is selected,
\item \emph{instruction placement variables} \var{place}{i,b} to
  indicate whether instruction $i$ can be placed in basic block $b$ by
  the selected restrictions, and
\item \emph{instruction linearization variables} \var{s}{i}
  (\var{instr}{i} in the original notation) representing the order of
  instruction $i$ within a linearization of the function's
  instructions which Barany and Krall find necessary to avoid circular
  data and control dependencies.
\end{inparaitem}
The linearization can be seen as a global instruction ordering.
The model includes linear constraints to ensure that:
\begin{inparaitem}[]
\item data and control dependencies are reflected in the instruction
  linearization,
\item all instructions can be placed in at least one basic block,
\item instruction placements and linearizations are compatible with
  the selected restrictions, and
\item the linearization of dependent instructions is compatible with
  their allowed placements.
\end{inparaitem}
The objective function maximizes primarily the potential of the
selected restrictions to decrease spill cost (by letting their
corresponding temporaries reuse registers) and secondarily the number
of allowed instruction placements (to preserve as much code motion
freedom as possible).
Unlike other instruction scheduling approaches, a solution does not
directly correspond to a valid schedule but gives instead a set of
code motion restrictions that might be useful to avoid costly spills.
The final choice of restrictions is left to a heuristic PBQP register
allocator (see Section~\ref{sec:basic-approaches}) where a restriction
is chosen if its associated temporaries are assigned to the same
register.
A small cost is associated with the choice to prevent unnecessary
restrictions.

Barany and Krall use LLVM's~\cite{Lattner2004} register pressure-aware
heuristic code motion and instruction scheduling algorithms as a
baseline, and evaluate dynamically their approach with SPEC2000
functions of up to 1000 instructions on an out-of-order ARM
Cortex-A9~\cite{ARM} processor.
The results show that their approach does not have a notable impact on
the execution time of the generated code: the code generated by LLVM
is actually slightly slowed down (by 0.4\%).
Reducing the scope of the restrictions to single basic blocks improves
to some degree the results, giving a speedup of~0.5\% over LLVM.
Barany and Krall conclude that the approach should be refined to
capture the trade-offs between spilling and restricting certain code
motions more accurately.
An open question is whether this refinement is possible without a
tighter integration between instruction scheduling and register
allocation.

\section{Integrated Code Generation}\label{app:full-integration}

\begin{figure}
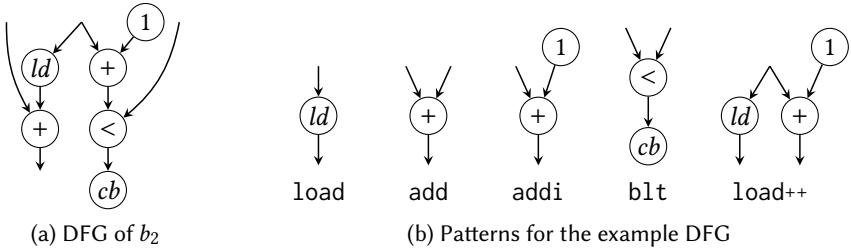

  \hfill
  \subfloat[DFG of $\B{2}$]{%
    \scalebox{1.0}{\input{./figures/styles}%

\begin{tikzpicture}

    \coordinate (t1source);
    \coordinate [left=of t1source] (t3source);
    \coordinate [right=1.3cm of t1source] (t4source);
    \node [pra storage, right=0.6cm of t1source] (one) {$1$};
    \coordinate [below=0.6cm of t1source] (ldmid);
    \node [pra storage, left= of ldmid] (ld) {$\textrm{{\emph{ld}}}$};
    \node [pra storage, right=of ldmid, xshift=-0.2cm] (inc) {$+$};
    \node [pra storage, below=of ld] (add) {$+$};
    \node [pra storage, below=of inc] (lt) {$<$};
    \node [pra storage, below=of lt] (cb) {$\textrm{\emph{cb}}$};
    \coordinate [below=0.3cm of add] (addsink);

  \begin{scope}[flow, pra label]
    \draw (t1source) -- (ld);
    \draw (t1source) -- (inc);
    \draw (one) -- (inc);
    \draw (t3source) to [bend right=15] (add);
    \draw (t4source) to [bend left=20] (lt);
    \draw (ld) -- (add);
    \draw (inc) -- (lt);
    \draw (lt) -- (cb);
    \draw (add) -- (addsink);
  \end{scope}

\end{tikzpicture}}
    \label{fig:dfg}
  }
  \hfill
  \subfloat[Patterns for the example DFG]{%
   \scalebox{1.0}{\input{./figures/styles}%

\begin{tikzpicture}

    \node (load) {$\loadInstruction{}$};
    \coordinate [above=0.1cm of load] (loadout);
    \node [pra storage, above=0.4cm of loadout] (ld) {$\textrm{{\emph{ld}}}$};
    \coordinate [above=0.4cm of ld] (loadin);
  \begin{scope}[flow, pra label]
    \draw (loadin) -- (ld);
    \draw (ld) -- (loadout);
  \end{scope}

    \node [right=0.6cm of load] (add) {$\addInstruction{}$};
    \coordinate [above=0.1cm of add] (addout);
    \node [pra storage, above=0.4cm of addout] (plus) {$+$};
    \coordinate [above=0.4cm of plus, xshift=-0.3cm] (addinleft);
    \coordinate [above=0.4cm of plus, xshift=0.3cm] (addinright);
  \begin{scope}[flow, pra label]
    \draw (addinleft) -- (plus);
    \draw (addinright) -- (plus);
    \draw (plus) -- (addout);
  \end{scope}

    \node [right=0.6cm of add] (addi) {$\addiInstruction{}$};
    \coordinate [above=0.1cm of addi] (addiout);
    \node [pra storage, above=0.4cm of addiout] (plusi) {$+$};
    \coordinate [above=0.4cm of plusi, xshift=-0.3cm] (addiinleft);
    \node [pra storage, above=0.4cm of plusi, xshift=0.3cm] (one) {$1$};
  \begin{scope}[flow, pra label]
    \draw (addiinleft) -- (plusi);
    \draw (one) -- (plusi);
    \draw (plusi) -- (addiout);
  \end{scope}

    \node [right=0.6cm of addi] (blt) {$\jumpIfLTInstruction{}$};
    \node [pra storage, above=0.1cm of blt] (cb) {$\textrm{\emph{cb}}$};
    \node [pra storage, above=0.4cm of cb] (lt) {$<$};
    \coordinate [above=0.4cm of lt, xshift=-0.3cm] (ltinleft);
    \coordinate [above=0.4cm of lt, xshift=0.3cm] (ltinright);
  \begin{scope}[flow, pra label]
    \draw (ltinleft) -- (lt);
    \draw (ltinright) -- (lt);
    \draw (lt) -- (cb);
  \end{scope}

    \node [right=0.6cm of blt] (load++) {$\loadPPInstruction{}$};
    \coordinate [above=0.1cm of load++] (load++out);
    \coordinate [left=0.4cm of load++out] (load++outleft);
    \coordinate [right=0.4cm of load++out] (load++outright);
    \node [pra storage, above=0.4cm of load++outleft] (ld++) {$\textrm{{\emph{ld}}}$};
    \node [pra storage, above=0.4cm of load++outright] (addi++) {$+$};
    \coordinate [above=0.4cm of ld++, xshift=0.4cm] (load++in);
    \node [pra storage, above=0.4cm of addi++, xshift=0.4cm] (one++) {$1$};
  \begin{scope}[flow, pra label]
    \draw (load++in) -- (ld++);
    \draw (load++in) -- (addi++);
    \draw (one++) -- (addi++);
    \draw (ld++) -- (load++outleft);
    \draw (addi++) -- (load++outright);
  \end{scope}

\end{tikzpicture}}
   \label{fig:patterns}
  }
  \hfill\hspace{0.2cm}
  \caption{Input to integrated code generation.}
  \label{fig:isel-input}
\end{figure}

This appendix reviews approaches that tackle one of the grand
challenges of combinatorial code generation: the integration of
instruction selection with combinatorial register allocation and
instruction scheduling.
Instruction selection transforms a program represented with
intermediate operations into an equivalent one that uses instructions
of a certain processor.
Thus, the input program is represented at a higher level of
abstraction, in terms of a \emph{data-flow graph} (DFG) with
intermediate \emph{operations} and \emph{data edges} (\emph{edges} for
short).
Figure~\ref{fig:dfg} shows the DFG corresponding to basic
block~$\B{2}$ in the running example.
Operations $\textrm{\emph{ld}}$ and $\textrm{\emph{cb}}$ represent a
load and a conditional branch.
The immediate value $1$ is represented as an operation with a single
outgoing edge.

Instruction selection involves selecting among different
\emph{instruction patterns} (\emph{patterns} for short) that can
\emph{cover} one or more nodes and edges of the
DFG~\cite{HjortBlindell2016}.
This problem interacts closely with register allocation and
instruction scheduling, since the selection of processor instructions
imposes constraints on the processor resources and register banks that
are used.
Figure~\ref{fig:patterns} shows some patterns corresponding to
MIPS32-like instructions.
The pattern $\loadPPInstruction{}$ corresponds to a load instruction
that increments the address after performing the memory access.
Selecting the $\loadInstruction{}$, $\addInstruction{}$,
$\addiInstruction{}$, and $\jumpIfLTInstruction{}$ patterns results in
the basic block $\B{2}$ shown in Figure~\ref{fig:sum-cfg} on
page~\pageref{fig:sum-cfg}, while selecting the $\loadPPInstruction{}$
pattern removes the need for the $\addiInstruction{}$ instruction.

\paragraph{Outline}

\begin{table}[b]%
  \newcommand{\integratedApproachRA}[9]{#7 & #6 & #2 & #1 & #3 & #8 & #4 & #5 & #9}
  \newcommand{\integratedApproachIS}[5]{#1 & #3 & #4 & #5}
  \renewcommand{\columnWidth}{0.51cm}
  \setlength{\tabcolsep}{2pt}
  \setlength{\tabulinesep}{0.03cm}
  \caption{Integrated code generation approaches: technique (TC), scope
    (SC), spilling (SP), register assignment~(RA), coalescing (CO),
    load-store optimization (LO), register packing (RP), live-range
    splitting (LS), rematerialization (RM), multiple register banks
    (MB), multiple allocation (MA), bundling (BD), multiple usage~(MU),
    two-dimensional usage (2D), asynchronous usage (AS),
    size of largest problem solved
    optimally (SZ) in number of instructions, and whether a dynamic evaluation is available (DE).\label{tab:fully-integrated}}
    \begin{tabu}{%
        l%
        c%
        c%
        >{\centering\arraybackslash}p{\columnWidth}%
        >{\centering\arraybackslash}p{\columnWidth}%
        >{\centering\arraybackslash}p{\columnWidth}%
        >{\centering\arraybackslash}p{\columnWidth}%
        >{\centering\arraybackslash}p{\columnWidth}%
        >{\centering\arraybackslash}p{\columnWidth}%
        >{\centering\arraybackslash}p{\columnWidth}%
        >{\centering\arraybackslash}p{\columnWidth}%
        >{\centering\arraybackslash}p{\columnWidth}%
        >{\centering\arraybackslash}p{\columnWidth}%
        >{\centering\arraybackslash}p{\columnWidth}%
        >{\centering\arraybackslash}p{\columnWidth}%
        >{\centering\arraybackslash}p{\columnWidth}%
        c%
        >{\centering\arraybackslash}p{\columnWidth}}\hline
      \rowfont{\bfseries}
      approach & TC & SC & \integratedApproachRA{LO}{CO}{RP}{RM}{MB}{RA}{SP}{LS}{MA} & \integratedApproachIS{BD}{MC}{MU}{2D}{AS} & SZ & DE \\
      \hline
      Wilson~1994 & IP & global & \integratedApproachRA{\no}{\yes}{\no}{\no}{\no}{\yes}{\yes}{\yes}{\no} & \integratedApproachIS{\yes}{\no}{\yes}{\no}{\no} & 30 & \no \\
      \rowcolor{tblrow} Gebotys 1997 & IP & local & \integratedApproachRA{\yes}{\no}{\no}{\no}{\yes}{\yes}{\yes}{\yes}{\no} & \integratedApproachIS{\no}{\no}{\yes}{\no}{\no} & 108 & \no \\
      ICG & CP & local & \integratedApproachRA{\yes}{\yes}{\no}{\no}{\yes}{\yes}{\yes}{\yes}{\no} & \integratedApproachIS{\yes}{\no}{\no}{\no}{\no} & 23 & \no \\
      \rowcolor{tblrow} OPTIMIST & IP & sw.~pip. & \integratedApproachRA{\yes}{\no}{\no}{\no}{\yes}{\no}{\yes}{\yes}{\yes} & \integratedApproachIS{\yes}{\no}{\yes}{\yes}{\yes} & 100 & \no \\
    \end{tabu}%
\end{table}

Table~\ref{tab:fully-integrated} classifies integrated code generation
approaches with information about their optimization technique, scope,
problem coverage, approximate scalability, and evaluation method.
The rest of the appendix discuss these approaches in further detail.

\paragraph{Wilson et al}

The first combinatorial approach to integrated code generation
is proposed by Wilson~\etal{}~\cite{Wilson1994,Wilson2002}.
The model is designed with the goal of generating code for DSPs with
very limited and irregular register banks such as Motorola's
DSP56k~\cite{Kloker1987} processor.
Remarkably, the IP model captures more register allocation subproblems
than many approaches without instruction selection.
Unfortunately, experimental results are not publicly available, which
has limited the impact of the approach.

Wilson~\etal{}'s IP model is rather compact in relation to the amount
of captured subproblems.
Both operations as well as edges of the DFG can be deactivated by the
solver to model internal operations of complex processor instructions,
optional spilling, and alternative implementations of certain
operations such as computations of memory addresses.
Unlike most studied approaches, the register allocation submodel is
based on edges rather than temporaries, which increases the model
granularity since multiple uses of the same temporary yield different
edges.
Similarly to UNISON (see
Section~\ref{sec:unison}),
Wilson~\etal{} extend the input program with optional stores, loads,
and register-to-register moves~\cite{Wilson2002}.

Similarly to Arya (Section~\ref{sec:local-instruction-scheduling}),
the model uses general integer variables \var{c}{o} ($y_o$ in the
original notation) representing the issue cycle of each operation $o$,
which eases modeling dependency constraints but limits the generality
of the resource model.
The model is completed with the following types of 0-1 variables:
\begin{inparaitem}[]
\item \emph{pattern selection variables} \var{s}{p} indicate whether
  pattern $p$ is selected,
\item \emph{operation and edge activation variables} \var{a}{o} and
  \var{a}{e} indicate whether operation $o$ and edge $e$ are active,
  and
\item \emph{register assignment variables} \var{a}{e,r}
  indicate whether edge $e$ is assigned to register $r$.
\end{inparaitem}
These variables correspond to $z_p$, $z_o$, $x_{ij}$, and $u_{ijr}$
in the original notation, where $e = \sequence{i,j}$.
Linear constraints enforce that:
\begin{inparaitem}[]
\item pattern coverings do not overlap,
\item edges and operations are active iff they are not internalized by
  a pattern covering,
\item active edges are assigned to registers,
\item edges related across basic blocks are assigned to the same
  register (similarly to UNISON),
\item operations covered by the same pattern are scheduled in
  parallel,
\item the definers of active edges are scheduled before their users,
\item active operations that use the same resource are not scheduled
  in parallel, and
\item the definer and users of active edges that are assigned to the
  same register are scheduled such that the live ranges do not
  overlap.
\end{inparaitem}
Additional constraints make it possible to model alternative
implementations such that only one out of a group of operations
appears in the generated code.
For example, a solution for the basic block from Figure~\ref{fig:dfg}
where the pattern $\loadPPInstruction{}$ is selected is represented by
setting the variables \var{s}{\addInstruction{}},
\var{s}{\jumpIfLTInstruction{}}, and \var{s}{\loadPPInstruction{}} to
one.
This forces the remaining \var{s}{p} variables to zero since no
pattern covering is allowed to overlap.
In this solution, operations $\textrm{{\emph{ld}}}$, $1$, and $+$ are
scheduled in parallel since they are covered by the
$\loadPPInstruction{}$ pattern, which also deactivates the
internalized edge from $1$ to $+$.
The objective function minimizes the maximum makespan across all basic
blocks.

Unfortunately, Wilson~\etal{} do not report detailed experimental
results for their integrated code generation approach.
The discussion included in the paper indicates that the approach is
able to generate optimal solutions for up to 30 operations, of the
same quality as hand-optimized code.
Wilson~\etal{} propose some techniques to scale up the approach,
including taking preliminary decisions heuristically, and decomposing
functions into regions of basic blocks to be solved
incrementally~\cite{Wilson2002}.

\paragraph{Gebotys}

An alternative IP approach is due to Gebotys~\cite{Gebotys1997}.
The approach solves \emph{instruction compaction} as opposed to full
instruction scheduling.
Instruction compaction assumes a given fixed sequence of instructions
and decides, for each instruction $i$ in the sequence whether $i$ is
scheduled in parallel with its successor.
For example, instruction compaction for $\B{2}$ in
Figure~\ref{fig:sum-cfg} assumes the sequence
$\sequence{\InstrLoad{},\InstrAdd{},\InstrInc{},\InstrBlt{}}$ and
decides whether $\InstrAdd{}$ is scheduled in parallel with
$\InstrInc{}$ (the other compaction decisions are precluded by
dependencies).
Solving only compaction reduces the search space dramatically but
limits the capacity to exploit instruction-level parallelism.

The model proposed by Gebotys is composed mostly of linear constraints
derived from \emph{Horn clauses}.
A Horn clause is a logical formula with at most one
conclusion~\cite{Horn1951}.
The linear relaxation of an IP problem with only linear constraints
derived from Horn clauses is the optimal solution to the IP problem.
Even IP models not only composed of Horn constraints (as in Gebotys'
model) become generally easier to solve, although they might still
resort to branch-and-bound search.

Gebotys' model contains two types of 0-1 variables:
\begin{inparaitem}[]
\item \emph{pattern covering variables} \var{c}{o,p} ($x_{o,p}$ in the
  original notation) indicate whether pattern $p$ covers operation
  $o$, and
\item \emph{compaction variables} \var{m}{o} ($p_o$ in the original
  notation) indicate whether operation $o$ is compacted with its
  successor $o'$ in the input instruction sequence, due to the same
  pattern covering both $o$ and $o'$\!.
\end{inparaitem}
For example, if the pattern $\loadPPInstruction{}$ is selected for
Figure~\ref{fig:dfg} then operations $\textrm{{\emph{ld}}}$, $1$, and
$+$ are compacted similarly to Wilson~\etal{}
Instead of using register assignment variables as in Wilson~\etal{},
Gebotys' model includes patterns for each different flow that a data
edge could follow through registers and memory.
Thus, selecting a pattern covering implicitly decides the register
assignment and possible spilling of the covered edges.
For example, the data edge between operations $+$ and $<$ in
Figure~\ref{fig:dfg} could follow the flow $\register{R1} \rightarrow
\register{R1}$ (defined and used directly from $\register{R1}$),
$\register{R1} \rightarrow \register{R2}$ (moved from $\register{R1}$
to $\register{R2}$), $\register{R1} \rightarrow \register{M}
\rightarrow \register{R1}$ (defined in $\register{R1}$, spilled, and
loaded into $\register{R1}$), etc.
This representation assumes that the target processor has a very small
register file, as the number of patterns would otherwise explode.
The model includes non-Horn constraints to enforce that pattern
coverings do not overlap, and Horn constraints to enforce that:
\begin{inparaitem}[]
\item coverings are selected that assign interfering edges to
  different registers, and
\item operations are compacted whenever their edge assignments are
  compatible.
\end{inparaitem}
The objective function is a weighted sum of three different costs:
makespan, code size, and energy consumption (approximated as the
number of memory accesses of the generated code).

Gebotys presents experimental results for six single-block signal
processing examples and Texas Instrument's (TI) TMS320C2x~\cite{C2x}
-- a simple, single-issue DSP with very few registers.
In the experiments, the objective function is limited to minimizing
the makespan, which in the case of TMS320C2x is directly proportional
to code size.
The results show a makespan improvement from~9\% of up to 118\%
relative to the code generated by TI's C compiler.
Interestingly, the six basic blocks can be solved optimally from the
first linear relaxation, without resorting to branch-and-bound search.
Gebotys also experiments with extending the model with complete
instruction scheduling by adding a time dimension to the pattern
selection variables, and observes a dramatic increase of the solving
time due to an explosion of variables in the model and the
insufficiency of the initial linear relaxations to deliver the optimal
solution.
Since TMS320C2x is a single-issue processor, the extended model does
not yield better solutions to any of the six single-block signal
processing examples.

\paragraph{ICG}

Bashford and Leupers propose \emph{ICG}, a code generator that
integrates instruction selection, register allocation and instruction
scheduling using an approach that is partially based on
CP~\cite{Bashford1999}.
Unlike the other approaches, ICG decomposes the solving process into
several stages, sacrificing global optimality in favor of solving
speed.
In Bashford and Leupers's paper, ICG is described as a procedural code
generator with limited search capabilities.
However, the model is presented in this survey as a monolithic CP
model for clarity and consistency.

The ICG model has the following types of variables:
\begin{inparaitem}[]
\item \emph{pattern covering variables} \var{v}{o} give the pattern
  that covers operation $o$,
\item \emph{register variables} \var{r}{o, op} give the register to
  which each use and definition operand $op$ of operation $o$ is
  assigned (register assignment in ICG is operand-based),
\item \emph{bundle type variables} \var{t}{o} give the type of
  bundle to which operation $o$ belongs,
\item \emph{functional unit variables} \var{f}{o} give the
  functional unit used by operation $o$, and
\item \emph{operation cost variables} \var{c}{o} give the number of
  issue cycles required to execute operation $o$.
\end{inparaitem}
These variables correspond to $frt_o$, $U_{o, op}$, and a triple
$\sequence{t, f, c}_o$ in the original notation.
The register assignment variables include special registers for memory
banks and \emph{internalized} operands covered by multi-operation
patterns (corresponding to Wilson~\etal{}'s deactivated edges).
The model lacks scheduling variables as instruction scheduling runs
in a procedural fashion~\cite[Section 7]{Bashford1999}.
The model contains constraints to ensure that:
\begin{inparaitem}[]
\item the registers assigned to operands of different operations are
  compatible;
\item operations can only be bundled together if they have the same
  bundle type and do not use the same functional unit; and
\item specific instruction constraints relating register assignments,
  bundle types, functional units and costs within an operation are
  satisfied.
\end{inparaitem}
The remaining usual constraints such as precedences among operations
or non-interference among temporaries are handled by procedural
algorithms with a limited form of search.

The solving procedure is split into three steps:
\begin{inparaitem}[]
\item first, a constraint problem is formulated and initial constraint
  propagation is performed to discard invalid registers for operands;
\item then, a procedural algorithm schedules operations and generates
  spill code. The latter is performed \emph{on-demand} by extending
  the problem with new variables and constraints corresponding to
  load, store, and register-to-register instructions. A certain level
  of search is permitted within each iteration, and reparation (where
  the problem is transformed into an easier one by means of, for
  example, spilling) is used as a last resort to handle search
  failures. The (implicit) goals are to minimize makespan and
  spill code.
\item Finally, search is applied to assign values to all variables
  that have not been decided yet.
\end{inparaitem}
A post-processing step is run to handle memory address computations.

Bashford and Leupers evaluate ICG on four basic blocks from
DSPstone~\cite{Vojin1994} benchmarks for the
ADSP-210x~\cite{SH} processor.
The results show that the generated code has as short makespan as
hand-optimized code, and noticeably shorter than traditionally
generated code.
However, the scalability of ICG seems limited as the time to generate
code for a small basic block of 17~operations grows more than three
times when the size is slightly increased to 23~operations.

\paragraph{OPTIMIST}

The most recent approach to integrated code generation is
\emph{OPTIMIST}~\cite{Eriksson2012}.
This project has explored the application of several search techniques
for local code generation, including dynamic
programming~\cite{Kessler2001,Kessler2006} and genetic
algorithms~\cite{Eriksson2008}.
However, IP has established itself as the technique of choice within
OPTIMIST as a natural yet formal method to extend the code generation
model to more complex processors~\cite{Eriksson2008} and software
pipelining~\cite{Eriksson2009,Eriksson2012}.

Compared to previous integrated approaches, the OPTIMIST IP model has
a rich resource model in which, similarly to the early approach of
Ertl and Krall presented in
Section~\ref{sec:local-instruction-scheduling}, the resource usage of
an instruction is specified per pipeline stage.
Another distinctive feature is that copies between different register
banks and memory are only modeled implicitly by means of variables
indicating their movement, rather than inserted in the input program
in a preprocessing stage and defined as optional but otherwise regular
operations.
Similarly to UNISON (see
Section~\ref{sec:unison}),
memory is modeled as a register bank and thus spilling is unified with
handling multiple register banks.
The model captures thus register allocation but leaves out register
assignment, which precludes coalescing and register packing as seen in
Table~\ref{tab:fully-integrated}.

OPTIMIST's IP model is described first for local integrated code
generation and extended later for software pipelining.
The model (slightly simplified for clarity) includes the following
types of 0-1 variables:
\begin{inparaitem}[]
\item \emph{pattern selection variables} \var{s}{p,k} indicate
  whether pattern $p$ is selected and issued in cycle~$k$;
\item \emph{operation and edge covering variables} \var{c}{p,o,k}
  and \var{c}{p,e,k} indicate whether pattern~$p$ covers operation
  $o$ and edge $e$ in cycle $k$;
\item \emph{register allocation variables} \var{a}{t,b,k} indicate
  whether temporary $t$ is allocated to register bank $b$ in cycle
  $k$; and
\item \emph{copy variables} \var{x}{t,s,d,k} indicate whether
  temporary $t$ is copied from source register bank $s$ to destination
  register bank $d$ in cycle $k$.
\end{inparaitem}
These variables correspond to $s_{p,k}$, $c_{o,p,k}$, $w_{i,j,p,k}$
(where $e = \sequence{i,j}$), $r_{b,t,k}$, and $x_{t,s,d,k}$ in the
original notation.
For simplicity, the original pattern node and edge dimensions in the
covering variables are omitted.
The model's instruction selection constraints ensure that all
operations are covered by exactly one pattern.
The register allocation constraints ensure that the program
temporaries:
\begin{inparaenum}
\item are assigned to the register banks supported by their defining
  and using instructions in their definition and use cycles;
  \label{con:def-use}
\item can only be allocated to a register bank $b$ in a cycle $k$ if
  they are defined in cycle~$k$, are already allocated to $b$ in cycle
  $k - 1$, or are copied from another register bank in cycle $k$;
  \label{con:continuation}
\item are not allocated to any register bank if they correspond to
  internalized edges; and
  \label{con:internal}
\item do not exceed the capacity of register banks in any cycle.
\end{inparaenum}
These constraints are related to earlier IP
models:~(\ref{con:def-use}) and~(\ref{con:continuation}) are similar
to those of Nagarakatte and Govindarajan (see
Section~\ref{sec:other-integrated})
and~(\ref{con:internal}) is similar to those in Wilson~\etal{}'s
model.
In both cases, the constraints are lifted from individual registers to
register banks.
Last, the model's instruction scheduling constraints ensure that:
\begin{inparaitem}[]
\item temporaries can only be used or copied after their definition
  cycle, and
\item resources such as functional units are not overused in any
  cycle.
\end{inparaitem}
The objective function, as is typical in local integrated approaches,
minimizes the basic block's makespan.

For example, to generate code for $\B{2}$ as in
Figure~\ref{fig:unison-sol} from the DFG shown in
Figure~\ref{fig:dfg}, OPTIMIST sets the variables
\var{s}{\loadInstruction{},0}, \var{s}{\addiInstruction{},1},
\var{s}{\addInstruction{},2}, and \var{s}{\jumpIfLTInstruction{},3} to
one and the remaining \var{s}{p,k} variables to zero.
Since all temporaries in the example are allocated to the same
register bank $\code{R32}$ (containing registers $\register{R1}$,
$\register{R2}$, \dots), OPTIMIST sets the register variables
\var{a}{t,\code{R32},k} to one for each temporary $t$ and cycle $k$
and which $t$ is live, and all copy variables to zero.

The model extension to single-block software pipelining preserves the
same variables and introduces only a few changes to handle two
fundamental characteristics of software pipelining: the existence
of loop-carried dependencies (see example in Figure~\ref{fig:dg-swp}),
and the fact that resource usage by an instruction and temporary live
ranges might overlap over time across different iterations.
As is common in combinatorial software pipelining, the initiation
interval $i$ is a fixed parameter and the IP problem is generated and
solved for increasing values of $i$ until the optimal solution is
found.

Eriksson~\etal{} present experimental results for both the local and
the software pipelining IP approaches.
The original experimental results are presented in two
papers~\cite{Eriksson2008,Eriksson2012}.
This survey presents updated results from a rerun reported in
Eriksson's doctoral dissertation~\cite{Eriksson2011} where a newer IP
solver and host architecture are used.
The local IP approach is compared to a heuristic code generator based
on genetic algorithms.
The input basic blocks are extracted from the \emph{jpeg} and
\emph{mpeg2} benchmarks in MediaBench~\cite{Lee1997}, and the targeted
processor is a slightly modified version of TI's
two-clustered TMS320C62x~\cite{C62x} processor.
The local IP approach improves the solutions of the heuristic in
around 50\% of the cases, where the largest basic block solved
optimally has 191 operations.
In the original results published three years before, the largest
basic block solved by the local IP approach had 52 operations, and
only 40\% of the basic blocks up to 191 operations could be
solved~\cite{Eriksson2008}.
This difference illustrates the speed with which solvers and
architectures evolve to enable solving increasingly harder
combinatorial problems.

Eriksson compares the fully-integrated software pipelining IP approach
to a separated approach where instruction selection is solved in
isolation prior to register allocation and instruction scheduling.
The experiments are run for the same processor as the local approach
and loop kernels from SPEC2000, SPEC2006~\cite{CPU}, MediaBench, and
FFmpeg benchmarks.
% note: the total average improvement is computed as 39% \times 20%
The results show that the integrated approach improves the initiation
interval for 39\% of the kernels, yielding a total average improvement
of around 8\%~\cite[Chapter 8]{Eriksson2011}.
The approach scales up to kernels of approximately 100
operations~\cite[Chapter 4]{Eriksson2011}, 40 operations larger
compared to the original experimental results~\cite{Eriksson2012}.

\end{document}